\documentclass[final,3p,times]{elsarticle}

\usepackage{amssymb}
\usepackage{amsmath}
\usepackage[x11names]{xcolor}
\usepackage{siunitx}
\DeclareSIUnit \pc {pc}
\DeclareSIUnit \mp {m_p}

\usepackage{tabularx}
\usepackage{cancel}

\usepackage{fancyref}
\usepackage{hyperref}
\hypersetup{colorlinks,breaklinks}
\usepackage{xspace}

\newcommand{\swift}{{\sc Swift}\xspace}
\newcommand{\woma}{{\sc WoMa}\xspace}

\newcommand{\pencil}{{\sc Pencil}\xspace}

\newcommand\nonumfootnote[1]{%
  \begingroup
  \renewcommand\thefootnote{}\footnote{#1}%
  \addtocounter{footnote}{-1}%
  \endgroup
}

\journal{Journal of Computational Physics}

\begin{document}

\begin{frontmatter}



\title{REMIX SPH -- improving mixing in smoothed particle hydrodynamics simulations using a generalised, material-independent approach}


\author[Durham]{T.~D.~Sandnes$^{*,\,}$}
\author[Durham]{V.~R.~Eke}
\author[SETI,Ames]{J.~A.~Kegerreis}
\author[Durham]{R.~J.~Massey}
\author[Durham]{S.~Ruiz-Bonilla}
\author[Lorentz,Leiden]{M.~Schaller}
\author[Oslo,Glasgow]{L.~F.~A.~Teodoro}

%\affiliation[Durham]{organization={Department of Physics, Institute for Computational Cosmology, Durham University},
\affiliation[Durham]{organization={Institute for Computational Cosmology, Department of Physics, Durham University},
             addressline={South Road},
             city={Durham},
             postcode={DH1 3LE},
             country={UK}}
 \affiliation[SETI]{organization={SETI Institute},
             addressline={339 Bernardo Avenue, Suite 200, Mountain View},
             postcode={94043},
             state={CA},
             country={USA}}
 \affiliation[Ames]{organization={NASA Ames Research Center},
             addressline={MS 245-3, Moffett Field},
             postcode={94035},
             state={CA},
             country={USA}}

\affiliation[Lorentz]{organization={Lorentz Institute for Theoretical Physics, Leiden University},
             addressline={PO Box 9506},
             city={NL-2300 RA Leiden},
             country={the Netherlands}}

\affiliation[Leiden]{organization={Leiden Observatory, Leiden University},
             addressline={PO Box 9513},
             city={NL-2300 RA Leiden},
             country={the Netherlands}}
             
 \affiliation[Oslo]{organization={Faculty of Mathematics and Natural Sciences, University of Oslo},
             addressline={Sem S{\ae}lands vei 24, 0371 Oslo},
             country={Norway}}
             
 \affiliation[Glasgow]{organization={School of Physics and Astronomy, University of Glasgow},
             postcode={G12 8QQ},
             state={Scotland},
             country={UK}}

\nonumfootnote{
\footnotesize{
$^*$Corresponding author, 
\href{mailto:thomas.d.sandnes@durham.ac.uk}{thomas.d.sandnes@durham.ac.uk}.}}
             
%% Abstract
\begin{abstract}

We present REMIX, a smoothed particle hydrodynamics (SPH) scheme designed to alleviate effects that typically suppress mixing and instability growth at density discontinuities in SPH simulations. We approach this problem by directly targeting sources of kernel smoothing error and discretisation error, resulting in a generalised, material-independent formulation that improves the treatment both of discontinuities within a single material, for example in an ideal gas, and of interfaces between dissimilar materials. This approach also leads to improvements in capturing wider hydrodynamic behaviour unrelated to mixing. We demonstrate marked improvements in three-dimensional test scenarios, focusing on cases with particles of equal mass across the simulation. This choice is particularly relevant for use cases in astrophysics and engineering -- specifically those in which particles are free to evolve over a large range of density scales -- where bespoke choices of unequal particle masses in the initial conditions cannot easily be used to address emergent and evolving density discontinuities. We achieve these improvements while maintaining sharp discontinuities; without introducing additional equation of state dependence in, for example, particle volume elements; and without contrived or targeted corrections. Our methods build upon a fully compressible and thermodynamically consistent core-SPH construction, retaining Galilean invariance as well as conservation of mass, momentum, and energy. REMIX is integrated in the open-source, state-of-the-art \swift code and is designed with computational efficiency also in mind, meaning that its improved hydrodynamic treatment can be used for high-resolution simulations without prohibitive cost to run-speed.   

\end{abstract}



%% Keywords

\end{frontmatter}

% \linenumbers


\section{Introduction}\label{sec:introduction}

Computational simulations are an invaluable tool for studying the inherently complex behaviour of fluids. Smoothed particle hydrodynamics (SPH), first developed by \citet{lucy1977numerical} and \citet{gingold1977smoothed}, is frequently utilised across a range of applications spanning astrophysics \citep{benz1988applications, monaghan1992smoothed, springel2010smoothed} and engineering \citep{libersky1993high, lind2020review}. In astrophysics, it is used in particular for its geometry-independent adaptive resolution, inherent conservation properties, and elegant coupling with gravity solvers \citep{springel2005cosmological}. For engineering applications, it offers advantages in the treatment of dynamic free surfaces, fluid--structure interactions, and in simulating multiphase flow \citep{violeau2016smoothed, wang2016overview}. This is in addition to its relatively simple construction, numerical stability, and low computational cost. In this work, our motivation and methods focus on conservative, fully-compressible, gravity-coupled SPH schemes, where particles have unchanging masses and material types throughout a simulation.

Two key concepts characterise SPH: the representation of a fluid as a discrete set of interpolation points, or `particles', that move with the fluid velocity; and the use of a kernel function to estimate fluid fields and their gradients at particle positions, by interpolation over neighbouring particles \citep{price2012smoothed}. However, specific errors are introduced with the assumptions that underpin these core concepts. The discretisation of the continuous underlying fluid results in leading-order error in the momentum equation, which is sensitive to disorder in the local particle distribution \citep{read2010resolving}. Additionally, the use of an extended kernel in the traditional, integral form of the SPH density estimate leads to inadvertent smoothing of interpolated densities \citep{price2012smoothed, monaghan1985extrapolating}. In regions where variations in the underlying density field are not well resolved by the instantaneous particle configuration, this can lead to the calculation of spurious particle pressures, and subsequently to spurious pressure gradients that are used in the equations of motion. 

These errors combine particularly strongly at density discontinuities in simulations where a fluid is represented by particles of fixed, equal mass. In such a case, a density discontinuity constitutes a sharp change in particle spacing. Both discretisation and kernel smoothing error combine to give rise to a spurious surface tension-like effect that greatly suppresses both the mixing of fluid across the interface and the growth of instabilities that would act to drive turbulent mixing \citep{agertz2007fundamental}. This is a well-established shortcoming of SPH, and a range of approaches have been developed to address these sources of error.

First, we consider methods to reduce discretisation error. Using higher-order kernel functions with more particle neighbours will generally reduce error \citep{dehnen2012improving}, and choices of free functions in the generalised form of the equations of motion can be exploited to mitigate zeroth-order error \citep{read2010resolving, wadsley2017gasoline2}. In conjunction with these, improved gradient estimates from, for example, reproducing kernels \citep{liu1995reproducing, liu1998multiple, frontiere2017crksph} or integral-based gradient estimates \citep{garcia2012improving, rosswog2015boosting, rosswog2020lagrangian} have been demonstrated to improve the treatment of fluid mixing and instability growth. These methods have no dependence on material or equation of state (EoS) in their construction or underlying assumptions. We therefore make use of some of these methods in this work.

Next, we consider ways to address kernel smoothing error at contact discontinuities, many of which explicitly assume the use of a single, ideal gas EoS. In particular, we note the use of artificial conduction for this purpose, by which particle internal energies are smoothed over a similar length scale to the inadvertent density smoothing \citep{price2008modelling}. This requires thermodynamic behaviour such that smooth density and internal energy fields result in a smooth pressure field. Therefore, this cannot reliably improve the treatment of interfaces between dissimilar materials, represented by different EoS. Alternatively, methods that use modified density estimates, weighted by a simple thermodynamic quantity such as specific internal energy, also assume a simple relationship between density and internal energy at constant pressure, typically the inversely proportional relationship of an ideal gas \citep{read2010resolving, ritchie2001multiphase}.

Dealing with kernel smoothing errors at interfaces between arbitrarily different materials is more challenging since the simplicity of the ideal gas equation cannot be exploited. A boundary between dissimilar materials in thermal and pressure equilibrium will in general result in a density discontinuity, so these problematic scenarios occur frequently in simulations with multiple materials. Additionally, the surface tension-like effects caused by the density smoothing are particularly strong for ``stiff'' EoS, for which small changes in particle densities can result in large changes in calculated pressures \citep{melosh1989impact}.

Methods to improve the treatment of material interfaces in fully-compressible SPH formulations have been explored in the context of planetary impacts \citep{benz1989origin}, where density discontinuities between multiple, stiff materials are common and can evolve across a range of thermodynamic phase space throughout the course of a single simulation. The treatment of discontinuous free surface interfaces is also important in this context \citep{reinhardt2017numerical}. \citet{hosono2013density} present a  ``density independent SPH'' (DISPH \citep{saitoh2013density}) scheme adapted for use with multiple materials. Here, rather than being calculated from particle masses and densities, volume elements are based on functions of pressure that are evolved in time and recalculated to satisfy kernel normalisation in an additional iterative step \citep{hosono2016giant}. Although this approach prevents spurious pressures at density discontinuities, specifically in regions with otherwise continuous pressures, the extension of this method to arbitrary EoS leads to material-dependent volume elements that intricately depend on fluid thermodynamics. \citet{pearl2022fsisph} present an advanced scheme that, among other improvements, makes use of Riemann solvers \citep{inutsuka2002reformulation} and an optional slip condition at material interfaces. Their choice of material-dependent density estimate effectively smooths volume rather than density at material interfaces, in simulations where particles of the same material have equal masses. If particles are deliberately set up with equal volumes, then this density estimate will significantly reduce both kernel-smoothing and discretisation errors. This improvement is evident in \citet{pearl2022fsisph}'s mixing tests, where particles start on a single, ordered grid. 
% But here, the improved treatment of mixing and instability growth is likely based primarily on the chosen particle configuration. 
%However, such a bespoke particle configuration is not representative of many science applications.
However, these tests do not validate mixing with their methods for cases with emergent and evolving density discontinuities, where particle configurations cannot easily be controlled in this way throughout the simulation. 
 
This approach of addressing smoothing error through the choice of particle masses is also taken by \citet{deng2019enhanced}, who demonstrate enhanced mixing in their meshless finite-mass (MFM) \citep{vila1999particle, gaburov2011astrophysical, ivanova2013common, hopkins2015new} simulations of planetary giant impacts. 
%MFM uses improved kernel gradient terms and partially makes use of solutions to the Riemann problem. However, densities are calculated with an interpolated estimate, that in this case smooths volume, and therefore is still subject to kernel smoothing error.
Although MFM includes Riemann solvers and more advanced gradient estimates that can improve on standard SPH formulations, densities are still calculated with an interpolated estimate that in this case smooths volume, and therefore is still subject to kernel smoothing error. 
Additionally, a range of SPH modifications specific to material boundaries, rather than arbitrary density discontinuities, have also been developed \citep{woolfson2007practical, reinhardt2020bifurcation, ruiz2022dealing}.
All the methods discussed here that address density smoothing directly, rather than in the construction of initial conditions, rely on EoS- or material-dependent treatments in, for example, the calculation of volume elements and density estimates. 

Here, we present the REMIX (Reduced Error MIXing) SPH scheme. REMIX is constructed with the following goals in mind: (1) to improve the treatment of density discontinuities and mixing in simulations with both one and multiple EoS, by directly addressing sources of error in traditional SPH methods; (2) to be able to achieve this for simulations with particles of equal mass; (3) to retain the key characteristics of the SPH formalism; (4) to introduce no additional EoS dependence in, for example, volume elements or density estimates; (5) for computational efficiency, to require no more than three loops over particle neighbours and no additional iterative steps compared with the traditional formulation. An implementation of REMIX is publicly available as part of the open-source \swift code\footnote{\swift is in open development including extensive documentation and examples at \href{www.swiftsim.com}{swiftsim.com}.} \citep{schaller2024swift}.

This paper is structured as follows: in \S\ref{sec:methods}, we describe elements of the core SPH formalism that we build on and key sources of error that we address in the construction of REMIX, as well as the methods used in practice to run our simulations; in \S\ref{sec:remix}, we present each component of the REMIX SPH scheme; in \S\ref{sec:hydrotests}, we validate REMIX in a range of hydrodynamic test simulations; and we summarise our findings in \S\ref{sec:conclusions}.

\section{Methods}\label{sec:methods}

\subsection{Smoothed particle hydrodynamics}\label{subsec:sph}

We first describe the key constituent components of SPH. Although REMIX includes many improvements to traditional SPH\footnote{We use a tSPH formulation based on that of \citet{price2012smoothed}, summarised in \ref{app:trad}, as a basis for our discussion and for comparisons throughout.} (tSPH), we do not deviate far from the core SPH formalism. Additionally, we take this as an opportunity to describe both the sources of error in SPH, whose reduction is central to the REMIX formulation, and the nomenclature and notation that is used throughout. An additional glossary of notation is included in \ref{app:notation}.

\subsubsection{Kernel interpolation and the SPH density estimate}\label{subsubsec:sph_kernels}

Kernel interpolation theory forms the framework for SPH estimates of fluid fields and their gradients. In particular, the integral form of the density estimate is a core component of many SPH schemes \citep{price2012smoothed}, by which a smoothed density field at the position of particle $i$, $ \langle\rho_i\rangle$, can be reconstructed from the local spatial distribution of neighbouring particles $j$, their masses, $m_j$, and a kernel function, $W_{ij}$ (described below), via 

\begin{equation}\label{eq:sph_rho_estimate}
    \langle\rho_i\rangle = \sum_j m_j W_{ij} \;.
\end{equation}

\noindent
Throughout the governing equations of these SPH schemes, the interpolated density, $ \langle\rho_i\rangle$, is used as an estimate of the underlying density field at the positions of particles, $\rho(\mathbf{r}_i)$. This density estimate is a specific application of kernel interpolation, which in general can be used to reconstruct an arbitrary field, $F$, from its value sampled at the positions of particle neighbours, via

\begin{equation}\label{eq:disc_kernel_interpolation}
    \langle F_i \rangle = \sum_j F_j W_{ij} V_j \;,
\end{equation}

\noindent
where $V_j$ are volume elements of particle $j$. In Eqn.~\ref{eq:sph_rho_estimate}, volume elements are taken to be $V_j = m_j / \rho_j$. Kernel interpolation can also be used in estimates of the gradient\footnote{We make the choice of notation, here and throughout, to express kernel gradients as total derivatives rather than with ``$\nabla$'' which is often used to imply derivatives with fixed smoothing length. In later sections, this allows us to more easily distinguish between gradient estimates with and without grad-$h$ terms \citep{hopkins2013general}.} of $F$,

\begin{equation}\label{eq:disc_grad}
    \left\langle \frac{dF}{d\mathbf{r}}\bigg|_{i}  \right\rangle = \sum_j F_j \frac{d W}{d \mathbf{r}}\bigg\rvert_{ij}  V_j \;,
\end{equation}

\noindent
such as in the calculation of pressure gradients and velocity divergences for the SPH equations of motion. 

The smoothing kernel, $W(\mathbf{r} - \mathbf{r'}, h(\mathbf{r}))$, is a weighting function with radial extent characterised by the smoothing length $h$. $W(\mathbf{r} - \mathbf{r'}, h(\mathbf{r}))$ approaches a delta function in the limit $h(\mathbf{r}) \rightarrow 0$. Traditionally, $W$ is a positive function with approximately a truncated Gaussian-like shape; the kernel is typically normalised, and spherically symmetric, as this ensures the exact interpolation of linear fields in the continuum limit of kernel sampling (number of neighbours, $N \rightarrow \infty$). For a particle pair $i$, $j$: $W_{ij} \equiv W(\mathbf{r}_{ij}, h_i) \equiv W(\mathbf{r}_{i} - \mathbf{r}_{j}, h(\mathbf{r}_{i}))$, where $\mathbf{r}_{ij} \equiv \mathbf{r}_{i} - \mathbf{r}_{j}$. Subscripts denote quantities either sampled at the position of, or associated with, a particle. Kernels with a compact support, $H \equiv H(h)$ such that $W(r>H) = 0$, are used to limit the number of neighbours to a finite number. We adopt the convention of defining the smoothing length $h$ as twice the standard deviation of the kernel\footnote{Despite the ubiquitous use of the nomenclature and notation of the ``smoothing length, $h$'', different definitions are frequently used for both the relationship between $H$ and $h$, and the method used to calculate $h$ (in our case, Eqn.~\ref{eq:smoothinglength}). Although the differences are subtle, we draw attention to this as an example of the difficulty of one-to-one comparisons between simulation codes, especially as methods become increasingly complex.} \citep{dehnen2012improving}. This relates the smoothing length to the compact support by a constant multiplication factor $H/h$.

In the SPH construction presented here, a particle's smoothing length is evaluated iteratively to satisfy 

\begin{equation}\label{eq:smoothinglength}
    h_i = \eta_{\rm kernel} \left(\frac{1}{\sum_j W(\mathbf{r}_{ij}, \, h_i)}\right)^{1/d} \;,
\end{equation}

\noindent
 where $d$ is the spatial dimensionality of the simulation and  $\eta_{\rm kernel}$ is a chosen constant. Eqn.~\ref{eq:smoothinglength} ensures that particles across the simulation have an approximately constant number of neighbours, determined by the form of the kernel function and the choice of $\eta_{\rm kernel}$. 

We use the Wendland $C^2$ kernel \citep{wendland1995piecewise} for validation of the REMIX SPH scheme:

\begin{equation}\label{eq:wendlandc2}
    W_{\text{WC2}}\left(|\mathbf{r} - \mathbf{r'}|, \, H(h)\right) = 
     \begin{cases}    
    \dfrac {C}{H^d}\left(1 - \dfrac {|\mathbf{r} - \mathbf{r'}|}{H}\right)^{4}\left( 1 + 4 \dfrac {|\mathbf{r} - \mathbf{r'}|}{H} \right) & \text{for $|\mathbf{r} - \mathbf{r'}| < H$} \;,\\
    0 & \text{otherwise}\;,
     \end{cases}
\end{equation}

\noindent
with $\eta_{\rm kernel} = 1.487$ (${\sim}100$ neighbours for $d=3$)   \citep{dehnen2012improving}. Here $C= 21 / (2\pi)$ is the normalisation constant for the Wendland $C^2$ kernel in 3D. Higher-order kernels can reduce error but require greater numbers of neighbours, which can come at a significant cost to code speed. This kernel offers a suitable compromise between improved accuracy and fast simulation run-time, a relevant consideration for science applications. In \ref{app:kernel} we demonstrate the effect of the choice of kernel function in REMIX simulations, finding that REMIX performs well even with the comparatively low-order cubic spline kernel with fewer neighbours.

\subsubsection{SPH equations of motion}\label{subsubsec:sph_eom}

The equations of motion govern the kinematic and thermodynamic evolution of SPH particles. The Euler equations are used as the basis for the SPH equations of motion for inviscid fluids. These consist of the continuity equation, momentum equation, and energy equation, which are closed by an ``equation of state'', discussed in \S\ref{subsec:eos}. The general, thermodynamically consistent SPH equations of motion, where we additionally use the same kernel function across the equations, take the form \citep{read2010resolving, hopkins2013general}

\begin{align}\label{eq:general_drhodt}
    \frac{d \rho_i}{d t} &\,=\, \sum_{j} m_j \frac{\zeta_i}{\zeta_j} \mathbf{v}_{ij} \cdot \frac{dW}{d\mathbf{r}}\bigg|_{ij} \;,
\\
\label{eq:general_a}
    \frac{d \mathbf{v}_i}{d t} &\,=\, -\sum_{j} m_j \left( \frac{P_i}{\rho_i^2}\frac{\xi_i}{\xi_j} + \frac{P_j}{\rho_j^2}\frac{\xi_j}{\xi_i} \right)  \frac{dW}{d\mathbf{r}}\bigg|_{ij} \;,
\\
\label{eq:general_dudt}
    \frac{d u_i}{d t} &\,=\, \frac{P_i}{\rho_i^2} \sum_{j} m_j \frac{\zeta_i}{\zeta_j} \mathbf{v}_{ij} \cdot \frac{dW}{d\mathbf{r}}\bigg|_{ij} \;,
\end{align}

\noindent
where particle densities, $\rho_i$, velocities, $\mathbf{v}_i$, and specific internal energies, $u_i$, are evolved in time based on gradients of pressure, $P_i$, and velocity divergences calculated using the relative velocity of particle pairs $\mathbf{v}_{ij} \equiv \mathbf{v}_{i} - \mathbf{v}_{j}$. The free functions $\zeta$ and $\xi$ are introduced in the process of discretisation. An SPH scheme that explicitly conserves energy and momentum requires antisymmetric kernel gradient terms in the exchange of particle pairs $i$ and $j$. The integral form of the density estimate (Eqn.~\ref{eq:sph_rho_estimate}) is equivalent to the differential form (Eqn.~\ref{eq:general_drhodt}) in the continuum limit, for $\zeta_i = \zeta_j$ \citep{read2010resolving}.

\subsubsection{Kernel smoothing error}\label{subsubsec:sph_smoothing_error}

 A fluid field reconstructed using an extended kernel with $h \neq 0$ will be affected by smoothing error, even when sampled in the continuum limit \citep{price2012smoothed}. In the continuum limit, a reconstructed field, $\langle F \rangle$, is the convolution of the underlying field, $F$, with a smoothing kernel $W$,

\begin{equation}\label{eq:cont_kernel_interpolation}
    \langle F (\mathbf{r}, h) \rangle = \int F (\mathbf{r'}) \, W(\mathbf{r} - \mathbf{r'}, h) \, dV' \;. 
\end{equation}

\noindent
Eqn.~\ref{eq:disc_kernel_interpolation} is the discretised form of this equation. Assuming a continuous, infinitely differentiable field $F$, we can Taylor expand about the point $\mathbf{r}$ to give

\begin{align}
    \langle F (\mathbf{r}, h) \rangle \,=\, &\;F(\mathbf{r}) \, \cancelto{1}{\int \!\! W(\mathbf{r} - \mathbf{r'}, h) \, dV'} ~~~+~~~ \frac{d F}{d r^{\,\alpha}}\bigg|_{\mathbf{r}} \,\, \cancelto{\mathbf{0}}{\int \!(\mathbf{r'} - \mathbf{r})^{\alpha} \,  W(\mathbf{r} - \mathbf{r'}, h) \, dV'}
    \nonumber
    \\
    \label{eq:smoothing_error_}
    & ~+~~  \frac{1}{2} \frac{d^2  \! F}{d r^{\,\alpha}d r^{\,\beta}}\bigg|_{\mathbf{r}} \int \! (\mathbf{r'} - \mathbf{r})^{\alpha}  \, (\mathbf{r'} - \mathbf{r})^{\beta}  \, W(\mathbf{r} - \mathbf{r'}, h) \, dV' ~~~+~~~\dots
    \\\nonumber\\
    \label{eq:smoothing_error}
    \,=\, &\;F (\mathbf{r}) ~~+~~  \frac{1}{2} \frac{d^2 \! F}{d r^{\,\alpha}d r^{\,\beta}}\bigg|_{\mathbf{r}} \int \! (\mathbf{r'} - \mathbf{r})^{\alpha} \, (\mathbf{r'} - \mathbf{r})^{\beta} \, W(\mathbf{r} - \mathbf{r'}, h) \, dV' ~~~+~~~\dots \;, 
\end{align}

\noindent
where Greek letter superscripts correspond to spatial dimensions, and like indices are summed over \citep{price2012smoothed, sigalotti2019new}. We separate the first two terms in Eqn.~\ref{eq:smoothing_error_} to demonstrate that, in the continuum limit, the choice of a normalised, spherically symmetric kernel results in the zeroth- and first-order integrals of the expansion taking values $1$ and $\mathbf{0}$ respectively. 

In the continuum limit, kernel interpolation will only reproduce $F(\mathbf{r})$ without error if the integrals of the second and higher order terms are all equal to zero. Due to the assumed symmetry properties of the kernel, integrals in odd terms of the expansion are trivially equal to 0, while in general even terms will be non-zero. For a positive kernel these non-zero terms act to smooth the reconstructed field. Although the integrals in Eqn.~\ref{eq:smoothing_error} will be of order $h^2$ and higher powers of $h$, with exponents corresponding to the term of the expansion \citep{monaghan1985extrapolating}, the errors become significant in regions where second and higher order derivatives of the underlying field are large over length scales of $h$ \citep{violeau2019calculating}. This is, in particular, the case for an underlying field that approaches a discontinuity relative to $h$-length scales. A discontinuity, where the field is not differentiable, will inevitably be erroneously smoothed by kernel interpolation.

The integral SPH density estimate, Eqn.~\ref{eq:sph_rho_estimate}, is an example of the discrete form of Eqn.~\ref{eq:cont_kernel_interpolation}. Through Eqn.~\ref{eq:smoothing_error}, we see how a quantity calculated by kernel interpolation in this way will experience smoothing error when the underlying field varies sharply over $h$-length scales. At density discontinuities, smoothing of the density field leads to spurious pressures that contribute to surface tension-like effects that impede particle mixing across the interface.

\subsubsection{Discretisation error}\label{subsubsec:sph_discretisation_error}

The kernel smoothing errors discussed above are in addition to, and separate from, errors introduced by discretisation \citep{price2012smoothed, spreng2020advanced}. Discretisation errors manifest themselves both through the choice of free functions in the equations of motion -- affecting how closely Eqns.~\ref{eq:general_drhodt}--\ref{eq:general_dudt} approximate their continuous Euler equation equivalents -- and through the imperfect sampling of the kernel by a finite number of particle neighbours, i.e. in the discretisation of integrals like Eqn.~\ref{eq:cont_kernel_interpolation}. 

The use of a normalised, spherically symmetric kernel leads to the exact reconstruction of linear fields in the continuum limit by Eqn.~\ref{eq:cont_kernel_interpolation}, as the higher-order derivatives in Eqn.~\ref{eq:smoothing_error} are zero by construction. However, in the process of discretisation of the fluid into a finite set of particles, the conditions

\begin{align}\label{eq:m0_condition}
    \sum_j  W_{ij} V_j &= 1 \;,\\
\label{eq:m1_condition}
     \sum_j \mathbf{r}_{ij}  W_{ij} V_j &= \mathbf{0} \;,
\end{align}

\noindent
are no longer enforced. The exact reconstruction of fluid fields is therefore lost, even to zeroth order. The amount of discretisation error is a function of the disorder in the local particle distribution. This also applies to gradient estimates, such as those used in the equations of motion. Furthermore, in the equations of motion, gradient estimates are typically modified to enforce conservation, so generally deviate further from exact reproduction of underlying linear fields.

In SPH simulations where a fluid is represented by particles of equal mass, a density discontinuity constitutes a sharp change in particle spacing and thus large local anisotropies in particle distribution. This leads to discretisation error also playing a considerable role in suppressing mixing at density discontinuities \citep{read2010resolving}.

\subsection{Equations of state}\label{subsec:eos}

The EoS characterises the thermodynamic behaviour of a material. In SPH simulations, hydrodynamical evolution is tied directly to pressures and sound speeds, calculated through the EoS. Many applications in astrophysics use simulations with only a single, ideal gas EoS. However, in some cases, multiple EoS are required to simulate dissimilar materials or phases, such as for planetary impacts, where EoS are often highly complex \citep{melosh1989impact}. The improvements offered by the REMIX SPH scheme are EoS-independent, and so our methods can be applied effectively to these simulations, as well as other applications with multiphase fluids.

For the hydrodynamic test simulations presented in \S\ref{sec:hydrotests}, we validate the REMIX scheme using both ideal gases and more complex EoS. For ideal gas simulations, the adiabatic index, $\gamma$, is problem-specific and chosen to draw comparisons with past work. For simulations using more complex materials, we use EoS typically used for planetary impact simulations. In most of these tests, we consider iron and rock in conditions representative of the core--mantle boundary in an Earth-like planet. We use the updated ANEOS Fe\textsubscript{85}Si\textsubscript{15} and forsterite EoS for these materials, respectively \citep{stewart2020shock}. For simplicity, we hereafter refer to these as ``iron'' and ``rock''. In \S\ref{subsec:planet}, we also consider a Jupiter-like planet. For these simulations, we use the hydrogen--helium EoS from \citet{chabrier2021new}, with a helium mass fraction of $Y = 0.245$, and the AQUA EoS from \citet{haldemann2020aqua} to represent heavy elements or ice.

We note that in the simulations we present here, these materials are treated as fluids without physical viscosities or strength properties.

\subsection{The \swift code}\label{subsec:swift}

\swift is a state-of-the-art, open-source hydrodynamics and gravity code that specialises in SPH simulations for planetary applications as well as galaxy formation and cosmology \citep{schaller2024swift, kegerreis2019planetary}. By using task-based parallelism, asynchronous communications, and graph-based decomposition of the work between compute nodes, \swift can perform high-resolution simulations efficiently on modern high-performance computing architectures \citep{schaller2016SWIFT}. REMIX is fully integrated into and was developed using the \swift code, and is therefore publicly available\footnote{\swift is available at \url{www.swiftsim.com} alongside extensive documentation and a large suite of examples.}. All simulations presented here were carried out using the \swift code and all tests shown below are shipped with the code package. The algorithms used for gravity and neighbour-finding are detailed in \citet{schaller2024swift} and are used identically for simulations with both REMIX and traditional SPH.

We use a kick--drift--kick time-stepping scheme, with time steps calculated with the CFL condition and $C_{\rm CFL} = 0.1$ \citep{springel2005cosmological}. In tests of Sod shock tubes, we find only small differences in REMIX simulations with $C_{\rm CFL}$ up to 0.2. Above this we find sharper spikes at the discontinuities of density and internal energy. For REMIX simulations, the signal velocity in time-step calculations is taken to be $v_{\text{sig}, \, i} = \max_j\,(c_i + c_j - 4 \min(\mu_{ij}, \mu_{ji}))$ where $c$ are particle sound speeds, the operation $\max_j$ is the maximum value for all neighbouring particles, and $\mu$ are calculated by Eqn.~\ref{eq:calcmu}, presented later. Future work will explore optimising time-step choices to further enhance performance.

\section{REMIX SPH}\label{sec:remix}

In this section, we detail the constitutive equations of the REMIX SPH scheme\footnote{The full set of the final equations used in the REMIX scheme are listed in \ref{app:remix}. The equations of the traditional SPH scheme that we use for comparison simulations are listed in \ref{app:trad}.}. We improve the treatment of mixing by directly addressing the sources of SPH error discussed in \S\ref{subsec:sph}. By targeting both smoothing and discretisation error, we alleviate spurious surface tension-like effects at density discontinuities, including in challenging cases with equal-mass particles and at interfaces between dissimilar, stiff materials. Note that we aim to address mixing at the particle scale and not below. Therefore, we do not consider diffusion of material type between particles, meaning that the material of each particle remains fixed for the duration of the simulation.

We target error by exploiting three key freedoms in the SPH equations of motion presented in \S\ref{subsubsec:sph_eom}: in the choice of density estimate (\S\ref{subsec:remix_densityest}); in the choice of free functions (\S\ref{subsec:remix_freefunc}); and in the form of the kernel function (\S\ref{subsec:remix_linearkernels}). Additionally, we develop a method that enables the appropriate treatment of free surfaces when using these improved kernels (\S\ref{subsec:remix_vacuum}), and we use improved artificial viscosity (\S\ref{subsec:remix_visc}) and artificial diffusion (\S\ref{subsec:remix_diff}) formulations. These include new approaches both for the treatment of shocks and to weakly smooth and mitigate accumulated noise on the particle scale. We also include a term in the density evolution that re-ties densities to the local particle distribution (\S\ref{subsec:remix_norm}). These components combine into the REMIX equations of motion, given by

\begin{align}
\label{eq:final_drhodt}
\frac{d \rho_i}{d t} &=  \sum\limits_{j} m_j  \, \frac{\rho_i}{\rho_j} \, v_{ij}^{\alpha} \, \frac{1}{2}\left(\frac{d \widetilde{\mathcal{W}}}{dr^{\alpha}}\bigg|_{ij} - \frac{d \widetilde{\mathcal{W}}}{dr^{\alpha}}\bigg|_{ji}\right) \: + \: \left(\frac{d \rho_i}{d t}\right)_{\text{difn}}  \: + \: \left(\frac{d \rho_i}{d t}\right)_{\text{norm}} \;,
\\
\label{eq:final_a}
\frac{d v_i^{\alpha}}{d t} &=  -\sum\limits_{j} m_j \, \frac{P_i + Q_{ij} + P_j + Q_{ji}}{\rho_i \, \rho_j} \, \frac{1}{2}\left(\frac{d \widetilde{\mathcal{W}}}{dr^{\alpha}}\bigg|_{ij} - \frac{d \widetilde{\mathcal{W}}}{dr^{\alpha}}\bigg|_{ji}\right) \;,
\\
\label{eq:final_dudt}
\frac{d u_i}{d t} &=  \sum\limits_{j} m_j \, \frac{P_i + Q_{ij}}{\rho_i \, \rho_j} \, v_{ij}^{\alpha} \, \frac{1}{2}\left(\frac{d \widetilde{\mathcal{W}}}{dr^{\alpha}}\bigg|_{ij} - \frac{d \widetilde{\mathcal{W}}}{dr^{\alpha}}\bigg|_{ji}\right) \: + \: \left(\frac{d u_i}{d t}\right)_{\text{difn}} \;,
\end{align}

\noindent
where $\left(d \widetilde{\mathcal{W}} \!/ d\mathbf{r}|_{ij} -  d \widetilde{\mathcal{W}} \!/ d\mathbf{r}|_{ji}\right) / 2$ are improved kernel gradient terms that are antisymmetric in the exchange of $i$ and $j$ for explicit conservation of momentum and energy; $Q_{ij}$ and $Q_{ji}$ are pairwise, artificial viscous pressures; $\left(d \rho_i / d t\right)_{\text{difn}}$ and $\left(d u_i / d t\right)_{\text{difn}}$ are artificial diffusion of density and internal energy; and $\left(d \rho_i / d t\right)_{\text{norm}}$ is the kernel normalising term. Each of these are discussed in detail in their corresponding sections below.

The equations of the REMIX scheme were developed to be implemented in just three loops over particle neighbours, and without introducing any additional iterative steps. In our test simulations, performed on the COSMA8 HPC system\footnote{Simulations carried out on COSMA8 used 1 node with 128 cores and those on COSMA7 used 1 node with 28 cores. These are both part of the DiRAC cluster hosted by Durham University (\url{https://dirac.ac.uk/memory-intensive-durham/}).}, using REMIX led to a run-speed ${\sim}1.3$--$1.6$ times longer than equivalent simulations performed with traditional SPH (and everything else unchanged). The exact amount of slowdown is problem-dependent: this range includes simulations both with and without gravity, and those using different kernel functions\footnote{Simulations used to investigate the runtime were: 3D Kelvin--Helmholtz instabilities (\S\ref{subsubsec:kh_idg_discontinuous}) and planets in hydrostatic equilibrium (\S\ref{subsec:planet}). These were tested with cubic spline and Wendland $C^2$ kernels.}. On the COSMA7 HPC system (which has fewer cores per node), simulations with the overhead of gravity take ${\sim}1.6$--$1.8$ times longer, and simulations without gravity take ${\sim}2$--$3$ times longer, depending on the test case. 
We find that REMIX, in addition to dealing with density discontinuities that are problematic in traditional SPH at all resolutions, is able to achieve an improved treatment of non-discontinuous regions in simulations with over an order of magnitude lower resolution compared with equivalent traditional SPH results (\S\ref{subsubsec:kh_idg_smooth}). 
The effective slowdown from using REMIX is therefore much smaller in practice than the ranges above suggest, since simulations with a lower resolution (fewer SPH particles) could be used to obtain equivalent results. As such, in many cases a science simulation with REMIX would run faster than a traditional SPH simulation that would require a higher resolution to achieve a comparable level of numerical convergence. For example, the $2.9 \times 10^{5}$ particle REMIX Kelvin--Helmholtz instability in \S\ref{subsubsec:kh_idg_smooth} runs over 20 times faster (on COSMA8) than the $4.7 \times 10^{6}$ particle traditional SPH simulation, and is closer to the converged solution\footnote{See REMIX, $N=128$ and tSPH,  $N=512$ in Fig.~\ref{fig:modes_idg}.}.

\subsection{Density estimate}\label{subsec:remix_densityest}

In the REMIX SPH scheme we use a differential form of the density estimate: we evolve the density in time with Eqn.~\ref{eq:final_drhodt} rather than recalculating it each timestep (e.g. Eqn.~\ref{eq:sph_rho_estimate}), similarly to internal energy in traditional SPH schemes. There are three key benefits of this treatment: (1) we directly address systematic smoothing error in particle densities, which is particularly significant at density discontinuities, including those at free surfaces; (2) it allows us to constrain zeroth-order error in the equations of motion while starting from a basis of thermodynamic consistency (\S\ref{subsec:remix_freefunc}); (3) we do not require an additional loop over particle neighbours to calculate a new density each timestep. We note that particle mass is fixed throughout the simulation, so the evolution of densities is equivalent to an evolution of volumes. In \S\ref{subsec:kh_earth}, we show the differences in Kelvin--Helmholtz instability simulations when using the full REMIX scheme, and the REMIX scheme modified to use a traditional integral density estimate. Using our evolved density estimate, both to calculate thermodynamic quantities and in volume elements, leads to a considerable improvement in addressing spurious surface tension-like effects that suppress instability growth and mixing on the particle scale. 

In practice, we set a density floor $\rho_{\rm min, \, i} \equiv \langle\rho_i\rangle_{\rm min} = m_i W(\mathbf{0}, h_i)$ such that $\rho_i =\rho_{\rm min, \, i}$ if the density would evolve below the minimum value. This prevents EoS extrapolation issues that arise for tiny densities in simulations involving a vacuum region.

Evolved density estimates are used frequently in SPH schemes developed for engineering applications \citep{antuono2010free} as well as in some astrophysical SPH schemes, in particular those that include material strength models \citep{benz1995simulations}. However, in most astrophysical SPH schemes, an integral density estimate is preferred for its robustness: the accumulation of error in an evolved density estimate is less predictable than the relatively controlled errors in a density estimate calculated each timestep from the instantaneous local particle distribution. For instance, if left to evolve freely over many timesteps, densities could in principle take values such that volume elements $m_j / \rho_j$ are far from normalising the kernel $W_{ij}$, despite the kernel being a normalised function\footnote{Volume elements that use the interpolated density, $V_j = m_j / \langle\rho_j\rangle$, are inherently tied to kernel normalisation. The equations for kernel normalisation, Eqn.~\ref{eq:m0_condition}, and the integral density estimate, Eqn.~\ref{eq:sph_rho_estimate}, are equivalent to each other in the limit of constant density on the kernel length scale, $\langle\rho_j\rangle \rightarrow \langle\rho_i\rangle$ for all $j$.}. We address these concerns with four approaches: (1) by introducing a novel term in the density evolution that re-ties densities to the local particle distribution (\S\ref{subsec:remix_norm}); (2) by using kernels that are normalised  to the evolved densities (\S\ref{subsec:remix_linearkernels}); (3) by including a weak density diffusion to smooth out accumulated noise (\S\ref{subsec:remix_diff}); (4) and by taking preventative measures in reducing error that could accumulate with time, reflected in the choices of our equations of motion (\S\ref{subsec:remix_freefunc}), the use of kernel functions constructed to reduce discretisation error (\S\ref{subsec:remix_linearkernels}), and our improved viscosity formulation (\S\ref{subsec:remix_visc}). 

Evolved densities are used wherever density appears in the equations of the REMIX scheme. This includes for calculating thermodynamic quantities, using the equation of state, and in all volume elements.

\subsection{Free functions in the equations of motion}\label{subsec:remix_freefunc}

In traditional SPH formulations, the free functions, $\zeta$ and $\xi$, in the equations of motions (Eqns.~\ref{eq:general_drhodt}--\ref{eq:general_dudt}) typically take equal values for all particles and cancel. An alternate formulation with $\zeta = \xi = \rho$, such that the equations of motion include ratios of the densities of particles $i$ and $j$, helps to constrain error in the equations of motion at density discontinuities and for irregular particle distributions on the kernel scale \citep{read2010resolving}. This choice avoids the use of gradients of density in the derivation of the momentum equation, by using the identity

\begin{equation}\label{eq:momentum_derivation_unity}
    \frac{\nabla P}{\rho} = \frac{\nabla P}{\rho} + \frac{P}{\rho}\nabla 1 \;, 
\end{equation}

\noindent
rather than

\begin{equation}\label{eq:momentum_derivation_trad}
    \frac{\nabla P}{\rho} = \nabla \left( \frac{P}{\rho} \right) + \frac{P}{\rho^2} \nabla \rho \;. 
\end{equation}

\noindent
SPH formulations using the density as the free functions have been shown to improve the treatment of mixing \citep{wadsley2017gasoline2}. For simulations using only a single ideal gas, the choice of  $1 / u$ as a free function is equivalent to this, with the additional assumption of constant pressure on the kernel scale \citep{ritchie2001multiphase}. 

Using density as the free function in the integral form of the density estimate (Eqn.~\ref{eq:sph_rho_estimate}) for simulations with arbitrary EoS is not possible without iteration, since the density would be needed in the density calculation. However, using the differential form to evolve the density (Eqn.~\ref{eq:final_drhodt}) enables us to develop the REMIX SPH scheme from a basis of full thermodynamic consistency with  $\zeta_i = \rho_i$. We also use $\xi_i = \rho_i$ to reduce zeroth-order error in the momentum equation \citep{read2010resolving, wadsley2017gasoline2}. All densities used are the evolved densities of particles.

In \S\ref{subsec:kh_earth}, we demonstrate the improvements in REMIX simulations of the Kelvin--Helmholtz instability from using $\zeta_i = \xi_i = \rho_i$, compared with the REMIX scheme modified to use traditional equal-valued free functions.

\subsection{Linear-order reproducing kernels}\label{subsec:remix_linearkernels}

To reduce discretisation error, we construct kernels that explicitly satisfy the conditions given by Eqns.~\ref{eq:m0_condition} and \ref{eq:m1_condition}. Therefore, these kernels reproduce exact values for fields that are spatially constant or that vary linearly with position. This methodology is largely based on that of \citet{frontiere2017crksph}. To account for spatial variations of the smoothing length, we include grad-$h$ terms that were previously neglected. These grad-$h$ terms take a non-standard form, compared with \citet{hopkins2013general}, since our evolved density is not tied directly to smoothing lengths through the instantaneous distribution of particles. We also modify our kernels to include a free-surface treatment (\S\ref{subsec:remix_vacuum}) to allow them to appropriately handle vacuum boundaries.

The modified kernel, $\mathcal{W}_{ij}$, is constructed so that the sum over neighbours always satisfies

\begin{align}
\label{eq:m0_condition_better}
    \sum_j  \mathcal{W}_{ij} V_j &= 1 \;,
\\
\label{eq:m1_condition_better}
     \sum_j \mathbf{r}_{ij}  \mathcal{W}_{ij} V_j &= \mathbf{0} \;.
\end{align}

\noindent
We use volume elements $V_j = m_j / \rho_j$, where $\rho_j$ are the evolved densities.  We stress that for use in the equations of motion we must undergo a necessary step to make the kernel gradient terms antisymmetric in exchanges of particle pairs, to enforce the conservation of energy and momentum, as is also done by \citet{frontiere2017crksph}. Therefore, the gradient estimates used in the equations of motion end up being not exactly first-order reproducing. Despite this, these gradient estimates show significant improvements when compared with unmodified kernels (as seen directly in \S\ref{subsec:kh_earth}).

To construct $\mathcal{W}_{ij}$, an unmodified SPH kernel is multiplied by a linear polynomial

\begin{equation}\label{eq:linear_kernel}
\mathcal{W}_{ij} \equiv A_i \left(1 + B_i^{\alpha} r_{ij}^{\alpha}\right) \overline{W}_{ij} \;,
\end{equation}

\noindent
where $\overline{W}_{ij} \equiv [W(\mathbf{r}_{ij}, h_i) + W(\mathbf{r}_{ji}, h_j)] / 2$ is a symmetrised kernel\footnote{We find this to be beneficial when we enforce the antisymmetrisation required for use in the equations of motion, as demonstrated in \ref{app:kernelconstruction}. Note that for certain computational steps, this choice extends the definition of particle $i$'s ``neighbours'', $j$, to be those that satisfy either $|\mathbf{r}_{ij}| < H_i$ or $|\mathbf{r}_{ij}| < H_j$ rather than just the first condition.}, and $A_i$ and $\mathbf{B}_i$ are coefficients that satisfy Eqns.~\ref{eq:m0_condition_better} and \ref{eq:m1_condition_better}, as shown in Appendix A of \citet{frontiere2017crksph}:

\begin{align}
\label{eq:A}
A_i &= \left( \overline{m}_{0,\,i} - \left(\overline{m}_{2,\,i}^{\, -1}\right)^{\alpha \beta} \overline{m}_{1,\,i}^{\, \alpha} \, \overline{m}_{1,\,i}^{\, \beta} \right)^{-1} \;,
\\
\label{eq:B}
B_i^{\alpha} &= - \left(\overline{m}_{2,\,i}^{\, -1}\right)^{\alpha \beta} \overline{m}_{1,\,i}^{\, \beta} \;,
\end{align}

\noindent
where the geometric moments are defined as

\begin{align}\label{eq:m0}
\overline{m}_{0,\,i} &= \sum_j  \overline{W}_{ij} V_j \;,\\
\label{eq:m1}
\overline{m}_{1,\,i}^{\,\alpha} &= \sum_j r_{ij}^{\alpha}  \overline{W}_{ij} V_j \;,\\
\label{eq:m2}
\overline{m}_{2,\,i}^{\,\alpha \beta} &= \sum_j r_{ij}^{\alpha} r_{ij}^{\beta}  \overline{W}_{ij} V_j \;.
\end{align}

\noindent
Greek letter indices correspond to spatial dimensions and like indices are summed over. Bars indicate the use of the symmetrised kernel in the kernel interpolation. This distinction becomes important since we use $m_{0,\,i}$, calculated similarly but using an unsymmetrised kernel, for alternative gradient estimates used later in this section and in \S\ref{subsec:remix_norm}.

To calculate gradient terms for the equations of motion, we require the spatial derivative of $\mathcal{W}$. We include terms that depend on the gradient of smoothing lengths, unlike \citet{frontiere2017crksph}. We find the effects of these to be small in practice, but include them for completeness of the method -- without assuming these to be negligible. 

The smoothing length dependence of Eqns.~\ref{eq:linear_kernel}--\ref{eq:m2} is contained within $\overline{W}_{ij}$. We therefore express the derivatives with the parameterisation $\mathcal{W}(\mathbf{r},\, \mathbf{r}_j) \equiv \mathcal{W}\left(\mathbf{r} - \mathbf{r}_j, \; \overline{W}(\mathbf{r} - \mathbf{r}_j, h(\mathbf{r}), h(\mathbf{r}_j)),  \; A(\mathbf{r}),  \;\mathbf{B}(\mathbf{r})\right)$, giving

\begin{align}\label{eq:total_derivative}
\frac{d \mathcal{W}}{dr^{\gamma}} &= A B^{\gamma} \overline{W}+ 
                                    \frac{\partial \mathcal{W}}{\partial \overline{W}} \frac{d \overline{W}}{d r^{\gamma}} +
                                    \frac{\partial \mathcal{W}}{\partial A} \frac{d A}{d r^{\gamma}} + 
                                    \frac{\partial \mathcal{W}}{\partial B^{\alpha}}  \frac{d B^{\alpha}}{d r^{\gamma}} \;.
\end{align}

\noindent
When evaluated for a particle pair $i, j$ this  becomes\footnote{We use the notation $\frac{d \mathcal{W}}{dr^{\gamma}}\big|_{ij} \equiv \frac{d\mathcal{W}}{dr^\gamma}\left(\mathbf{r}_i - \mathbf{r}_j, \; \overline{W}(\mathbf{r}_i - \mathbf{r}_j, h(\mathbf{r}_i), h(\mathbf{r}_j)), \; A(\mathbf{r}_i), \; \mathbf{B}(\mathbf{r}_i)\right) \equiv\frac{d\mathcal{W}}{dr^\gamma}\left(\mathbf{r}_{ij}, \; \overline{W}_{ij}, \; A_i, \; \mathbf{B}_i\right)$.}

\begin{align}\label{eq:total_derivative_pair}
 \frac{d \mathcal{W}}{dr^{\gamma}}\bigg|_{ij}  &=  A_i B_i^{\alpha} \overline{W}_{ij} +
                                                    A_i \left(1 + B_i^{\alpha} r_{ij}^{\alpha}\right) \frac{d \overline{W}}{d r^{\gamma}}\bigg|_{ij} 
                                                    + \left(1 + B_i^{\alpha} r_{ij}^{\alpha}\right) \overline{W}_{ij} \frac{d A}{d r^{\gamma}}\bigg|_{i} + 
                                                    A_i r_{ij}^{\alpha} \overline{W}_{ij} \frac{d B^{\alpha}}{d r^{\gamma}}\bigg|_{i} \;. 
\end{align}

\noindent
Equations to calculate the gradients of $A$, $B$, and the geometric moments are included in \ref{app:remix}. The derivative of the symmetrised kernel is given by\footnote{Note that we are taking the derivative of the continuous function $\overline{W}(\mathbf{r}, \mathbf{r}_j) = [W(\mathbf{r} - \mathbf{r}_j, h(\mathbf{r})) + W(\mathbf{r}_j - \mathbf{r}, h(\mathbf{r}_j))] / 2$ with respect to $\mathbf{r}$, with fixed neighbour positions $\mathbf{r}_j$, and evaluating it at $\mathbf{r}_i$. Therefore, there is only a grad-$h$ term associated with the first term in the brackets.}

\begin{equation}\label{eq:dW_dr}
\frac{d \overline{W}}{d r^{\gamma}}\bigg|_{ij} = \frac{1}{2}\left(\frac{\partial W}{\partial r^{\gamma}}\bigg|_{ij}  + \frac{\partial W}{\partial h}\bigg|_{ij} \frac{d h}{d r^{\gamma}}\bigg|_{i} - \frac{\partial W}{\partial r^{\gamma}}\bigg|_{ji} \right) \;,
\end{equation}

\noindent
and so the inclusion of grad-$h$ terms in the gradient calculations in practice only takes the form of the additional term in Eqn.~\ref{eq:dW_dr}. Both $\partial W / \partial r^{\gamma}$ and $\partial W / \partial h$ can be calculated directly from the kernel function \citep{price2018phantom}. 

Finally, we require $dh/dr^{\gamma}$. In SPH schemes that use the traditional density estimate, $dh/dr^{\gamma}$ do not need to be calculated explicitly \citep{hopkins2013general}, since smoothing lengths and densities are inherently linked. However, for the scheme presented here, where we use an evolved density estimate, we must calculate this explicitly. One approach is to directly differentiate Eqn.~\ref{eq:smoothinglength}. However, we find that zeroth-order error from calculating grad-$h$ terms in this way leads to spurious behaviour in simulations. We therefore calculate these by kernel interpolation. Since, in practice, $d \overline{W} / d r^{\gamma}$ has not been constructed yet due to the order of these operations in the loops over particle neighbours, we are unable to use these improved gradient terms for $dh/dr^{\gamma}$ if we want to avoid introducing a 4\textsuperscript{th} loop. This also applies for gradient estimates in our viscosity (\S\ref{subsec:remix_visc}) and diffusion (\S\ref{subsec:remix_diff}) schemes, discussed later. We therefore require an alternative gradient estimate for these calculations. However, we must be mindful of kernel normalisation in these alternative gradient estimates, since we use evolved densities for volume elements throughout. We therefore use the kernel gradient term

\begin{equation}\label{eq:normalise_kernel_gradient}
\partial_i^{\gamma} \hat{W}_{ij} \equiv \frac{\partial_i^{\gamma} W_{ij}}{m_{0,\,i}} - \frac{W_{ij}}{m_{0,\,i}^2} \partial_i^{\gamma} m_{0} \;,
\end{equation}

\noindent
where we note that the lack of bars throughout indicates the use of a standard (e.g. Wendland $C^2$) kernel, rather than one symmetrised by averaging with neighbouring kernels, and `$\partial$', rather than total derivatives, indicates a lack of grad-$h$ terms. These choices allow us to calculate these kernel gradients in two loops over particle neighbours, so they can be used here and in the artificial viscosity and diffusion schemes. Circumflexes, here and throughout, indicate the use of the normalised kernel $\hat{W}_{ij} \equiv W_{ij} \, / \, m_{0,\,i}$.

We then calculate

\begin{equation}\label{eq:dh_dr}
\partial_i^{\gamma} \hat{h}  =  \sum_j (h_j - h_i) \, \partial_i^{\gamma} \hat{W}_{ij}  \frac{m_j}{\rho_j} \;,
\end{equation}

\noindent
and use this in place of $dh/dr^{\gamma}$.

All these equations combine in Eqn.~\ref{eq:total_derivative_pair} to give the gradients of the linear-order reproducing kernels. The use of these kernels reduces discretisation error in the equations of motion. In \S\ref{subsec:kh_earth}, we show the effect of these kernels on simulations of the Kelvin--Helmholtz instability by using either the full REMIX scheme or the REMIX scheme with unmodified, Wendland $C^2$ kernels.

\subsection{Vacuum boundary treatment}\label{subsec:remix_vacuum}

We present a method to switch the kernel gradients constructed in \S\ref{subsec:remix_linearkernels} to the unmodified spherically symmetric kernel gradients in regions identified as vacuum boundaries. We stress that this method is not applying a targeted correction to vacuum boundaries as done by, for example, \citet{reinhardt2017numerical}. In fact, our evolved density estimate corrects density smoothing at discontinuous free surfaces without any need for a targeted approach. Instead, the vacuum treatment we present here is just an expansion of the form of the linear-order reproducing kernels (\S\ref{subsec:remix_linearkernels}) to allow them to capture free surfaces as vacuum boundaries, a case not considered -- rather than handled poorly -- in their general construction.

Similar approaches, in which a kernel gradient correction matrix is used to revert to the gradient of the uncorrected kernel function at free surfaces, have been previously developed for schemes utilising different forms of kernel corrections.  \citet{oger2007improved} presented a method that switches to uncorrected kernel gradients based on a discrete threshold. Similar approaches have been demonstrated in the context of breaking wave simulations, by \citet{zago2021overcoming} and \citet{lyu2023derivation}. \citet{ren2023efficient} presented a scheme more similar to our own, in which the switch to the uncorrected kernel gradient function is treated smoothly.

A region with no SPH particles is not trivially equivalent to the representation of a vacuum. Since SPH particles are moving interpolation points, a region not sampled by SPH particles can be seen as analogous to a region in a grid-based code where the grid points have been removed. There is therefore no inherent information associated with these regions that would make them equivalent to a region with zero pressure, rather than a region to extrapolate into. However, if a spherically symmetric kernel, normalised to the continuum, is used to calculate pressure gradients in the equation of motion, vacuum-like behaviour is achieved. At a free surface, a particle with a spherically symmetric kernel will calculate pressure gradients equivalent to those calculated if the vacuum region were built up of particles with appropriate volumes but zero pressure\footnote{These gradients may not be fully equivalent in the equations of motion where the additional condition of antisymmetry in exchange of neighbours is imposed, however, they remain closely related.}.

This is not the case for the linear-order reproducing kernels described in \S\ref{subsec:remix_linearkernels}. Since kernels are constructed to satisfy Eqns.~\ref{eq:m0_condition_better} and \ref{eq:m1_condition_better} for volumes built up by particles only, the vacuum region is treated as a region to extrapolate into. SPH applications typically require the treatment of a region without SPH particles as a vacuum, or a region with negligible pressure. We therefore switch our kernel gradient terms to gradients of unmodified kernels at free surfaces:

\begin{equation}\label{eq:vactreatment}
\frac{d \widetilde{\mathcal{W}}}{dr^{\gamma}}\bigg|_{ij} = s_i\frac{d \mathcal{W}}{dr^{\gamma}}\bigg|_{ij} + \left(1 - s_i\right)\frac{d W}{d r^{\gamma}}\bigg|_{ij} \;, 
\end{equation}

\noindent
where $s$ is a function that switches from 1 in regions where no vacuum boundary is detected, to 0 in regions near a vacuum boundary. Note that we smoothly switch between kernel gradients rather than the kernels themselves. This is to avoid terms with gradients of $s$. A switch that is accurate in identifying vacuum boundary particles \textit{only} will inevitably have sharp spatial gradients, which could significantly influence the evolution of particles. Since we do not calculate densities by Eqn.~\ref{eq:sph_rho_estimate}, we do not require the direct calculation of the function whose derivative is given by Eqn.~\ref{eq:vactreatment} to maintain thermodynamic consistency. 

We modify the kernel gradient terms based only on parameters of the kernel function itself. Therefore, conceptually, we adapt the kernel \textit{function} rather than making the kernel  respond to the physical system simulated. For $s$, we use a Gaussian switch,

\begin{equation}\label{eq:vacswitch}
s\left(h_i |\mathbf{B}_i|\right) = 
   \begin{cases}
      \exp{\left[ -\,\dfrac{\left(0.8 - h_i |\mathbf{B}_i|\right)^2}{0.08} \right] } & \;\;\text{for $h_i |\mathbf{B}_i| \ge 0.8$} \;,\\
      1 & \;\;\text{otherwise}\;,\\
    \end{cases}     
\end{equation}

\noindent
where the offset, $0.8$, and denominator, $0.08$, of the switch are chosen empirically to identify boundary particles as those with a large $|\mathbf{B}_i|$ (Eqn.~\ref{eq:B}) greater than ${\sim}1/h_i$. These are particles whose kernels would have to drastically change shape to deal with large anisotropies in the volume elements of particle neighbours. We find that using these values allows the switch to identify particles near free surfaces reliably without misidentifying particles in non-vacuum regions, as we show in \S\ref{subsec:planet}, where we also demonstrate the need for this vacuum boundary treatment. In the example presented, the free surface of a Jupiter-like planet in hydrostatic equilibrium is unstable when the vacuum boundary treatment is not included. As well as its use in switching the kernel function, $s$ is also used in the kernel normalisation term in the density evolution, as detailed in \S\ref{subsec:remix_norm}.

\subsection{Artificial viscosity}\label{subsec:remix_visc}

Artificial viscosity is required to capture shocks in SPH simulations, whose constituent equations otherwise model adiabatic and dissipationless evolution \citep{monaghan1983shock}. A difficulty faced by artificial viscosity constructions is over-dissipation in regions not affected by a shock. Artificial viscosity switches, like the Balsara switch \citep{balsara1995neumann},

\begin{equation}\label{eq:Balsara}
   \mathcal{B}_i = \frac{\left| \nabla \cdot \mathbf{v}_i \right|}{\left| \nabla \cdot \mathbf{v}_i \right| \,+\, \left| \nabla \times \mathbf{v}_i \right| \,+\, 0.0001\, c_{i} / h_i} \;,
\end{equation}

\noindent
where $c$ is the sound speed, or higher-order switches like that of \citet{read2012sphs} are used to switch artificial viscosity off in shearing regions. Time-dependent viscosity parameters have also been developed \citep{morris1997switch, cullen2010inviscid, borrow2022sphenix} to reduce over-dissipation. 

Recently, the limiting of artificial viscosity by the use of reconstructed velocities at particle-pair midpoints has been demonstrated to be an effective alternative approach \citep{frontiere2017crksph, rosswog2020lagrangian, pearl2022fsisph}. For each particle pair, two velocities are estimated at the midpoint of the pair based on Taylor expansions from each particle using their individual velocities and estimated velocity gradients. The difference between these velocities is then used in the viscosity scheme instead of the relative velocity of the particles themselves. This is the approach taken in REMIX. We use linear reconstruction as we find further improvements due to quadratic reconstruction to be small, as also noted by \citet{rosswog2020lagrangian}. If the velocity field is locally linear, artificial viscosity would effectively be switched off with linear reconstruction. For schemes that use linear reconstruction, the viscosity in shearing regions where the velocity field is not exactly locally linear is not negligible and will still influence the fluid behaviour. However, this results in a helpful effect, acting as a weak artificial diffusion of momentum that smooths particle noise in the velocity field by guiding it towards being locally linear on the particle scale.

Our artificial viscosity treatment is largely based on those of \citet{frontiere2017crksph} and \citet{rosswog2020lagrangian}, with some additional, novel approaches. As detailed below, a slope limiter is used to prevent reconstruction at discontinuities, thereby increasing artificial viscosity where it is required for shock capturing. However, we find that a slope limiter alone does not effectively switch off reconstruction, because the velocity gradients used to construct it are inherently smoothed by their calculation using a smoothing kernel. Therefore, they do not identify sharp discontinuities well. We introduce a Balsara switch (Eqn.~\ref{eq:Balsara}) into the slope limiter term to switch off reconstruction at shocks more effectively. Here we calculate $\left| \nabla \cdot \mathbf{v}_i \right|$ and $\left| \nabla \times \mathbf{v}_i \right|$ in the Balsara switch using the kernel gradient term given by Eqn.~\ref{eq:normalise_kernel_gradient}, and also use these same gradient estimates for the velocity gradients used in the linear reconstruction,

\begin{equation}\label{eq:grad_v}
    \partial_i^{\gamma}  \hat{v}^\alpha = \sum_j (v_j^\alpha - v_i^\alpha)  \, \partial_i^{\gamma} \hat{W}_{ij} \frac{m_j}{\rho_j} \;. 
\end{equation}

\noindent
The velocity reconstructed to the midpoint of a particle pair is given by

\begin{equation}\label{eq:vtilde}
  \tilde{v}_{ij}^{\,\alpha} = v_{i}^\alpha + \frac{1}{2}\left(1 - \mathcal{B}_{i}^{\text{SL}}\right) \Phi_{v,\;ij} \left(r_j^\gamma - r_i^\gamma \right)  \partial_i^{\gamma}  \hat{v}^\alpha \;,
\end{equation}

\noindent
where $\mathcal{B}_{i}^{\text{SL}}$ is the standard Balsara switch (Eqn.~\ref{eq:Balsara}), and the SL (slope limiter) superscript just indicates its use in conjunction with the slope limiter. $\Phi_{ij}$ is the van Leer slope limiter \citep{van1974towards}, given by

\begin{equation}\label{eq:calcphi}
   \Phi_{ij} = 
   \begin{cases}
      0 & \;\;\text{for $A_{ij} < 0$}\;,\\
      \dfrac {4 A_{ij}}{(1 + A_{ij})^2} \exp{\left[-\left( \dfrac {\eta_{ij}^{\rm min} - \eta_{\text{crit}}}{0.2} \right)^2\right]} & \;\;\text{for $\eta_{ij}^{\rm min} < \eta_{\text{crit}}$}\;,\\
      \dfrac {4 A_{ij}}{(1 + A_{ij})^2} & \;\;\text{otherwise}\;, 
    \end{cases}       
\end{equation}

\noindent
where the additional Gaussian term in Eqn.~\ref{eq:calcphi} switches the slope limiter to 0 for particle pairs with a small separation. $\eta_{ij}^{\rm min}$ is the smaller value of $|\boldsymbol\eta_{ij}|$ and $|\boldsymbol\eta_{ji}|$, where $\boldsymbol\eta_{ij} = (\mathbf{r}_i - \mathbf{r}_j) / h_i$ and similarly for the exchanged particle indices.
$\eta_{\rm crit}$ represents a separation closer than one would expect from the distribution of the rest of the particle's neighbours. For viscosity calculations, we use the ratio of projected velocity gradients $A_{ij} \equiv A_{v, \;ij}$ given by

\begin{equation}\label{eq:calcA_v}
   A_{v, \;ij} = \frac{\partial_i^\alpha \hat{v}^{\,\beta} (\mathbf{r}_j - \mathbf{r}_i)^\beta (\mathbf{r}_j - \mathbf{r}_i)^\alpha }{\partial_j^\gamma \hat{v}^{\,\phi} (\mathbf{r}_j - \mathbf{r}_i)^\gamma (\mathbf{r}_j - \mathbf{r}_i)^\phi} \;.
\end{equation}

\noindent
For $\eta_{\rm crit}$ we use
 
 \begin{equation}\label{eq:eta_crit}
   \eta_{\rm crit} = \frac{1}{h_i} \left(\frac{1}{\sum_j W_{ij}}\right)^{1/d} \equiv \frac{1}{\eta_{\rm kernel}} \;,
\end{equation}

\noindent
where the equivalency is due to the definition of the smoothing length in Eqn.~\ref{eq:smoothinglength}. Note that the term in brackets is an approximation of the particle volume assuming neighbours with equal volumes. 

The reconstructed velocities appear in the artificial viscosity formulation through

\begin{equation}\label{eq:calcmu}
   \mu_{ij} = 
   \begin{cases}
   \dfrac {\tilde{\mathbf{v}}_{ij} \cdot \boldsymbol\eta_{ij}}{\boldsymbol\eta_{ij} \cdot \boldsymbol\eta_{ij} + \epsilon^2}& \;\text{for $\tilde{\mathbf{v}}_{ij} \cdot \boldsymbol\eta_{ij} < 0$}\;,\\
   0 & \;\text{otherwise}\;,
   \end{cases}
\end{equation}

\noindent
and similarly for $\mu_{ji}$ with all particle indices exchanged throughout the calculations. $\epsilon = 0.1$ is a small constant. Similarly to the artificial viscosity of \citet{monaghan1983shock}, each pressure term in the equations of motion is modified with the addition of a pairwise viscous pressure\footnote{$P_i$ becomes $P_i + Q_{ij}$ and $P_j$ becomes $P_j + Q_{ji}$.}. The viscous pressure terms $Q_{ij}$ combine a linear bulk viscosity term and a quadratic Von Neumann--Richtmyer viscosity term \citep{vonneumann1950method},

\begin{equation}\label{eq:calcQ}
   Q_{ij} = \frac{1}{2}\left(a_{\mathrm{visc}} + b_{\mathrm{visc}}\mathcal{B}_{i}^{\text{visc}}\right)\rho_i \left(-\alpha c_{i} \mu_{ij} + \beta \mu_{ij}^2\right) \;,
\end{equation}

\noindent
and similarly for $Q_{ji}$ with all particle indices exchanged throughout the calculations. The constants $\alpha$ and $\beta$ set the strengths of the bulk and Von Neumann--Richtmyer terms. The constants $a_{\mathrm{visc}}$ and $b_{\mathrm{visc}}$ set the strength of the viscosity in regions of different flow, based on the Balsara switch, $\mathcal{B}_{i}^{\text{visc}}$.

The REMIX artificial viscosity scheme differs from those of \citet{frontiere2017crksph} and \citet{rosswog2020lagrangian} in some notable aspects: firstly, the Balsara switch, $\mathcal{B}_{i}^{\text{SL}}$, is included in the slope limiter term (in Eqn.~\ref{eq:vtilde}). This avoids reducing the artificial viscosity where it is needed, leading to a more effective targeting of shocks. This allows us to introduce a factor of $1/2$ in Eqn.~\ref{eq:calcQ} to recover equations more closely equivalent to those in \citet{price2012smoothed}. Otherwise, the contributions from both $Q_{ij}$ and $Q_{ji}$ would effectively lead to this being a factor of $2$ stronger, which is to some extent mitigated by those schemes being ineffective at switching off velocity reconstruction in shocks. Secondly, we use $\alpha = 1.5$ and $\beta = 3$ as we find that these slightly larger constants, compared with $\alpha = 1$ and $\beta = 2$ as used by \citet{frontiere2017crksph} and \citet{rosswog2020lagrangian}, help to dissipate spurious oscillations in shocks in 3D. This is consistent with typical values used in planetary impact simulations \citep[e.g.][]{reinhardt2017numerical, canup2004simulations}. Thirdly, we use an additional Balsara switch directly in Eqn.~\ref{eq:calcQ}, which, combined with the values we use for $a_{\mathrm{visc}} = 2/3$ and $b_{\mathrm{visc}} = 1/3$, acts to switch between $\alpha = 1.5$ and $\beta = 3$ in shocks and $\alpha = 1$ and $\beta = 2$ in shearing regions. Here we make relatively conservative choices to limit the effect of artificial viscosity in smoothing particle noise in shearing regions, despite finding it to be a useful effect, owing to the velocity reconstruction to particle midpoints. Our artificial viscosity scheme is constructed to be less dissipative in shearing regions and to target shocks more effectively than similar schemes. These choices are all discussed in more detail in \ref{app:viscdiff}.

Note that REMIX does not include methods to explicitly redistribute particles towards more isotropic or otherwise error-reducing distributions, such as particle shifting techniques \citep{sun2017deltaplus, gao2023new} or regularisation approaches \citep{borve2001regularized}. Instead, similarly to \citet{frontiere2017crksph}, the artificial viscosity in our scheme helps smooth perturbations below the resolution scale that can otherwise lead to emerging anisotropic particle structures. 
%Alternative approaches would be valuable to investigate in the future.

\subsection{Artificial diffusion}\label{subsec:remix_diff}

Artificial diffusion of internal energy, or ``artificial conduction\footnote{In later sections, we use ``artificial diffusion'' to refer to cases that include the diffusion of both density and internal energy and ``artificial conduction'' where there is only diffusion of internal energy.}'', is frequently used to smooth accumulated noise in particle internal energies \citep{monaghan1997sph} or entropies \citep{schaller2015eagle}, and to improve the treatment of density discontinuities in ideal gas-only simulations \citep{price2008modelling}. As with artificial viscosity, a targeted approach is desirable to avoid artificial diffusion playing a dominant role in the thermodynamic evolution, instead of acting as a correction on the particle scale \citep{read2012sphs}. 

In some SPH schemes, relatively strong artificial conduction is used to address kernel smoothing at density discontinuities by smoothing particle internal energies over kernel length scales \citep{price2018phantom}. For a single equation of state, with no phase transitions, this leads to a smooth pressure field in the continuous limit. However, this is not an appropriate treatment in simulations with multiple and/or complex materials, where smooth density and internal energy fields do not necessarily lead to smooth pressures. Additionally, even in ideal gas-only simulations, this does not completely solve the issue, since (1) artificial conduction becomes a less effective correction at large density discontinuities; (2) in simulations with gravity, strong diffusion will disturb a system's hydrostatic equilibrium; (3) artificial conduction does not attempt to address the source of kernel smoothing error directly, instead it alters the physical system itself to one without discontinuities.

In simulations that use an evolved density estimate (Eqn.~\ref{eq:general_drhodt}), a similar artificial diffusion term can be used in the evolution of densities, for example, in the $\delta$-SPH formulation, used predominantly for engineering applications \citep{antuono2010free, antuono2012numerical, sun2021accurate}.

In REMIX, we include artificial diffusion of specific internal energy and of density, both to improve the treatment of shocks and to smooth accumulated noise on the particle scale, using reconstruction to particle midpoints \citep{rosswog2020lagrangian, antuono2010free}. Similarly to the phase dependence in the diffusion schemes of \citet{sun2021accurate} and \citet{pearl2022fsisph}, we only allow diffusion between particles of the same material type. Without this distinction, artificial diffusion of internal energy between different materials would cause unphysical evolution, since smoothing would be based on \textit{internal energy} and not \textit{temperature}. Diffusing density between different materials would lead to density discontinuities at material interfaces returning to a similar, smoothed state as in simulations with smoothing error in the density estimate.  

The diffusion terms in the equations of motion take the form

 \begin{align}\label{eq:u_diff}
\left(\frac{du_i}{dt}\right)_{\text{difn}} &=  \sum\limits_{j} \kappa_{ij} \left(a_{u} + b_{u} \mathcal{B}_{ij}^{\text{difn}} \right) v_{\text{sig}, ij} \: (\tilde{u}_{j} - \tilde{u}_{i}) \frac{m_j}{\rho_{ij}} \frac{1}{2}\left|\frac{d \widetilde{\mathcal{W}}}{d\mathbf{r}}\bigg|_{ij} - \frac{d \widetilde{\mathcal{W}}}{d\mathbf{r}}\bigg|_{ji}\right| \;,
\\
\label{eq:rho_diff}
\left(\frac{d\rho_i}{dt}\right)_{\text{difn}} &=  \sum\limits_{j} \kappa_{ij} \left(a_{\rho} + b_{\rho} \mathcal{B}_{ij}^{\text{difn}} \right) v_{\text{sig}, ij} \: (\tilde{\rho}_{j} - \tilde{\rho}_{i})  \frac{\rho_i}{\rho_j} \frac{m_j}{\rho_{ij}} \frac{1}{2}\left|\frac{d \widetilde{\mathcal{W}}}{d\mathbf{r}}\bigg|_{ij} - \frac{d \widetilde{\mathcal{W}}}{d\mathbf{r}}\bigg|_{ji}\right| \;,
\end{align}

\noindent
where $\kappa_{ij} = 1$ for particles of the same material and $\kappa_{ij} = 0$ otherwise. The average Balsara switch for each particle pair is used, $\mathcal{B}_{ij}^{\text{difn}} = (\mathcal{B}_{i} + \mathcal{B}_{j}) / 2$, for conservation. We take the signal velocity to be $v_{\text{sig},\; ij} = \left|\tilde{\mathbf{v}}_{i} - \tilde{\mathbf{v}}_{j}\right|$ and do not draw any distinctions between simulations with and without gravity (unlike some previous works \citep{price2018phantom, rosswog2020lagrangian}), since we aim to validate the full REMIX formulation independently of specific simulation properties. The parameters $a_{u}$ and $a_{\rho}$ set the strength of the artificial diffusion in shearing regions (where $\mathcal{B}_{ij}^{\text{difn}} \rightarrow 0$) and are increased to $a_{u} + b_{u}$ and $a_{\rho} + b_{\rho}$ in shocks. In shearing regions we choose to have low amounts of diffusion to avoid this strongly influencing thermodynamic evolution, and to allow for persisting and emergent discontinuities. We therefore use $a_{u} = a_{\rho} = 0.05$, similarly to \citet{rosswog2020lagrangian}. In the presence of shocks we find that we need a much larger amount of diffusion to prevent spikes in density and internal energy, and so we use $b_{u} = b_{\rho} = 0.95$. We motivate and test the sensitivity of these choices in \ref{app:viscdiff}. The volume elements in Eqn.~\ref{eq:u_diff} are chosen to conserve energy. In Eqn.~\ref{eq:rho_diff}, they include an additional ratio of densities, to conserve volume in each pairwise interaction\footnote{Substituting $\frac{d\rho_i}{dt} = -\frac{\rho_i^2}{m_i} \frac{dV_i}{dt}$ and solving for $\frac{dV_i}{dt}$ gives an equation antisymmetric in the exchange of particles.}. Although conserving volume in a pairwise interaction between particles is not strictly necessary, we find that it improves the treatment of the density diffusion in shocks.

When calculating the artificial diffusion terms, internal energies and densities are reconstructed to particle midpoints similarly to the velocities in the artificial viscosity scheme via

\begin{equation}\label{eq:utilde}
  \tilde{u}_{i} = u_{i} + \frac{1}{2}\Phi_{u,\;ij} \left(r_j^\gamma - r_i^\gamma \right)  \partial_{\kappa, \,i}^{\gamma} \, \hat{u} \;,
\end{equation}

\begin{equation}\label{eq:rhotilde}
  \tilde{\rho}_{i} = \rho_{i} + \frac{1}{2}\Phi_{\rho,\;ij} \left(r_j^\gamma - r_i^\gamma \right) \partial_{\kappa, \,i}^{\gamma} \, \hat{\rho} \;.
\end{equation}

\noindent
The derivatives are calculated using only particles of the same material species as

\begin{equation}\label{eq:deriv_samemat_u}
    \partial_{\kappa, \,i}^{\gamma} \, \hat{u} = \sum_j \kappa_{ij}  \, (u_j - u_i)  \,\partial_i^{\gamma} \hat{W}_{ij} \frac{m_j}{\rho_j} \;, 
\end{equation}

\begin{equation}\label{eq:deriv_samemat_rho}
    \partial_{\kappa, \,i}^{\gamma} \, \hat{\rho} = \sum_j \kappa_{ij}  \, (\rho_j - \rho_i)  \, \partial_i^{\gamma} \hat{W}_{ij}  \frac{m_j}{\rho_j} \;.
\end{equation}

\noindent
The material dependence of these gradients helps to preserve real discontinuities at material boundaries.

The slope limiter is calculated in the same way as for the viscosity, Eqn.~\ref{eq:calcphi}, but with $A_{ij} = A_{u, \;ij}$ and $A_{ij} = A_{\rho, \;ij}$ given by

\begin{equation}\label{eq:calcA_u}
   A_{u, \;ij} = \frac{\partial_{\kappa, \,i}^\alpha \, \hat{u} (\mathbf{r}_j - \mathbf{r}_i)^\alpha }{\partial_{\kappa, \,j}^\beta \, \hat{u} (\mathbf{r}_j - \mathbf{r}_i)^\beta} \;,
\end{equation}

\begin{equation}\label{eq:calcA_rho}
   A_{\rho, \;ij} = \frac{\partial_{\kappa, \,i}^\alpha \, \hat{\rho} (\mathbf{r}_j - \mathbf{r}_i)^\alpha }{\partial_{\kappa, \,j}^\beta \, \hat{\rho} (\mathbf{r}_j - \mathbf{r}_i)^\beta} \;.
\end{equation}

Although our diffusion scheme technically includes material dependence, this is not a correction targeted at material boundaries, nor with any dependence on the actual EoS. Rather, we actively turn off these parts of our method for particles of different species. Our artificial diffusion scheme is used to improve the treatment of shocks, and to weakly smooth accumulated noise. It is not used to address surface tension-like effects that prevent mixing and instability growth at density discontinuities, even in our ideal gas-only simulations. 

\subsection{Normalising term}\label{subsec:remix_norm}

We add a normalising term to the density evolution equation. This aims to evolve densities to reflect the distribution of mass in nearby particles, particularly in regions where particle volume elements systematically fail to satisfy the normalisation of the kernel. Since error accumulates in the evolution of densities based on timescales set by the divergence operator used in the equations of motion, we set the normalising term to act over timescales determined by the motion of particles. This also allows particles to move in response to changes in density caused by the normalising term.

Particle volume elements should approximately satisfy $\sum_j  W_{ij} V_j = 1$ (Eqn.~\ref{eq:m0_condition}) for a normalised kernel function. However, this condition will not be satisfied either if particle densities are poor estimates of the underlying field or if particle masses do not appropriately represent the mass distribution of the fluid. Our methods inherently conserve mass, as particle masses do not evolve during the simulation, and are fully Lagrangian. Therefore, we choose to maintain the simplicity and computational stability of this construction, and address discrepancies in volume elements through particle densities rather than through particle masses or their distribution. We do this by including an additional term in the density evolution, which we refer to as the ``normalising term'', that evolves densities towards a set of volume elements that aim to appropriately build up the continuous simulation volume. We note that the role of this term is not to obtain volume elements that exactly satisfy normalisation for all particle kernels at any given time, but rather to keep volume elements loosely tied to kernel normalisation and to address regions with systematic discrepancies.

To construct our normalising term, we consider the zeroth geometric moment of the unmodified kernel, 

\begin{equation}\label{eq:m0_not_averaged}
    m_{0,\,i} = \sum_j  W_{ij} V_j \;,
\end{equation}

\noindent
where $m_{0,\,i} = 1$ if the kernel $W_{ij}$ is normalised over the volume elements $V_j = m_j / \rho_j$. For a single particle $i$, we could trivially satisfy this condition by modifying the density of the particle and all its neighbours, $j$, by replacing $\rho_j$ with $m_{0,\,i} \,\rho_j$. However, this does not imply that $m_{0,\,j} = 1$ for all $j$, which will all have different $m_{0,\,j}$ and different sets of neighbours. But if there are systematic discrepancies in $m_{0}$ for many neighbouring particles, then modifying densities in a similar way for all these particles \textit{will} move them closer to $m_{0,\,j} = 1$. For instance, consider a region where particles have systematically too low density, leading to a local trend of $m_{0,\,j} > 1$. Here, increasing the densities will evolve these particles towards $m_{0,\,j} = 1$ and towards a density field that better represents the local mass distribution. In practice, we capture this behaviour with a smooth evolution in time. Unlike in the initial na\"ive example of modifying the densities of all $j$ to satisfy $m_{0,\,i} = 1$ for $i$ only, we evolve the density of $i$ only, based on its own $m_{0,\,i}$. This reduces the risk of emergent chaotic behaviour and still captures the desired behaviour in regions of systematic trends away from kernel normalisation. The normalising term in the density evolution equation takes the form

\begin{equation}\label{eq:drho_dt_normalisation_term}
    \left(\frac{d\rho_i}{dt}\right)_{\text{norm}} = \alpha_{\text{norm}} \, s_i \, (m_{0,\,i} - 1) \, \rho_{i} \sum\limits_{j}  v_{\text{norm}, \; ij} \:   \frac{m_j}{\rho_{ij}} \frac{1}{2}\left|\frac{d \widetilde{\mathcal{W}}}{d\mathbf{r}}\bigg|_{ij} - \frac{d \widetilde{\mathcal{W}}}{d\mathbf{r}}\bigg|_{ji}\right| \;,
\end{equation}

\noindent
where $\alpha_{\text{norm}} = 1$ is a constant and $v_{\text{norm},\; ij} = \left|\mathbf{v}_{i} - \mathbf{v}_{j}\right|$ is the effective signal velocity. Eqn.~\ref{eq:drho_dt_normalisation_term} aims to contribute to a weak evolution of $\rho_{i}$ towards $m_{0,\,i} \, \rho_{i}$. We include the vacuum switch, $s$, described in \S\ref{subsec:remix_vacuum}, since the kernel should not be normalised by particle volume elements at vacuum boundaries\footnote{At vacuum boundaries, one would instead expect $m_{0,\,i} \approx 1/2$.}. Here, we use the same volume elements and kernel gradient terms as are used in the diffusion of internal energy (Eqn.~\ref{eq:u_diff}), despite not being motivated by conservation in this term, since it does not represent the exchange of a quantity between particles. We use these so that the timescale of the normalising evolution is based on terms in the sum that are equal for both particles in each pairwise interaction. This prevents individual particles dominating in the corrective evolution. Using a timescale that depends on particle motion $v_{\text{norm},\; ij}$ rather than, for example the sound speed, allows particles to react and move in response to changes in density caused by the normalising term. We find that using an effective signal velocity that depends on the sound speed, even with a small multiplication factor, can lead to spatial oscillations in density, because densities change to attempt to satisfy normalisation faster than particles can respond to these changes.

We show the effect of this term in simulations in \S\ref{subsubsec:kh_idg_discontinuous} and \S\ref{subsec:planet}. In particular, we show that without this term, an example Jupiter-like planet in hydrostatic equilibrium will develop numerical instabilities as particles with low evolved densities, but are in regions of high particle number density, move from the planet's surface towards its core (\S\ref{subsec:planet}). In less extreme cases, the normalising term does not have a significant effect on hydrodynamics, although it does lead to particle densities that are generally closer to satisfying kernel normalisation, $m_{0,\,i} = 1$.

\section{Hydrodynamic Tests}\label{sec:hydrotests}

In this section, we validate REMIX in simulations to test its ability to capture physically realistic fluid behaviour. The primary tests are performed with particles of equal mass across the simulation, though we also include a subset of additional simulations for direct comparisons with past work, where particles are placed onto a regular grid but have different masses.  We refer to these two cases as ``equal mass'' and ``equal spacing'' throughout the following sections. The choice to focus on simulations with equal mass is made to validate our methods for science applications where particle densities and configurations can evolve significantly from their initial states, so particle masses cannot be easily chosen in the initial configuration to address errors. All simulations are performed in 3D, to account for effects that do not change predictably when increasing the number of dimensions, such as due to more freedom in particle configurations, or the change in scaling between neighbour number and length scale of particle interactions\footnote{A 2D simulation will have a lower number of neighbours for a given smoothing length than the equivalent 3D simulation. Increasing $\eta_{\rm kernel}$ to compensate for this would lead to kernel smoothing over a larger length scale.}. Additionally, in figures showing simulation snapshots, we deliberately plot individual particles rather than the smooth, reconstructed fields shown in some works. It is particularly important to visualise small-scale behaviour of simulations that aim to improve the treatment of density discontinuities where the effects that suppress mixing act on the particle scale.

We present results for the following hydrodynamic test scenarios: 
\begin{itemize}
  \setlength\itemsep{0.1em}

  \item {the square test (\S\ref{subsec:square}), where we investigate the treatment of density discontinuities in static equilibrium;}

  \item {the Sod shock tube (\S\ref{subsec:sod}), where we investigate the treatment of shocks;}

  \item{the Kelvin--Helmholtz instability both with an ideal gas EoS (\S\ref{subsec:kh_idg}) and between different, stiff materials set up to be representative of iron \& rock material boundaries in an Earth-like planet (\S\ref{subsec:kh_earth});}

  \item{the Rayleigh--Taylor instability, also both with an ideal gas EoS (\S\ref{subsec:rt_idg}) and with iron \& rock (\S\ref{subsec:rt_earth});}

  \item{the blob test (\S\ref{subsec:blob}), with which we investigate the onset of turbulence due to unseeded instabilities in both subsonic and supersonic regimes;}

  \item{the Evrard collapse (\S\ref{subsec:evrard}), which is used to test the interaction of our hydrodynamic treatment with gravity and shocks;}

  \item{and finally, planets in hydrostatic equilibrium (\S\ref{subsec:planet}), which we consider as a test scenario that combines gravity, complex-material boundaries, and a vacuum boundary.}

\end{itemize}

\noindent
The initial conditions needed to perform these tests are included as examples in the open-source \swift code.

We include comparisons with simulations carried out both using a traditional SPH formulation (``tSPH'') and a traditional formulation that includes artificial conduction of internal energy (``tSPH $+$ cond.''), with full details in \ref{app:trad}. These are used to demonstrate the motivation and need for many of the improvements in REMIX. Some comparison simulations carried out using MFM and MFV are also included in \ref{app:mfm_mfv}. We note that in most ideal gas tests, we follow the convention of past work and leave quantities unitless.

\subsection{Square test}\label{subsec:square}

The ``square test'' is used to investigate spurious surface tension-like effects from sharp discontinuities in a system that should be in static equilibrium \citep{saitoh2013density}. Here we test both an equal spacing scenario, i.e., with different particle masses in the two regions, and an equal mass scenario. The significant contributions from both smoothing and discretisation error (\S\ref{subsubsec:sph_smoothing_error}, \S\ref{subsubsec:sph_discretisation_error}) at the density discontinuity make the equal mass test particularly challenging for SPH.

A square (or cube) of fluid of higher density is initiated in pressure equilibrium with the surrounding region of low density fluid. Since the fluid experiences no gradients in pressure, other than those created by numerical errors, the shape of the square should not distort with time. In tSPH simulations, spurious surface tension-like effects at the density discontinuity leads to non-zero accelerations and a deformation of the square \citep{springel2010smoothed}. Typically, this test is carried out in 2D however, here we simulate a more challenging 3D cube with its effectively ``sharper'' higher-dimension corners, similarly to \citet{rosswog2020lagrangian}.

First, for the equal spacing scenario, we use initial conditions set up to match those of \citet{rosswog2020lagrangian}. $40^3$ particles are placed in a simple cubic lattice with spacing $1/40$ between adjacent particles. The simulation box is periodic and has length 1 in each of the $x$, $y$, $z$ directions. Masses are chosen such that $m_i = \rho(\mathbf{r}_i) / 40^3$, with densities $\rho = 4$ in the region $-0.25 < x, y, z < 0.25$ and $\rho = 1$ otherwise. An ideal gas EoS with $\gamma = 5/3$ is used for all particles. Initial internal energies are set to give a uniform pressure\footnote{We note the use of the unsmoothed density $\rho$ rather than the smoothed $\langle\rho\rangle$ used to set the internal energies of the initial conditions. Therefore tSPH simulations are not initialised in pressure equilibrium, due to smoothing error in the density estimate.} of $P(\rho, u)=2.5$. 

\begin{figure}[t]
	\centering
{\includegraphics[width=\textwidth, trim={2.5mm 0mm 2.5mm 0mm}, clip]{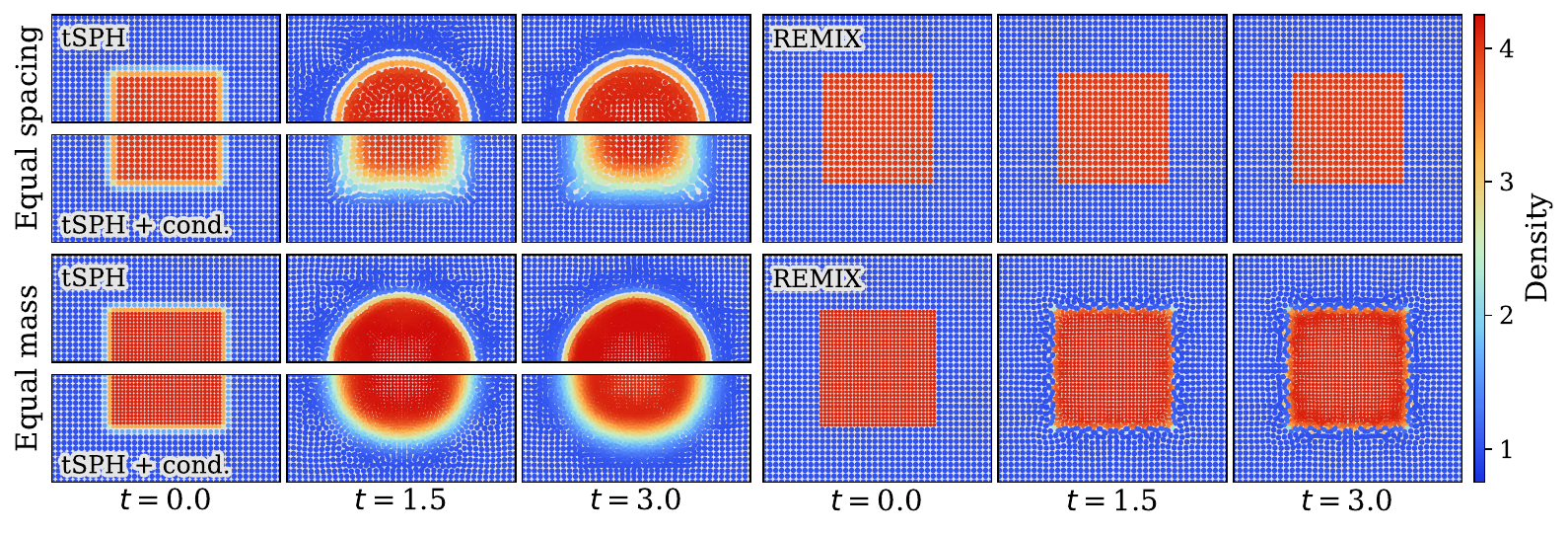}
}\hfill
	\vspace{-2em}
	\caption{Central cross-sections from 3D ``square test'' simulations. Snapshots of the cube are shown at three times from simulations with equal initial particle spacing (top full-row) and equal particle mass across the simulation (bottom full-row). Simulations were carried out using three SPH formulations: tSPH, tSPH with artificial conduction, and REMIX. Individual particles are plotted at their positions in the $x$--$y$ plane on a grey background, and coloured by their density. Particles in a slice of thickness 0.1 are plotted, so the grey background is visible in regions that have maintained their grid alignment in $z$ from the initial conditions.}
	\label{fig:square}
\end{figure}

The evolution of the equal spacing square test, carried out using each of tSPH; tSPH with artificial conduction; and REMIX, is shown in the top panels of Fig.~\ref{fig:square}. In the equal spacing scenario, the major contribution to spurious surface tension is due to the smoothing of the density field. The contribution of discretisation error is small, due to the well-ordered particle distribution and use of a relatively high-order kernel. With tSPH, the cube quickly deforms to a more stable, spherical shape, as illustrated by the upper, top-left panels of Fig.~\ref{fig:square}. Artificial conduction reduces the effects of smoothing error and so a square shape persists for longer, although the sharpness of the discontinuity is not maintained (Fig.~\ref{fig:square} lower, top-left panels). With REMIX, particle motion is negligible, relative to the particle separation, and the cube retains its shape  (Fig.~\ref{fig:square} top-right panels). This is in large part due to the use of the evolved density estimate, which prevents density smoothing -- and therefore spurious pressures -- at the discontinuities. Our choice of the free functions in the equations of motion and kernel construction also helps in reducing discretisation error to achieve these results.

Next, we consider the more challenging case for SPH: the use of equal mass particles, which leads to particles set up in considerably different grid-spacings interacting at the density discontinuity. Particles in the low density region are placed in the same configuration as in the equal spacing scenario. Then, instead of increasing particle masses in the high density region, the particle spacing is decreased and masses are kept the same as in the low density region. To satisfy these conditions while closely matching the density ratio in the equal spacing test, the high density region is given a grid-spacing of a factor 0.625 finer than the grid-spacing of the low density region. This corresponds to a density of 4.096. The new spacing of high density particles is chosen such that the layers of particles on either side of discontinuities are separated by the mean of the two grid-spacings, for all cube faces.

The evolution of this square test with equal mass particles is shown in the bottom panels of Fig.~\ref{fig:square}. There is now a large contribution of both smoothing and discretisation error in both of the traditional SPH formalisms. As such, the cube quickly deforms, even with conduction acting to reduce smoothing error. In the REMIX formulation, some minor deformation can be observed over these timescales. However, the general shape is maintained (Fig.~\ref{fig:square} bottom-right panels). We note that although past work typically shows 2D square test evolution over longer timescales than those of our plotted snapshots, our plots show times later than the comparable 3D tests in \citet{rosswog2020lagrangian}, beyond the time at which their equal spacing cubes have deformed. Reducing the effects of artificial surface tension requires all of (1) a density estimate that does not smooth density discontinuities, (2) our choice of equations of motion, and (3) improved gradient estimates. In the REMIX simulation, artificial diffusion is not the dominant source of correction, as discontinuities in both density and internal energy remain sharp.

If the linear-order reproducing kernels are used in the equations of motion without the antisymmetrisation, which is needed to enforce conservation, the square will remain undisturbed over much longer timescales, even in the equal mass case. The difference in outcome between using the conservative, antisymmetric construction and the exactly linear reproducing construction is sensitive to the kernel function used to construct the linear reproducing kernel. Therefore, reducing the additional error introduced in antisymmetrisation becomes an important consideration when choosing the form of the kernel from which the linear-order reproducing kernels are constructed. This can be seen in \ref{app:kernelconstruction}, where we present sensitivities in these results to different elements of the REMIX construction.

\subsection{Sod shock tube}\label{subsec:sod}

\begin{figure}[t]
	\centering
{\includegraphics[width=\textwidth, trim={2.5mm 0mm 2.5mm 0mm}, clip]{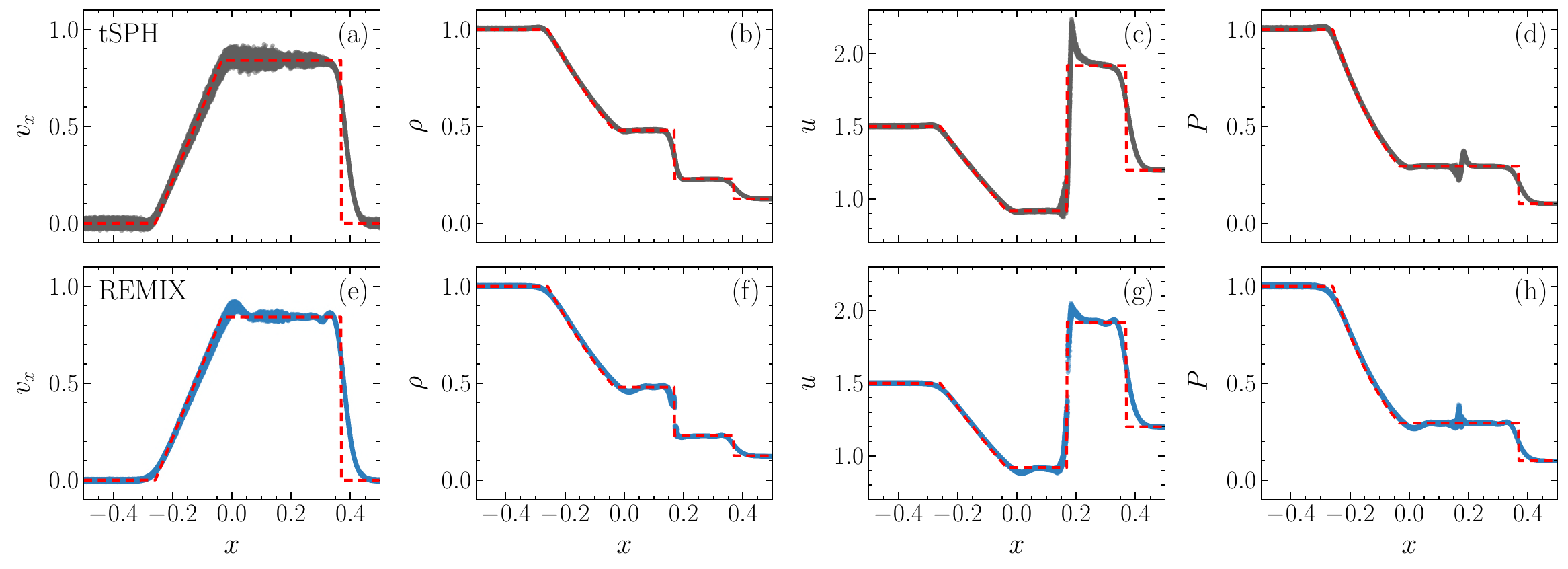}
}\hfill
	\vspace{-2em}
	\caption{3D Sod shock tube at time $t = 0.2$ simulated using tSPH and REMIX. Plots show velocity in the $x$-direction, $v_x$, density, $\rho$, specific internal energy, $u$, and pressure, $P$, of individual particles plotted against their $x$-position for tSPH (a--d) and REMIX (e--h) respectively. The red, dashed line shows a reference solution, solved for directly by using a Riemann solver. All particles are plotted.}
	\label{fig:sodshock}
\end{figure}

The ``Sod shock tube'' \citep{sod1978survey} is used to assess the shock capturing capabilities of our hydrodynamic scheme. This is a classic Riemann problem with a known analytic solution. Since the inclusion of artificial viscosity and diffusion are necessary to deal with shocks in the REMIX scheme, we also use this test to motivate choices made in the artificial viscosity and diffusion formulations, as detailed in \ref{app:viscdiff}. The choices made in the viscosity scheme relating to this test focus on reducing ringing oscillations behind the shock. The diffusion scheme focuses on reducing the size of spikes in density and internal energy at the discontinuity. 

Ideal gas, $\gamma = 5/3$, particles of equal mass are placed in a periodic 3D domain with size 2 in each of $x, y, z$ directions, centred at (0, 0, 0). We use two glass configurations, scaled appropriately for the two regions of different initial density: $\rho_1 = 1$ in the region $x < 0$ and $\rho_2 = 1/8$ in the region $x > 0$. Initial internal energies are set such that $P_1 = 1$ and  $P_2 = 0.1$. Simulations have a total of 589,824 equal mass particles.

The particle velocities in the $x$-direction, densities, internal energies, and pressures at a time $t = 0.2$ are shown in Fig.~\ref{fig:sodshock}. The shock is captured well with REMIX, and the particles align with the reference solution. Noise in particle velocities is reduced compared with tSPH. The size of the spike in internal energy is also reduced. The pressure blip could be further smoothed by increasing the strength of our artificial diffusion scheme, through choices of the $a$ and $b$ factors. However, we choose to take a conservative approach to artificial diffusion to avoid deviating far from the thermodynamically consistent core equations of motion.

\subsection{Kelvin--Helmholtz instability -- ideal gas}\label{subsec:kh_idg}

The Kelvin--Helmholtz instability (KHI) is the first test we use to investigate the treatment of mixing and dynamic instability growth in our simulations. 
The KHI arises at shearing interfaces in fluids \citep{chandrasekhar1961hydrodynamic}. Perturbations at the interface grow to form vortices that act to cascade energy to shorter length scales. As such, the KHI plays a significant role in the onset of turbulence in physical systems. Capturing the growth of the KHI has therefore been widely adopted as a benchmarking test to assess a numerical method's ability to simulate turbulence-driven mixing, as well as mixing on the particle scale. However, unlike the other tests above, an analytical solution does not exist for the KHI. 

Here we first consider the growth of these instabilities at shearing density contrasts in an ideal gas. All simulations presented are carried out in 3D, with a thin  $z$ direction depth relative to the other dimensions, similarly to \citet{hopkins2013general}, \citet{read2010resolving}, and \citet{rosswog2020lagrangian}. We focus primarily on cases with a sharp density discontinuity and equal mass particles. This is in contrast with an alternative setup with which we directly compare our results with a reference solution \citep{mcnally2012well}, where we consider an initially smoothed discontinuity and equal particle spacing. Although the use of this second form of initial conditions with smooth initial densities and velocities is motivated by the existence of a converged solution, these choices change the physical system to one with inherently less smoothing and discretisation error, which are the main effects of interest that normally suppress instability growth in SPH simulations. These smooth initial conditions therefore do not give the full picture of an SPH scheme's ability to capture KHI growth at sharp density discontinuities, where these sources of error can play a dominant role. This is particularly important at material boundaries, where smoothing the density discontinuity between different materials may lead to particles of both materials occupying extreme regions of their EoS phase space, so considering deliberately smoothed, equilibrium initial conditions would not be representative of a physical system.

Traditional formulations of SPH struggle to capture the KHI \citep{agertz2007fundamental}, with the growth of the instability being strongly suppressed. In particular, for shearing density discontinuities, smoothing in the density estimate leads to surface tension-like effects that act to artificially stabilise the interface. Additionally, for simulations where SPH particles in both density regions have equal mass, or configurations that give similarly anisotropic local particle distributions at the interface, leading-order error in the momentum equation will also contribute significantly to this spurious surface tension-like effect. Not only do these effects act to suppress mixing by hampering the large-scale evolution of naturally arising instabilities that should act to drive mixing, but they will also impede particles crossing interfaces, thereby suppressing mixing both indirectly and directly.

The growth of a mode of wavelength $\lambda$ is characterised by the timescale \citep{rosswog2020lagrangian,price2008modelling}

\begin{equation}\label{eq:tau_KH}
   \tau_{\rm KH} = \frac{(\rho_1 + \rho_2) ~\lambda}{\sqrt{\rho_1 \rho_2} \, |v_1 - v_2|} \;,
\end{equation}

\noindent
where $\rho_1$ and $\rho_2$ are the densities in regions separated by the shearing interface and $|v_1 - v_2|$ is their relative speed. We use this parameterisation so that comparisons can be drawn at the same $\tau_{\rm KH}$ between simulations with different initial conditions, since we consider KHIs with both smoothed and sharp interfaces, for different density ratios, and between different materials. We note that initial conditions with and without initial smoothing of fields at the interface are physically different systems, so we do not expect converged results between the two.

In the absence of stabilising influences such as physical surface tension or gravity, a shearing discontinuity is unstable to perturbation modes of all wavenumbers \citep{chandrasekhar1961hydrodynamic}. In a realistic system satisfying these conditions, instability will always be triggered, as even the smallest local inhomogeneities will seed mode growth. Similarly, in a simulation, numerical error will inevitably trigger instability at shearing discontinuities. The wavenumbers of error-seeded modes are sensitive not only to the numerical methods used and the construction of initial conditions, but also to the resolution of the simulation: a higher resolution simulation will be able to resolve the excitation of a wider range of mode wavelengths \citep{robertson2010computational}. The growth of KHIs at sharp discontinuities can therefore not be used reliably for convergence studies.

\citet{mcnally2012well} and \citet{robertson2010computational} construct KHI initial conditions with smooth initial velocities and densities across the shearing interface. They show that the inclusion of a well-resolved transition region acts to stabilise the system, suppressing modes other than those deliberately seeded in the initial conditions. They demonstrate convergence and present a well-posed method to benchmark the early evolution of KHI simulations. In \S\ref{subsubsec:kh_idg_smooth} below, we present REMIX simulations using the initial conditions of \citet{mcnally2012well}, including quantitative comparisons of mode growth with their converged reference solution. In \S\ref{subsubsec:kh_idg_discontinuous}, we present KHI simulations with sharp discontinuities in density and velocity across the interface. Although we cannot make quantitative comparisons of this more challenging case with converged reference solutions, useful comparisons can still be drawn between simulations and the expected qualitative behaviour of the instability, with a motivation of reducing the clear suppression of the KHI observed when using traditional SPH. We additionally use equal mass particles across the simulation, making this setup particularly challenging for SPH schemes, but more applicable to most science applications. In \S\ref{subsubsec:kh_idg_1_10} we present KHI simulations with a larger density ratio, a discontinuous interface, and equal mass particles. This system is even more challenging again for SPH schemes: both smoothing and discretisation errors are increased here due to the larger density-smoothing effects and the even more extreme local anisotropy in particle distribution at the interface. After considering these ideal gas scenarios, we present KHI simulations at interfaces between dissimilar, stiff materials in \S\ref{subsec:kh_earth}.

\subsubsection{KHI with smooth initial conditions}\label{subsubsec:kh_idg_smooth}

% maybe special lettering for PENCIL

%[throughout this and later sections, make sure $v_1$, $v_2$ have signs that match figures]

\begin{figure}[t]
	\centering
{\includegraphics[width=\textwidth, trim={2.5mm 0mm 2.5mm 0mm}, clip]{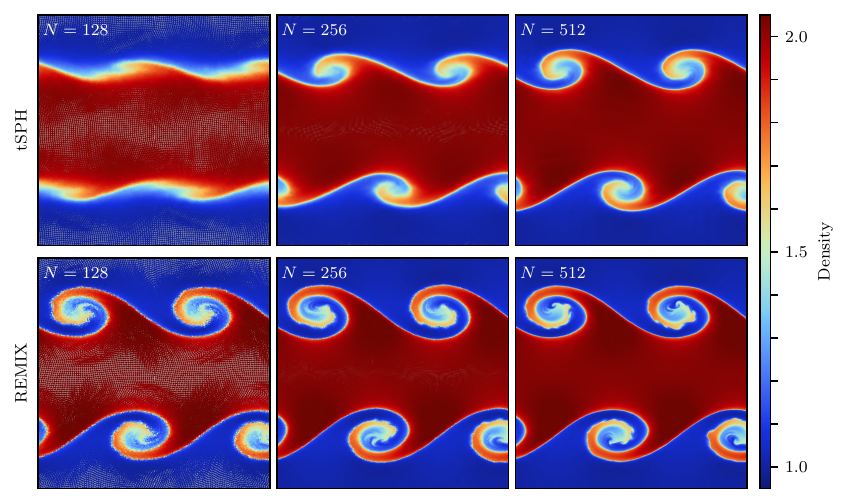}
}\hfill
	\vspace{-2em}
	\caption{Ideal gas Kelvin--Helmholtz instabilities with smoothed initial density and velocity profiles. Columns correspond to simulations of different resolutions, with the top row showing results from simulations using tSPH and the bottom row from simulations using REMIX. We plot snapshots at $t =2~\tau_{\rm KH}$ from 3D, ideal gas KHI simulations. The density ratio between the two regions is 1:2. Here particles of different mass are used to match consistent initial particle spacing and volume across the simulation. Individual particles are plotted on a grey background and coloured by their density. Particles at all $z$ are plotted, so the grey background is visible in regions that have maintained their grid alignment in $z$ from the initial conditions.}
	\label{fig:kh_idg_smooth}
\end{figure}

\begin{figure}[t]
	\centering
{\includegraphics[width=0.55\textwidth]{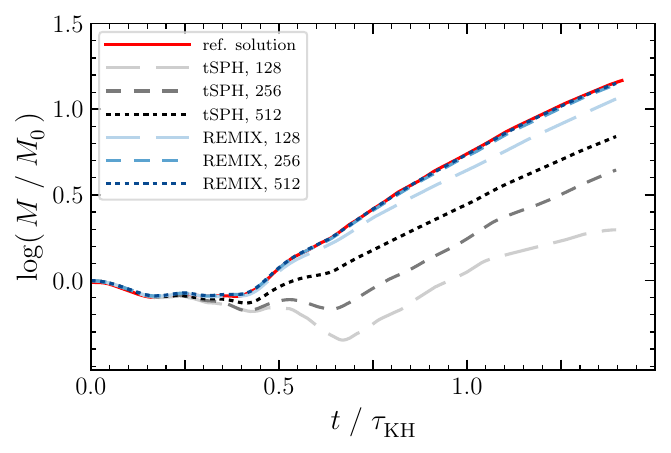}
}\hfill
	\\\vspace{-1em}
	\caption{Evolution of mode amplitude, $M$, in Kelvin--Helmholtz simulations with smoothed initial density and velocity profiles. We plot mode growth for simulations with three different resolutions ($N = 128$, $256$, $512$) using both tSPH (grey, dashed) and REMIX (blue, dashed). Mode amplitude is normalised to the initial amplitude of the excited mode, $M_0$, and time is normalised to the characteristic timescale of KHI growth, $\tau_{\rm KH}$. The reference solution (red, solid) corresponds to the 4096 $\times$ 4096 simulation of \citet{mcnally2012well} using  the \pencil code.}
	\label{fig:modes_idg}
\end{figure}

\citet{mcnally2012well} present converged, high-resolution simulations of the early linear growth of the KHI. They use initial conditions with smooth initial velocity and density fields across the shearing interface. Similarly to \citet{rosswog2020lagrangian}, here we use these smooth initial conditions, adapted to 3D, and use the mode growth of the reference solution of \citet{mcnally2012well} to quantitatively assess the accuracy of our numerical methods.

Particles are initialised in a 3D cubic lattice in a periodic box with $N \times N \times 18$ particles in $x, y, z$ directions (i.e. a thin slice in the $z$ direction relative to $x$ and $y$). We run simulations with resolutions $N = 128$, $256$, $512$. Spatial dimensions are normalised to the size of the simulation box length in the $x$ and $y$ directions. A low density region of  $\rho_1 = 1$ shears against a high density region of $\rho_2 = 2$ with speeds in the $x$ direction of $v_1 = -0.5$ and $v_2 = 0.5$ such that the relative velocity is $|v_1 - v_2| = 1$. The regions are layered in $y$ and have relative velocities in $x$. However, both density and shearing velocity are smoothed at the shearing interface such that initial densities are given by

\begin{equation}\label{eq:KHidgsmoothdensities}
  \rho(y) = \begin{cases}
      \rho_1 - \rho_m e^{(y - 0.25)/\Delta} & \text{for $0.00 \leq y < 0.25$} \,, \\
      \rho_2 + \rho_m e^{(0.25 - y)/\Delta} & \text{for $0.25 \leq y < 0.50$} \,,\\
      \rho_2 + \rho_m e^{(y - 0.75)/\Delta} & \text{for $0.50 \leq y < 0.75$} \,,\\
      \rho_1 - \rho_m e^{(0.75 - y)/\Delta} & \text{for $0.75 \leq y < 1.00$} \;,
    \end{cases}    
\end{equation}

\noindent
and initial velocities in the $x$ direction are given by

%[make sure that changing signs of $v_1$, $v_2$ hasn't cause problems here]
\begin{equation}\label{eq:KHidgsmoothvelocities}
  v_x(y) = \begin{cases}
      v_1 - v_m e^{(y - 0.25)/\Delta} & \text{for $0.00 \leq y < 0.25$} \,,\\
      v_2 + v_m e^{(0.25 - y)/\Delta} & \text{for $0.25 \leq y < 0.50$} \,,\\
      v_2 + v_m e^{(y - 0.75)/\Delta} & \text{for $0.50 \leq y < 0.75$} \,,\\
      v_1 - v_m e^{(0.75 - y)/\Delta} & \text{for $0.75 \leq y < 1.00$} \;.
    \end{cases}    
\end{equation}

\noindent
Here $\rho_m = (\rho_1 - \rho_2)/2$, $v_m = (v_1 - v_2)/2$, and $\Delta = 0.025$. Since particle positions are initialised in a single cubic lattice, particle masses are set by $m_i = \rho(y_i) / N^3$. Particle internal energies are set to give a pressure of $P(\rho, u) = 2.5$ across the simulation for an ideal gas with $\gamma = 5/3$. A small velocity perturbation,  $v_y = 0.01 \sin{(2 \pi x / \lambda)}$, is added in the $y$ direction with wavelength $\lambda = 0.5$, to seed the primary instability. 

The simulated KHI with these initial conditions is shown in Fig.~\ref{fig:kh_idg_smooth}. We plot particle densities at particle positions for simulations of resolution $N = 128$, $256$, $512$. Top row plots correspond to tSPH and bottom to REMIX. All snapshots are shown at simulation time $t = 2~\tau_{\rm KH}$. Traditional SPH struggles to capture this instability at low resolutions. In REMIX simulations the seeded mode is not suppressed and grows at a close to resolution-independent rate. We find, however, that at later times secondary modes will eventually grow and disturb the evolution of the primary mode, so we do not observe strict convergence over long timescales. For an SPH scheme aiming to model an inviscid fluid with realistic turbulence-driven mixing, a compromise on this is difficult to avoid.

The evolution of the amplitude of the seeded mode is shown in Fig.~\ref{fig:modes_idg}, for these simulations. This quantity, $M$, is calculated from Eqns.~10--13 of \citet{mcnally2012well}. We normalise the mode amplitude to $M_0 \equiv M(t=0)$ to allow for more direct comparisons between simulations with different initial conditions, presented later. The reference solution is from the high-resolution $4096^2$ cell KHI simulation performed by \citet{mcnally2012well} using the Eulerian mesh, finite-difference code \pencil. The mode growth of the tSPH simulations falls
significantly short of the reference solution.  This result is consistent with the SPH simulations used for comparisons by \citet{mcnally2012well}. In contrast, the mid- and high-resolution REMIX simulations closely match the reference solution, and even the lowest resolution REMIX simulation is considerably closer to the reference solution than the highest resolution tSPH simulation.

\subsubsection{KHI with discontinuous initial conditions}\label{subsubsec:kh_idg_discontinuous}

\begin{figure}[t]
	\centering
{\includegraphics[width=\textwidth, trim={2.5mm 0mm 2.5mm 0mm}, clip]{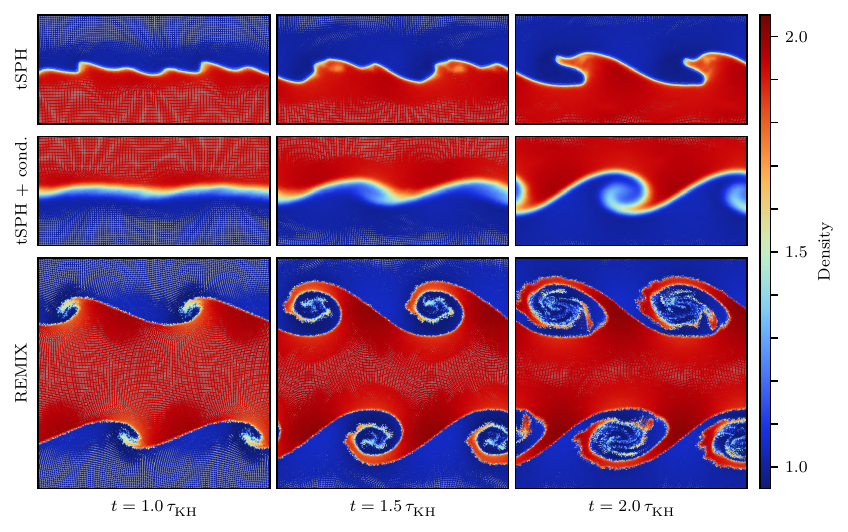}
}\hfill
	\vspace{-2em}
	\caption{Growth of 3D Kelvin--Helmholtz instabilities in the more challenging case of discontinuous initial density and velocity profiles and equal mass particles. Columns show snapshots at different times, with the top rows showing results from simulations using tSPH -- without and with artificial conduction -- and the bottom row from simulations using REMIX. These simulations are both relatively low resolution, with $N_1 = 128$. The density ratio between the two regions is 1:1.91.}
	\label{fig:kh_idg}
\end{figure}

\begin{figure}[t]
	\centering
{\includegraphics[width=\textwidth, trim={2.5mm 0mm 2.5mm 0mm}, clip]{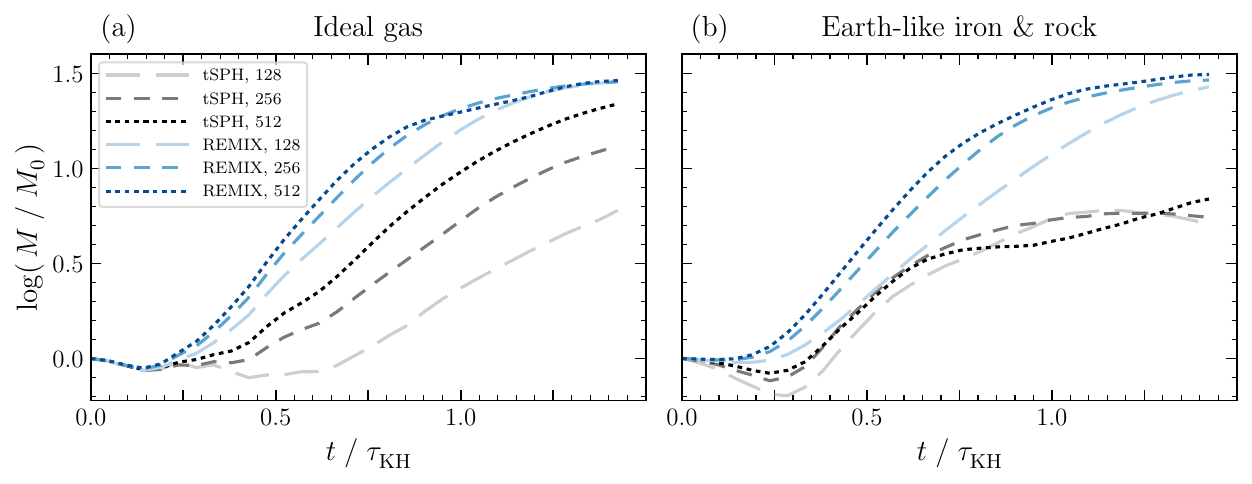}
}\hfill
	\vspace{-2em}
	\caption{Evolution of Kelvin--Helmholtz instability mode amplitude, $M$, in sharp-discontinuity, equal mass KHI simulations with (a) an ideal gas EoS (\S\ref{subsubsec:kh_idg_discontinuous}) and (b) between dissimilar, stiff EoS (\S\ref{subsec:kh_earth}). Mode growth in simulations using both tSPH (grey, dashed) and REMIX (blue, dashed) is shown for three different resolutions ($N_1 = 128$, $256$, $512$). The mode amplitude is normalised to the initial amplitude of the excited mode, $M_0$, and time is normalised to the characteristic timescale of KHI growth, $\tau_{\rm KH}$, which are both different for each case.}
	\label{fig:modes_sharp}
\end{figure}

Next we consider KHI growth at an interface that is discontinuous in density and velocity. A shearing discontinuous interface is unstable to modes of all wavelengths, so noise- or error-seeded secondary modes will inevitably lead to a turbulent system and preclude numerical convergence. Although no converged reference solution exists for this problem, we use this system to qualitatively demonstrate the suppression in tSPH of both instability growth at, and mixing across, density discontinuities, and the effectiveness of REMIX in alleviating these issues.  We deliberately constrain our analysis to low-resolution simulations, where the primary, intentionally seeded mode remains relatively undisturbed by secondary modes during the early growth of the instability (as discussed further in \ref{app:kh_secondary}).

Similarly to in \S \ref{subsubsec:kh_idg_smooth}, we consider shearing between a low density region of  $\rho_1 = 1$ and a high density region of $\rho_2 \approx 2$ with relative speeds of $v_1 = -0.5$ and $v_2 = 0.5$. Here we initialise particles with a sharp discontinuity in both density and shearing velocity. The low density region is set up in the same cubic lattice as in the smoothed simulations of the previous section.  We use particles of equal mass across the simulation. We refer to the resolution of these simulations by the effective resolution of the low density region: $N_{1} = 128$, $256$, $512$; if the box were filled with a cubic lattice of the low density material only, then this lattice would consist of $N_{1} \times N_{1} \times 18$ particles in $x, y, z$ directions. Particles in the high density region are arranged in a cubic lattice of shorter spacing. A cubic lattice corresponding to a density $\rho_2 = 2$ is adjusted to allow a continuous grid in the $x$ dimension of the periodic box\footnote{We also enforce that the effective resolution in this region, $N_{2}$, is divisible by $4$ (the number of vortices formed by the evolution of the seeded mode) to avoid the possibility of asymmetric evolution of individual vortices triggering an early onset of secondary modes. In practice this has no noticeable effect here, but similar considerations do matter for the Rayleigh--Taylor instability simulations we examine in \S\ref{subsec:rt_idg} and \S\ref{subsec:rt_earth}.}. The spacing of particles in $z$ is slightly adjusted away from a perfectly cubic lattice such that particle spacing in this dimension is also continuous across the boundary of the box. The regions are shifted in the $y$ direction such that the layers of particles across the interface from each other, directly adjacent to, and parallel with, the discontinuity are separated by the mean of the two grid-spacings. The size of the simulation box is adjusted in the $y$ direction to compensate for this and to maintain continuity across boundaries of the periodic box. The density $\rho_2$ is recalculated based on these grid modifications and the use of equal mass particles. To satisfy these conditions, in the high density region we use $\rho_2 = 1.91$ in a lattice with, for example,  $N_{2} = 160$ and $22$ particles in the $z$ direction for $N_{1} = 128$. Initial internal energies are calculated such that particles have a uniform initial pressure\footnote{We note here that the density used in these initial conditions is unsmoothed, so the tSPH simulations will not be in pressure equilibrium due to their smoothing of the densities at the discontinuities.} of $P(\rho, u) = 2.5$ by the ideal gas equation with $\gamma = 5/3$. We seed a small velocity perturbation,  $v_y = 0.01 \sin{(2 \pi x / \lambda)}$, in the $y$ direction with $\lambda = 0.5$. 

In Fig.~\ref{fig:kh_idg} we show the growth of these KHIs simulated using tSPH, tSPH with artificial conduction, and REMIX. We plot individual particles, coloured by their densities, at three times through the evolution of the instability. In the tSPH simulations, surface tension-like effects act both to suppress the growth of the instability and to prevent mixing of particles across the interface. As noted by  \citet{agertz2007fundamental}, particles form ordered bands with large gaps at the interface, which act as barriers to mixing. Artificial conduction helps to enable some mixing on the particle scale, allowing the boundary to become diffuse with time. However, the evolution of the instability is slow, as can be seen when comparing with the similar sharp-interface KHI simulations of \citet{hopkins2015new}, performed with their improved methods, at comparable scaled times (their Fig.~21). While we note that differences in the construction of initial conditions mean that we cannot make direct comparisons, the growth of the instability in both traditional cases is clearly too slow. The REMIX simulation shows a clear improvement: not only do the characteristic vortices of the KHI form without impedance by surface tension-like effects, but interfaces are maintained as sharp discontinuities as the system evolves. Particles do not align themselves in bands separated by gaps that would prevent mixing across the discontinuity.

The mode amplitude growth of these KHIs and equivalent higher resolution simulations are plotted in Fig.~\ref{fig:modes_sharp}(a). Since this system is constructed differently from that in \S\ref{subsubsec:kh_idg_smooth}, we cannot draw direct comparisons between these results and the converged reference solution for a smoothed interface. For example, the instability grows more quickly in this case where the shearing velocity is discontinuous. However, we do observe qualitatively similar behaviour when comparing Fig.~\ref{fig:modes_sharp}(a) with Fig.~\ref{fig:modes_idg}. The seeded mode grows more quickly in REMIX simulations than in those using tSPH. The early growth rate of modes in REMIX simulations is slightly steeper as resolution is increased, mirroring the behaviour of the analogous simulations in Fig.~\ref{fig:modes_idg}. The approach of the mode evolution of tSPH simulations towards the REMIX simulations is also similar here, and again, the lowest resolution REMIX simulation grows more quickly than the highest resolution tSPH simulation. Despite this behaviour with increased resolution, high-resolution tSPH simulations still fail to form spiralling plumes, as surface tension-like effects continue to dominate, as shown in \ref{app:kh_secondary}.

\begin{figure}[t]
	\centering
{\includegraphics[width=\textwidth, trim={2.5mm 0mm 2.5mm 0mm}, clip]{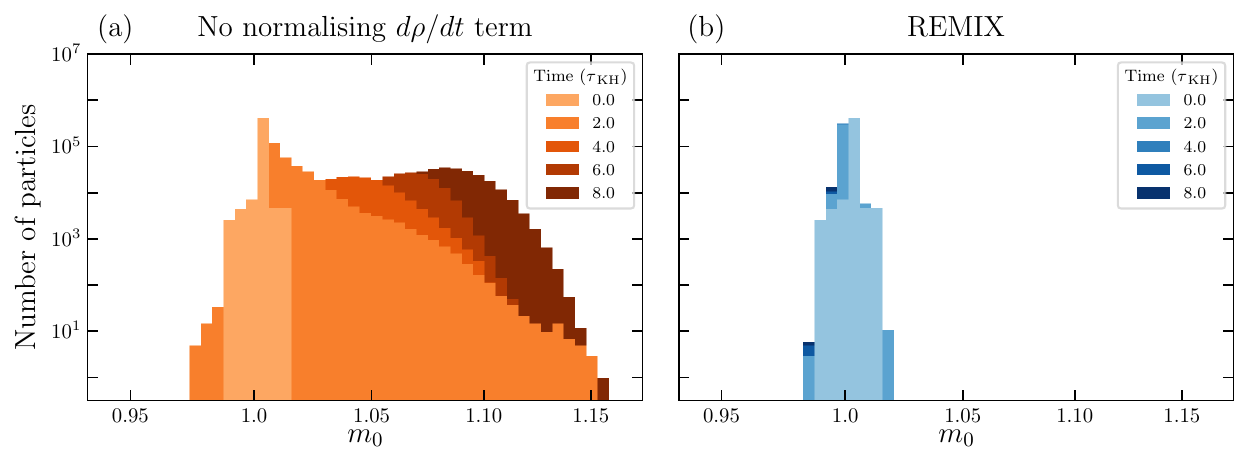}
}\hfill
	\vspace{-2.5em}
	\caption{The evolution of the distribution of the $m_0$ kernel geometric moment in ideal gas KHI simulations with sharp discontinuities. $m_0 \approx 1$ corresponds to a local distribution of densities that reflects the particle configuration. Plots show results from simulations with resolution $N_1 = 128$ using the REMIX scheme both (a) without and (b) with the normalising term in the density evolution.}
	\label{fig:kh_m0}
\end{figure}

The effect of the normalising term (\S \ref{subsec:remix_norm}) in a KHI simulation with sharp discontinuities is demonstrated in Fig.~\ref{fig:kh_m0}. Without it, as the simulation evolves, $m_0$ of some particles  drifts away from 1, the value corresponding to normalisation of the unmodified kernel (see Eqn. \ref{eq:m0_not_averaged}). The normalising term ties the density evolution to kernel normalisation, so as the system evolves, volume elements continue to accurately build up the continuum over which the kernel function is normalised. In these simulations, the drift in $m_0$ does not noticeably affect the simulation outcome, however, in \S\ref{subsec:planet}, we show an example where the inclusion of the normalising term is necessary to simulate a system in hydrostatic equilibrium.

\begin{figure}[h!]
	\centering
    \vspace{-2em}
{\includegraphics[width=\textwidth, trim={2.5mm 0mm 2.5mm 0mm}, clip]{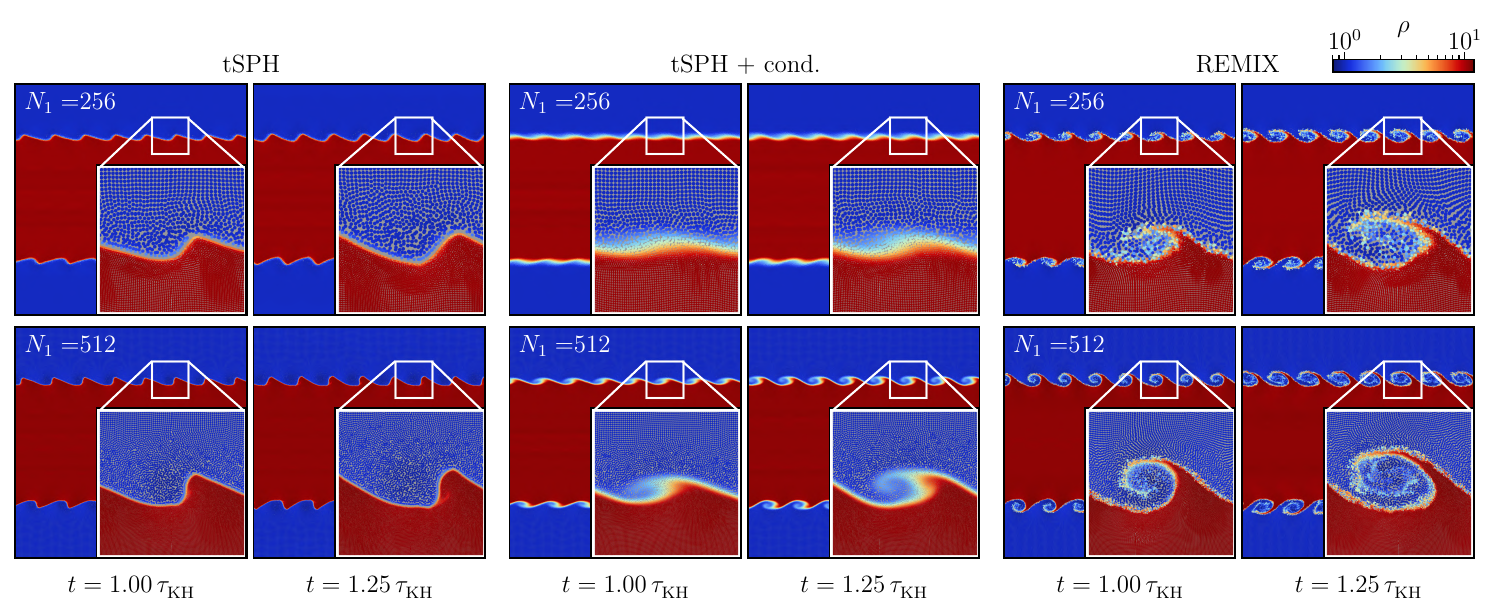}
}\hfill
	\vspace{-2em}
	\caption{Growth of Kelvin--Helmholtz instabilities between layers with a large density contrast. We plot snapshots from 3D, ideal gas KHI simulations with discontinuous initial density and velocity profiles. We show results from simulations carried out using tSPH, tSPH with artificial conduction, and REMIX. Snapshots are plotted at two times, from simulations with two resolutions. The density ratio between the two regions is 1:10.4 and particles have the same mass across the simulation. Insets show a magnified view of a KHI plume.}
	\label{fig:kh_idg_1_10}
\end{figure}

\subsubsection{KHI with a large density ratio}\label{subsubsec:kh_idg_1_10}

Capturing the KHI at interfaces in fluids with a large density jump is additionally challenging for SPH. More smoothing in the density estimate and larger discretisation error, in equal mass particle simulations, will make surface tension-like effects stronger at larger density contrasts. Additionally, using artificial conduction as a method for correcting density discontinuity is not as effective at larger jumps in density \citep{price2008modelling}. Our initial conditions aim to follow those of \citet{price2008modelling} with a density ratio of 1:10, however, we continue to use 3D simulations to validate our methods for more typical applications.

Here we construct initial conditions similarly to \S \ref{subsubsec:kh_idg_discontinuous}: sharply discontinuous in both density and shearing velocity. The low density region is constructed exactly equivalently with $\rho_1 = 1$, while resolution is increased in the high density region, following the same method as outlined previously, such that this region has a density of $\rho_2 = 10.4$. Speeds in the $x$ direction are again set to $v_1 = -0.5$ and $v_2 = 0.5$, however the wavelength of the initial perturbation in the $y$ direction is decreased to $\lambda = 0.128$, although with the same amplitude of $0.01$ \citep{price2008modelling}. Comparisons of resolution can not be directly drawn to the previous section, as here fewer particles will make up individual vortices at a given time due to the decreased perturbation wavelength.

In Fig.~\ref{fig:kh_idg_1_10}, we plot snapshots showing the evolution of these initial conditions in tSPH, tSPH with conduction, and REMIX simulations for two resolutions. Due to the lower wavelength of the seeded mode compared with that in previous sections, we consider simulations with overall higher resolutions, although this does not necessarily correspond to higher resolution in each individual vortex, which now occupies a smaller region in the simulation box. The instability fails to grow with tSPH and grows only slowly in the higher resolution simulation with conduction. However, the instability is captured successfully with REMIX, in particular at the higher-resolution, where we capture spiralling within the plume.

\subsection{Kelvin--Helmholtz instability -- Earth-like iron \& rock}\label{subsec:kh_earth}

\begin{figure}[t]
	\centering
{\includegraphics[width=\textwidth, trim={2.5mm 0mm 2.5mm 0mm}, clip]{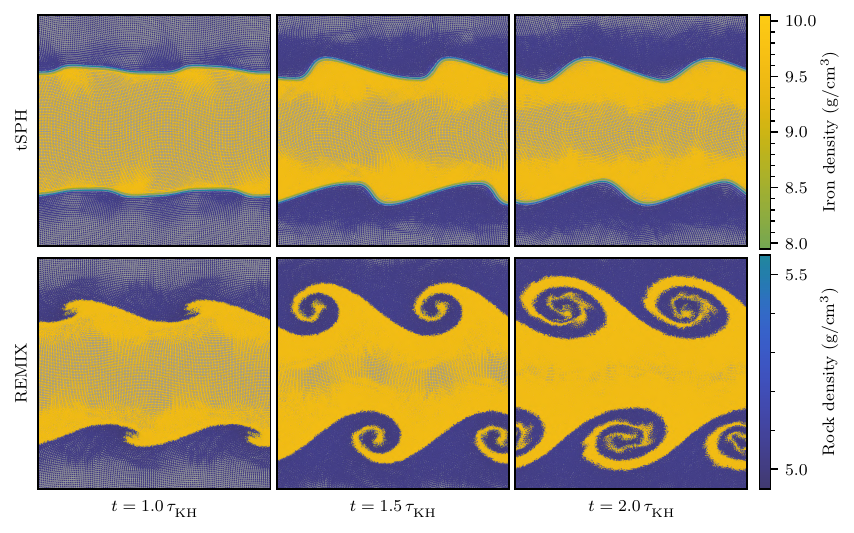}
}\hfill
	\vspace{-2.5em}
	\caption{Kelvin--Helmholtz instability growth between dissimilar, stiff materials. We plot snapshots from 3D KHI simulations with multiple, complex equations of state at densities and pressures representing those at material boundaries within the Earth. Columns show snapshots at different times with the top row showing results from simulations using tSPH and the bottom row using REMIX. The initial density and velocity profiles are discontinuous and particles have equal mass. These simulations are both relatively low resolution, with $N_1 = 128$.  Individual particles are plotted on a grey background and coloured by their material type and density. Particles at all $z$ are plotted, so the grey background is visible in regions that have maintained their grid alignment in $z$ from the initial conditions.}
	\label{fig:kh_earth}
\end{figure}

\begin{figure}[t]
	\centering
{\includegraphics[width=0.8\textwidth]{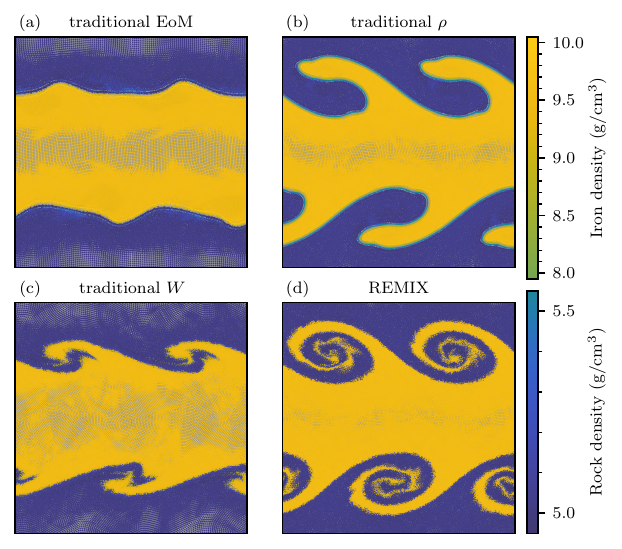}
}\hfill
	\\\vspace{-1em}
	\caption{Kelvin--Helmholtz instability simulations illustrating the interplay between multiple component methods of the REMIX scheme. Here we show results from REMIX simulations when using: (a) a more traditional form of the equations of motion; (b) the traditional, integral-form of the density estimate, instead of the evolved density; (c) an unmodified Wendland $C^2$ kernel, instead of the linear-order reproducing kernels; (d) the full REMIX SPH scheme. Removing any one of these affects the growth of the instability significantly. We plot snapshots at $t =2~\tau_{\rm KH}$ from 3D simulations with multiple, stiff equations of state at densities and pressures representative of those at the core-mantle boundary within the Earth.
 }
	\label{fig:kh_earth_methods}
\end{figure}

Since the evolution of the KHI is predominantly inertial, we expect instabilities to grow similarly between shearing fluids of different materials, represented in our simulations as inviscid fluids only differing in the calculation of pressures and sound speeds through the EoS (\S\ref{subsec:eos}). We construct similar initial conditions to those used in \S\ref{subsubsec:kh_idg_discontinuous}, but using the ANEOS Fe\textsubscript{85}Si\textsubscript{15} (iron) and forsterite (rock) EoS with densities and pressures comparable with those of the Earth's core-mantle interface \citep{stewart2020shock}.

We simulate the KHI at a discontinuity between low-density rock at $\rho_1 = 5000$~kg~m$^{-3}$ and high-density iron at $\rho_2 = 9550$~kg~m$^{-3}$. Particles are placed in a periodic box in a configuration exactly matching that of \S\ref{subsubsec:kh_idg_discontinuous}. These simulations use particles of equal mass. Spatial dimensions are scaled such that the box spans a length of $1~R_\oplus = 6371$~km in the $x$ and $y$ dimensions. The velocities in $x$ of the two layers are initialised to $v_1 = -10^{-4}~R_\oplus$~s$^{-1}$, $v_2 = 10^{-4}~R_\oplus$~s$^{-1}$ and the seeded mode has the form $v_y = 0.01 |v_1 - v_2| \sin{(2 \pi x / \lambda)}$ with $\lambda = 0.5~R_\oplus$. Initial internal energies are calculated through each material's EoS such that the regions are in pressure equilibrium with $P(\rho, u) = 1.2\times10^{11}$~Pa.

In Fig.~\ref{fig:kh_earth} we show the evolution of a KHI with these initial conditions using tSPH and REMIX. In the tSPH simulation, surface tension-like effects are strong. Undesired smoothing of the discontinuity in the SPH density estimate combined with the stiff equations of state leads to strong artificial forces at the interface, which both prevent mixing of particles of different materials and strongly suppress the growth of the instability. These effects as well as their contributions from zeroth-order error in the momentum equation are addressed in the construction of the REMIX SPH scheme, so the instability is allowed to grow and particles of different materials are able to intermix in a qualitatively similar way to the ideal gas cases.

The mode amplitude growth of these simulations is plotted in Fig.~\ref{fig:modes_sharp}(b). We find strong quantitative similarities between these and the mode growth of the ideal gas simulations potted in Fig.~\ref{fig:modes_sharp}(a). Although we have no experimental or analytical predictions for the growth of the KHI in these conditions and with these materials, we find that: (1) spurious surface tension analogous to that in tSPH KHI simulations with ideal gas is also clearly visible and strong in tSPH simulations with multiple materials; (2) the construction of the REMIX scheme is general in, and shown to be effective in, its reduction of established sources of error in the SPH formalism; (3) without any tuning of the method to material-specific boundaries, improvements that alleviate surface tension-like effects in ideal gas KHI simulations also allow the KHI to form in a qualitatively similar manner in the multi-material case.

To achieve these improved results of the REMIX scheme demonstrated in Fig.~\ref{fig:kh_earth}, we require interplay between a combination of its constitutive methods (\S\ref{sec:remix}). We use this KHI with iron \& rock to highlight the importance of individual methods included in the REMIX SPH scheme as, while their effects are visible in all simulations, they present particularly clearly in this case. Fig.~\ref{fig:kh_earth_methods} shows Earth-like KHI simulations that use the REMIX SPH scheme with different ones of its constituent methods removed from the construction and reverted to its traditional SPH analogue in each panel. We show simulations that: (a) use a more standard form of the equations of motion with equal-valued free functions (\S\ref{subsec:remix_freefunc}); (b) use the integral rather than differential form of the density estimate (\S\ref{subsec:remix_densityest}); (c) use an unmodified Wendland $C^2$ kernel rather than linear-order reproducing kernels (\S\ref{subsec:remix_linearkernels}); (d) the full REMIX SPH scheme. Taking a more traditional approach in any one of these methods leads to much stronger surface tension-like effects, such that only the full scheme enables the expected spirals to form. The improvements of the REMIX scheme are in many cases due to interplay between its constitutive methods all together, rather than individual components solving separate issues.

\subsection{Rayleigh--Taylor instability -- ideal gas}\label{subsec:rt_idg}

\begin{figure}[t]
	\centering
{\includegraphics[width=\textwidth, trim={2.5mm 0mm 2.5mm 0mm}, clip]{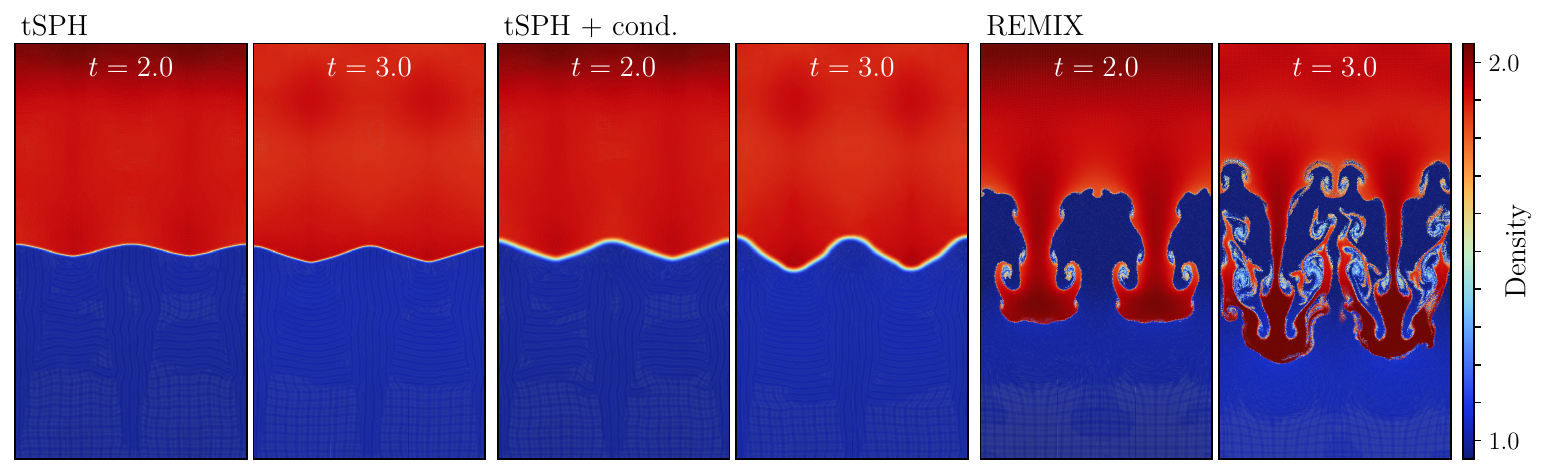}
}\hfill
	\vspace{-2em}
	\caption{Rayleigh--Taylor instabilities in an ideal gas. The RTIs are shown at two times, in simulations using tSPH, tSPH with artificial condition, and REMIX. These simulations have a resolution of $N_1=256$, equal mass particles, and are performed in 3D. The regions of fixed particles at the top and bottom of the simulations have been cropped from the figure; their positions and densities do not change.
 }
	\label{fig:rt_idg}
\end{figure}

We next consider the Rayleigh--Taylor instability (RTI) as an additional scenario to test the treatment of instability growth and mixing, which, unlike the previous tests, also includes gravity.

The RTI arises at the interface between a high density fluid being displaced by a low density fluid \citep{chandrasekhar1961hydrodynamic}.  We simulate the gravity-driven growth of the RTI, where a layer of high density fluid is initially positioned above a layer of low density fluid (relative to the downward direction of gravity). Hydrostatic equilibrium is disturbed by a small velocity perturbation. Similarly to the KHI, surface tension-like effects in traditional SPH formulations strongly suppress the growth of the RTI.

Our initial conditions are based on those of \citet{frontiere2017crksph}. However, as with the KHI tests, these simulations are carried out in 3D, with particles of equal mass, and without deliberate smoothing of the initial density discontinuity. Particles are placed in a periodic simulation domain in two cubic lattices, each a square in the $x, y$ dimensions and thin in $z$. The box has dimensions of 0.5, 1 in the $x$ and $y$ directions, with a thin and resolution-dependent $z$ box size. The low density region has $N_{1} \times N_{1} \times 18$ particles with density $\rho_1 = 1$ and occupies the bottom half of the domain. The high density region is constructed similarly to that in \S\ref{subsubsec:kh_idg_discontinuous}, giving a density of $\rho_2 = 1.91$ for the upper region while also ensuring a lattice that is consistent across the periodic simulation box edges. Particles in the top and bottom 0.05 of the box are fixed in place throughout the course of the simulation. Initial internal energies are set to satisfy hydrostatic equilibrium using an ideal gas EoS with $\gamma = 7/5$, constant gravitational acceleration $g = -0.5$, and a pressure at the interface of $P_0 = \rho_2 / \gamma$. Particles are initially at rest, other than an initial velocity perturbation that seeds the instability,

\begin{equation}\label{eq:rt_seed}
    v_y(x, y) = 
    \begin{cases}
      \delta_y \left[ 1 + \cos\left(8\pi\left(x + 0.25\right) \right) \right] \left[1 + \cos\left( 5\pi \left(y - 0.5\right) \right) \right] & \text{for $0.3 < y < 0.7$} \;,\\
      0 & \text{otherwise.} 
    \end{cases} 
\end{equation}

\noindent
We use a perturbation amplitude of $\delta_y = 0.025$.

In Fig.~\ref{fig:rt_idg}, we show snapshots from RTI simulations with resolution $N_{1} = 256$, simulated using tSPH, tSPH with artificial conduction, and REMIX. The growth of this instability is strongly suppressed, even with artificial conduction. REMIX is able to capture the growth of the RTI well. Additionally, we are able to maintain discontinuities as the simulation evolves. As in the KHI, these discontinuities are inherently unstable to perturbation modes of all wavelengths and so we see the growth of secondary, unseeded KHIs and RTIs that contribute to an onset of turbulent mixing. As the simulation progress, we observe turbulence driving mixing on the particle scale, the scale of the primary instability, and in between.

\subsection{Rayleigh--Taylor instability -- Earth-like iron \& rock}\label{subsec:rt_earth}

\begin{figure}[t]
	\centering
{\includegraphics[width=\textwidth, trim={2.5mm 0mm 2.5mm 0mm}, clip]{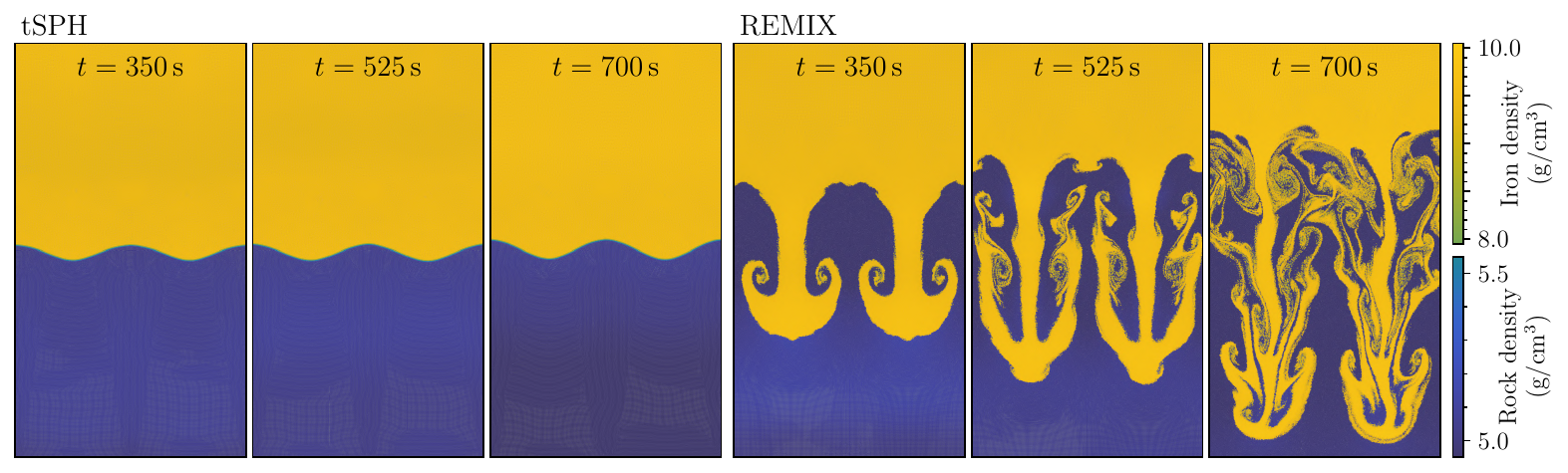}
}\hfill
	\vspace{-2em}
	\caption{Rayleigh--Taylor instabilities between dissimilar, stiff materials. The RTIs are shown at three times, in simulations using tSPH and REMIX. These simulations have a resolution of $N_1=256$, equal mass particles, and are performed in 3D.}
	\label{fig:rt_earth}
\end{figure}

We now consider the treatment of the RTI at an interface between different materials. The stiff iron \& rock EoS that we use makes this an even more challenging scenario for traditional SPH.

The high density iron layer is placed above the low density rock layer. The particle configuration is constructed just as in the ideal gas case above, since the density ratio is taken to be the same. However the box is scaled to have dimensions $0.05~R_\oplus$ and $0.1~R_\oplus$ in the $x$ and $y$ dimensions. The velocity perturbation is similar, although scaled to the box size and with an amplitude $\delta_y = 2.5 \times 10^{-5}~R_\oplus$~s$^{-1}$. Again, particles are initially in hydrostatic equilibrium, other than due to the seeded perturbation. Internal energies are chosen to satisfy this for the constant gravitational acceleration $g = -9.9$~m~s$^{-2}$  and interface pressure $P_0 = 120$~GPa, representative of the gravitational acceleration and pressure at the Earth's core-mantle boundary.

In Fig.~\ref{fig:rt_earth} we show snapshots from RTI simulations with Earth-like materials with resolution $N_{1} = 256$, with tSPH and REMIX. The RTI does not grow in the tSPH simulation. In contrast, the behaviour of the REMIX simulation is similar to the equivalent ideal gas case: unimpeded evolution of the instability, mixing at different length scales, onset of turbulence, and growth of unseeded secondary modes.

\begin{figure}[t]
	\centering
{\includegraphics[width=\textwidth, trim={2.5mm 0mm 2.5mm 0mm}, clip]{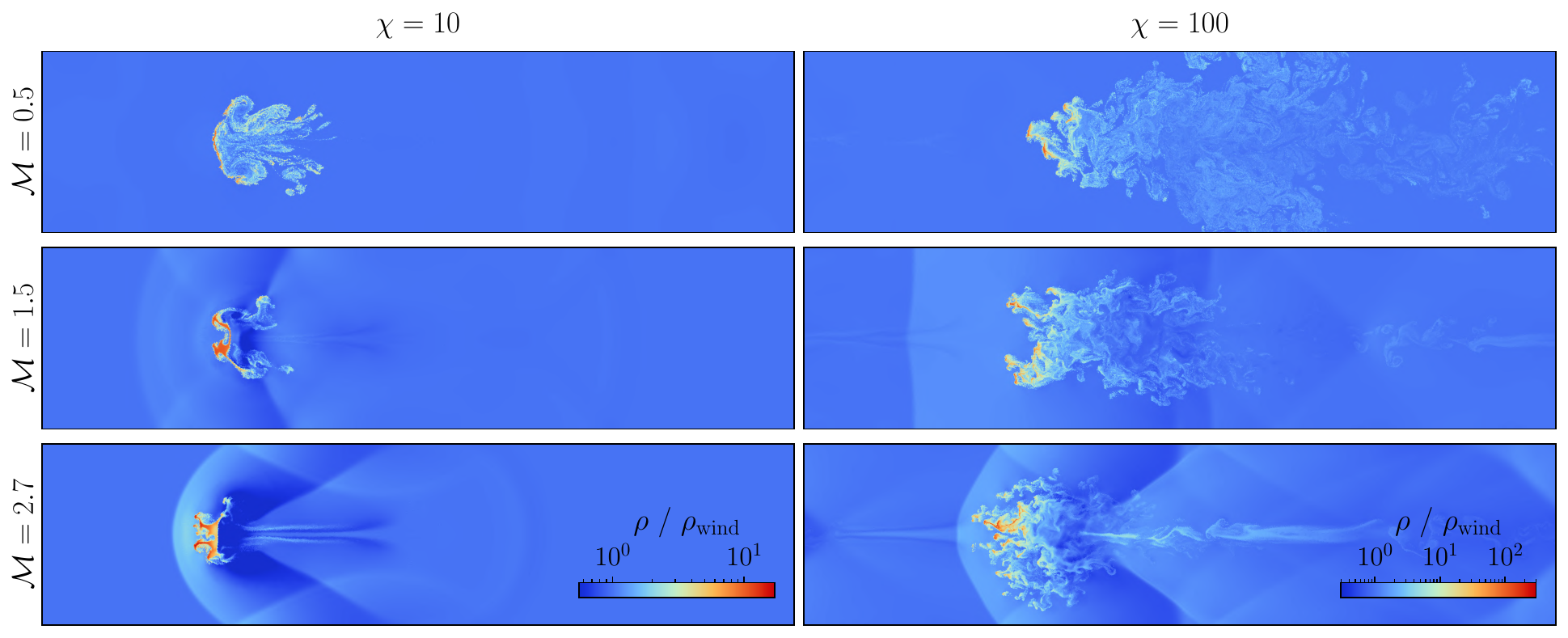}
}\hfill
	\vspace{-2em}
	\caption{REMIX simulations of a high-resolution ($N=256$) 3D blob test, at time $t \approx 5~t_{\rm cc}$. The disruption of the cloud is shown in scenarios with two different initial density contrasts (columns) and three Mach numbers (rows). Individual particles in a central cross-section in $z$ are plotted, and coloured by their density, $\rho$, relative to the wind density, $\rho_{\rm wind}$.}
	\label{fig:blob_snapshots}
\end{figure}

\subsection{Blob test}\label{subsec:blob}

%[what is the blob test? Past work. end with Braspenning paper and e.g. M=1.5]

In a physical system, mixing due to fluid instabilities is typically much less controlled and isolated than in the deliberately seeded scenarios of the previous sections. The ``blob test'' \citep{agertz2007fundamental} is used to investigate the treatment of turbulent mixing at density discontinuities due to unseeded instabilities. 

A spherical cloud of high-density fluid, initially at rest, is placed in an uniform flow of low-density fluid. Emergent Kelvin--Helmholtz and Rayleigh--Taylor instabilities at the interface, as well as ram-pressure stripping, should act to break up the cloud, driving its evolution to a well-mixed state. As with the instability tests presented in previous sections, traditional SPH schemes struggle to capture instability growth at density discontinuities and so the mixing of the cloud into the surrounding fluid is strongly suppressed. Typically, blob tests are carried out in a supersonic regime, where interactions between shock waves and the cloud can also be assessed, applicable to a range of astrophysical scenarios. However, here we additionally simulate blob tests in a subsonic regime to demonstrate the ability of the REMIX SPH scheme in capturing subsonic turbulent mixing, which is even more strongly suppressed in the tSPH formalism \citep{bauer2012subsonic}.

%[ICs and tcc]
 \citet{braspenning2023sensitivity} compare blob test simulations using seven hydrodynamical solvers, including SPH schemes and mesh-based methods. We reproduce their initial conditions to allow direct comparisons with their simulations. Particles are placed in a 3D periodic box with axes aligned such that the initially uniform wind flows in the $x$ direction. The length of the box in $y$ and $z$ is chosen to be 1~pc, and the length in the $x$ direction is 4~pc. Particles in the ambient wind are initially placed in a cubic lattice with $4N \times N \times N$ in the $x, y, z$ directions, where we parameterise the simulation resolution by $N$. We carry out simulations with $N = 16,\,32,\,64,\,128,\,256$. Particles in both the cloud and surrounding wind have equal masses and so particles in the cloud are placed in a cubic lattice of higher number density corresponding to the chosen density contrast. We simulate blob tests with initial density contrasts $\chi = 10,\,100$ and the initial density of the surrounding medium is $10^{-4}$~m$_{\rm p}$~cm$^{-3}$, where m$_{\rm p}$ is the proton mass. Clouds are spherical and have a radius of $R_{\rm cloud} = 0.1$~pc. Both the cloud and surrounding medium are an ideal gas with $\gamma = 5/3$ and internal energies are chosen so that the cloud and surrounding medium are in pressure equilibrium with each other and the cloud has an initial temperature of $10^4$~K. We carry out simulations with three wind speeds, characterised by the Mach number $\mathcal{M} \equiv v_{\rm wind} / c_{\rm wind}$: $\mathcal{M} = 1.5$ for a direct comparison to the simulations of \citet{braspenning2023sensitivity}, $\mathcal{M} = 2.7$ the value used most frequently in validating hydrodynamic methods \citep{agertz2007fundamental, frontiere2017crksph}, and $\mathcal{M} = 0.5$ to test mixing in the subsonic regime. We use units of the cloud crushing timescale

\begin{equation}\label{eq:t_cc}
   t_{\rm cc} = \frac{\sqrt{\chi} \,R_{\rm cloud}}{v_{\rm wind}} \;,
\end{equation}

\noindent
to compare simulations with different initial density contrasts and wind speeds and for direct comparisons with the results of \citet{braspenning2023sensitivity}.

\begin{figure}[t]
	\centering
{\includegraphics[width=\textwidth, trim={2.5mm 0mm 2.5mm 0mm}, clip]{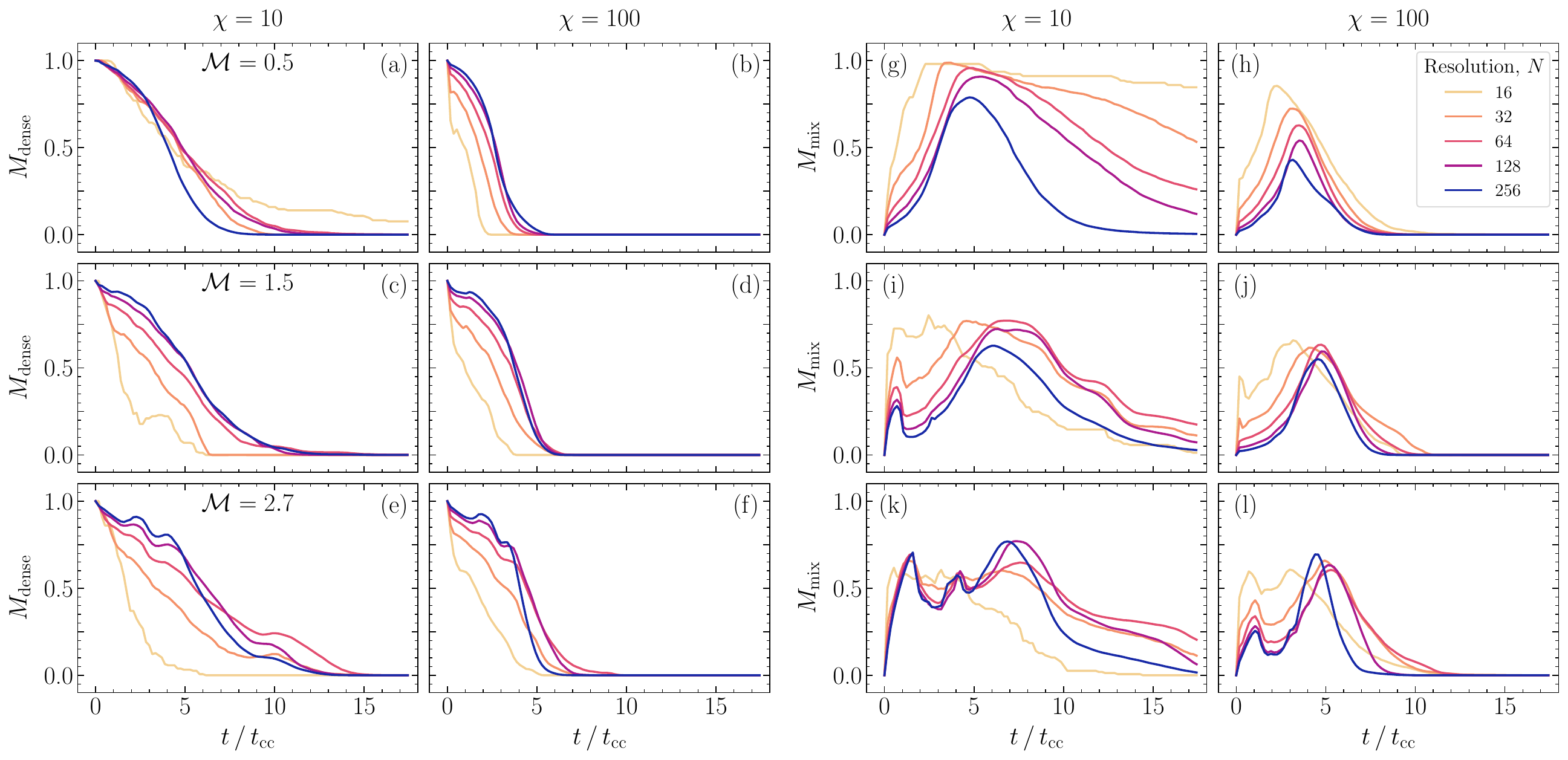}
}\hfill
	\vspace{-2em}
	\caption{Time evolution of the mass of dense gas, $M_{\rm dense}$ (a--f), and the mass of intermediate-temperature gas, $M_{\rm mix}$ (g--l), in blob test simulations. These quantities are plotted for simulations for two different initial density contrasts ($\chi$; columns) and three Mach numbers ($\mathcal{M}$; rows). Line colour corresponds to simulation resolution.}
	\label{fig:blob_graphs}
\end{figure}

%[Fig 1, good mixing]

In Fig.~\ref{fig:blob_snapshots} we plot particle densities from a central cross-section of REMIX blob test simulations with $N = 256$. Results from simulations with three initial wind speeds and two initial density contrasts are shown for a time $t \approx 5~t_{\rm cc}$. The middle row therefore corresponds directly to results from simulations plotted in  Fig.~1 of \citet{braspenning2023sensitivity}. REMIX captures disruption of the cloud in both a subsonic and supersonic regime. With time, the cloud reaches a well-mixed state with the surrounding medium. This contrasts with the \citet{braspenning2023sensitivity} SPH simulations, in which clouds with $\chi = 100$ do not fully mix (their Fig.~A1). Additionally, the onset of turbulence does not produce a highly symmetric structure like that provoked by the use of a regular grid in simulations using an adaptive mesh refinement (AMR) method (seen most clearly in  Figs.~5 and 6 \citet{braspenning2023sensitivity}). 

To facilitate direct quantitative comparisons to the simulations of \citet{braspenning2023sensitivity}, we consider the evolution of the mass of dense gas and the mass of intermediate-temperature gas. The mass of dense gas, $M_{\rm dense}$, is defined as the sum of masses of particles with density above a threshold of $\rho_{\rm cloud} / \, 3$, where $\rho_{\rm cloud}$ is the initial cloud density. The mass of intermediate-temperature gas, $M_{\rm mix}$, is defined as the sum of particle masses, $m_i$, of particles whose temperature, $T_i$, lies within half the logarithmic temperature range between the cold cloud and the hot wind, centred on the
geometric mean temperature, i.e.

\begin{equation}\label{eq:M_mix}
 M_{\rm mix} = \sum_i m_{\text{mix},\; i} \quad \text{where} \quad  m_{\text{mix},\; i} = 
\begin{cases}
    m_i & \text{for } \log(T_{\rm mix}) - \frac{1}{4}\log(\chi)  <  \log(T_i)  < \log(T_{\rm mix}) + \frac{1}{4}\log(\chi) \;,\\
    0              & \text{otherwise,}
\end{cases}
\end{equation}

\noindent
where $T_{\rm mix}$ is the geometric mean of the cloud and wind temperatures, $T_{\rm mix} = \sqrt{T_{\rm cloud} T_{\rm wind}}$. We normalise both $M_{\rm dense}$ and $M_{\rm mix}$ to the initial cloud mass.

%[Fig 2, timescales and compare with Braspenning]
The evolution of $M_{\rm dense}$ and $M_{\rm mix}$ for simulations varying initial wind speed, initial density contrast, and resolution is plotted in Fig.~\ref{fig:blob_graphs}. We find that REMIX is able to capture the disruption of the cloud in all these simulations, as shown in Fig.~\ref{fig:blob_graphs}(a--f). The middle row corresponds directly to Figs.~2 and 3 of \citet{braspenning2023sensitivity}. We see strong similarities between the behaviour of our REMIX simulations and the simulations of \citet{braspenning2023sensitivity} with hydrodynamic solvers that they find demonstrate good mixing. The evolution of these quantities is well parameterised by the cloud crushing timescale for this range of Mach numbers, with features appearing at approximately the same scaled time for all rows. Increasing resolution results in behaviour that indicates an approach towards numerical convergence for both $M_{\rm dense}$ and $M_{\rm mix}$, despite the scenario itself being highly turbulent with no true converged solution.

\subsection{Evrard collapse}\label{subsec:evrard}

The Evrard collapse \citep{evrard1988beyond} considers the collapse of an isothermal, spherical cloud of gas under its self-gravity. A shock is formed and moves outwards as the cloud collapses. We use this test to investigate the coupling of gravity and hydrodynamics, with large transformations of energy between gravitational, kinetic and thermal forms.

\begin{figure}[t]
	\centering
{\includegraphics[width=\textwidth, trim={2.5mm 0mm 2.5mm 0mm}, clip]{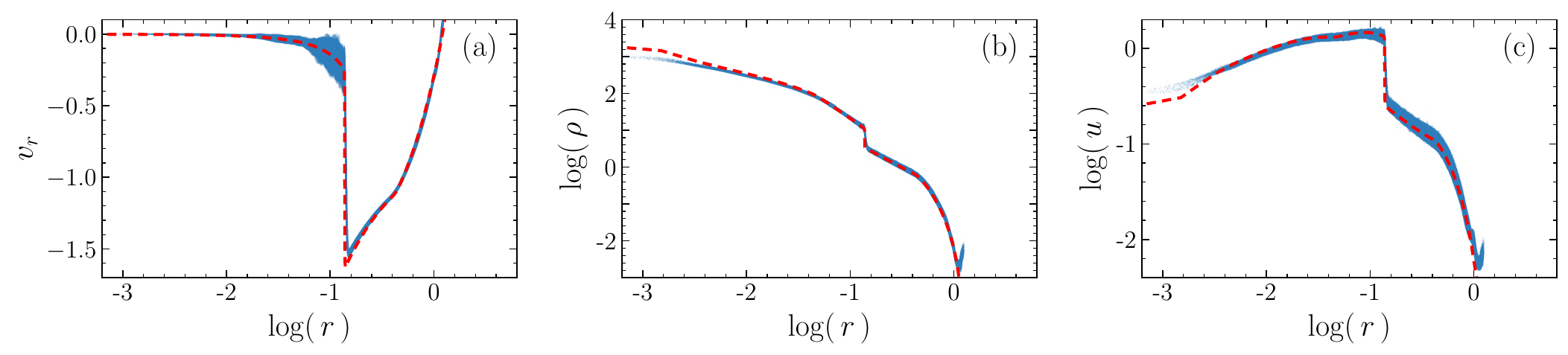}
}\hfill
	\vspace{-2em}
	\caption{Evrard collapse with a resolution of ${\sim}10^7$ SPH particles at time $t=0.8$, simulated using REMIX. Plots show (a) radial velocity, $v_r$, (b) density, $\rho$, and (c) specific internal energy, $u$, plotted against radial distance from the cloud centre, $r$. Individual particles are plotted in blue, and the dashed red line shows a reference solution from a high-resolution grid code simulation \citep{borrow2022sphenix}.}
	\label{fig:evrard}
\end{figure}

\begin{figure}[t]
	\centering
{\includegraphics[width=\textwidth, trim={2.5mm 0mm 2.5mm 0mm}, clip]{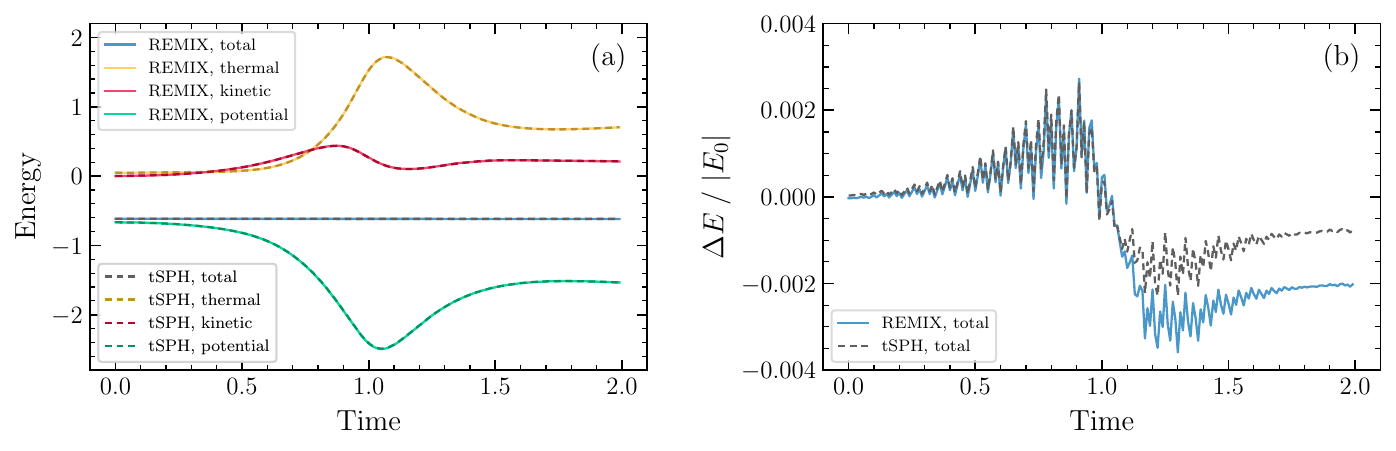}
}\hfill
	\vspace{-2em}
	\caption{Energy evolution in the Evrard collapse. (a) Different forms of energy, and (b) fractional deviation of total energy from its initial value, are shown as functions of time for tSPH and REMIX simulations, both with a resolution of ${\sim}10^7$ particles.}
	\label{fig:evrard_energy}
\end{figure}

Initial conditions are constructed similarly to \citet{borrow2022sphenix}. We place ${\sim}10^7$ equal mass particles of ideal gas with $\gamma = 5/3$ and $u = 0.05$ in a spherical cloud of density profile $\rho(r) = 1 / (2 \pi r)$, where $r$ is the radial distance from the cloud centre. The total cloud mass and radius are given by $M = 1$ and $R = 1$, and the gravitational constant is set to $G = 1$. Particle positions are chosen randomly, following \citet{borrow2022sphenix}, to satisfy the initial density profile. This method of choosing positions results in particles quickly readjusting to a glass-like structure, therefore experiencing divergences that lead to a seeding of noise in internal energies and densities. Both artificial diffusion and the normalising density evolution term
act to smooth this noise over time.

The Evrard collapse is captured well by REMIX, as shown in Fig.~\ref{fig:evrard}. We observe sharp shocks and evolution that closely follows the reference solution. The scatter in internal energy around the reference solution could be reduced by increasing the strength of artificial diffusion of internal energy, through choices of $a_u$ and $b_u$. However, we choose to maintain a conservative approach to artificial diffusion so as not to deviate far from the thermodynamically consistent basis of our equations of motion. We therefore judge this amount of scatter to be sufficiently small. There is less scatter in density than in internal energy, since the normalising term is also contributing to smoothing the density.

At the vacuum boundary, we see a slight upturn in density and internal energy. Since divergence estimates in the evolution of these quantities revert to using kernels that are normalised to the continuum at vacuum boundaries, bulk expansion at vacuum boundaries may be underestimated. This is because for a region of locally isotropic expanding gas, a spherically symmetric kernel that is sampled by diverging particles in only approximately half its volume will underestimate the local velocity divergence. We note however that the logarithmic scales in Fig.~\ref{fig:evrard} perhaps overemphasise the upturning features in terms of their importance in a typical science application.

The evolution of energy in Evrard collapse simulations is shown in Fig.~\ref{fig:evrard_energy}. The exchange of energy between different forms is closely aligned between REMIX and tSPH simulations, as demonstrated in Fig.~\ref{fig:evrard_energy}(a). These curves are consistent with those shown in Fig.~42 of \citet{springel2010pur}. The fractional deviation of the total energy from its initial value is plotted in Fig.~\ref{fig:evrard_energy}(b). Both REMIX and tSPH are constructed to explicitly conserve energy. Fluctuations of total energy are of the same order of magnitude in both cases, with REMIX showing variations of less than $0.4\%$ during the simulation. Small deviations of energy of this size are expected for SPH schemes with non-reversible timesteps, even in formulations like ours, whose governing equations are explicitly conservative.

\subsection{Planets in hydrostatic equilibrium}\label{subsec:planet}

\begin{figure}[t]
	\centering
{\includegraphics[width=\textwidth, trim={2.5mm 0mm 2.5mm 0mm}, clip]{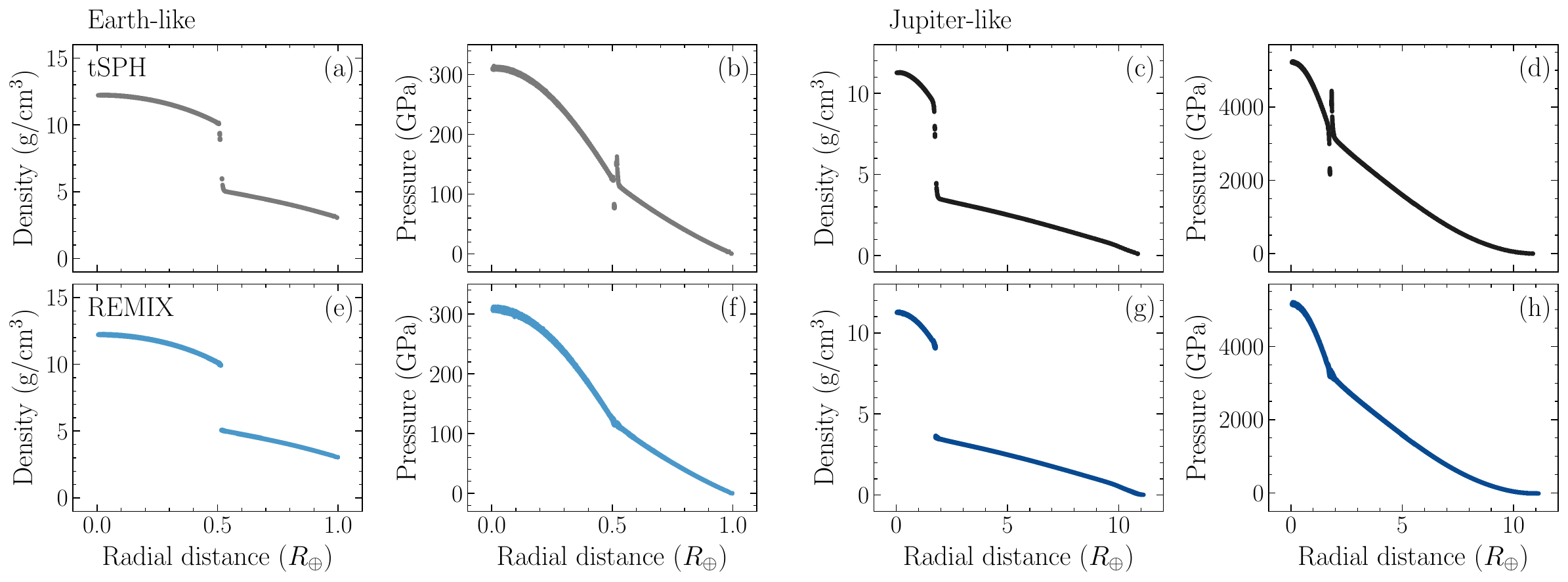}
}\hfill
	\vspace{-2em}
	\caption{Radial profiles from simulations of an Earth-like (a, b, e, f) and a Jupiter-like (c, d, g, h) planet at time $t = 10,\!000\,$s, simulated using tSPH (a--d) and REMIX (e--h). Particle densities and pressures are plotted against radial distance from the centre of the planet. REMIX corrects density discontinuities in simulations of planets in hydrostatic equilibrium.}
	\label{fig:planet_profiles}
\end{figure}

\begin{figure}[t]
	\centering
{\includegraphics[width=\textwidth, trim={2.5mm 0mm 2.5mm 0mm}, clip]{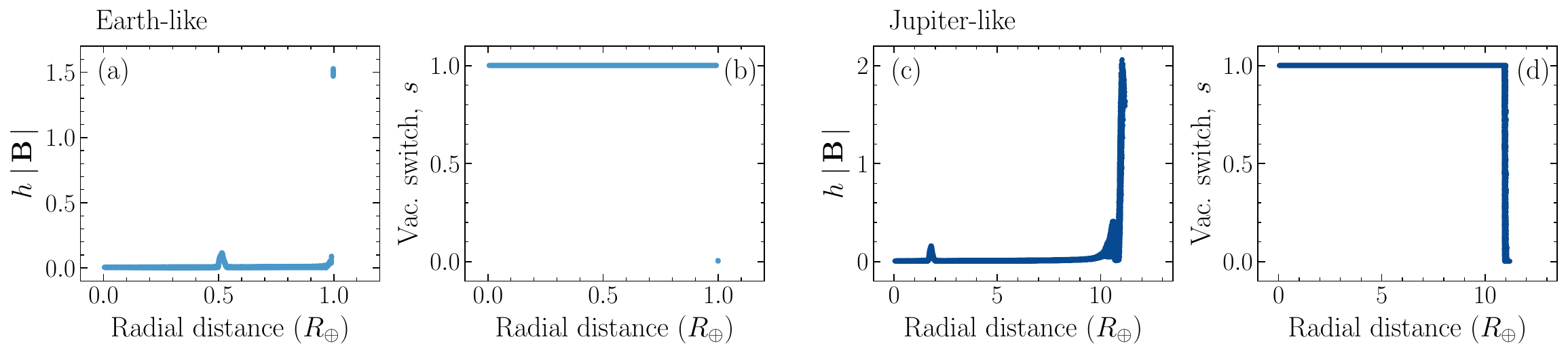}
}\hfill
	\vspace{-2em}
	\caption{Identification of planetary vacuum boundaries in REMIX simulations. Plots correspond to Earth-like (a, b) and Jupiter-like (c, d) planets at time $t = 10,\!000\,$s. We plot the quantity $h\,|\mathbf{B}|$ for individual particles, which is used in our vacuum boundary switch (Eqn.~\ref{eq:vacswitch}), and the vacuum boundary switch, $s$, itself.}
	\label{fig:planet_vacterm}
\end{figure}

\begin{figure}[t]
	\centering
{\includegraphics[width=\textwidth, trim={2.5mm 0mm 2.5mm 0mm}, clip]{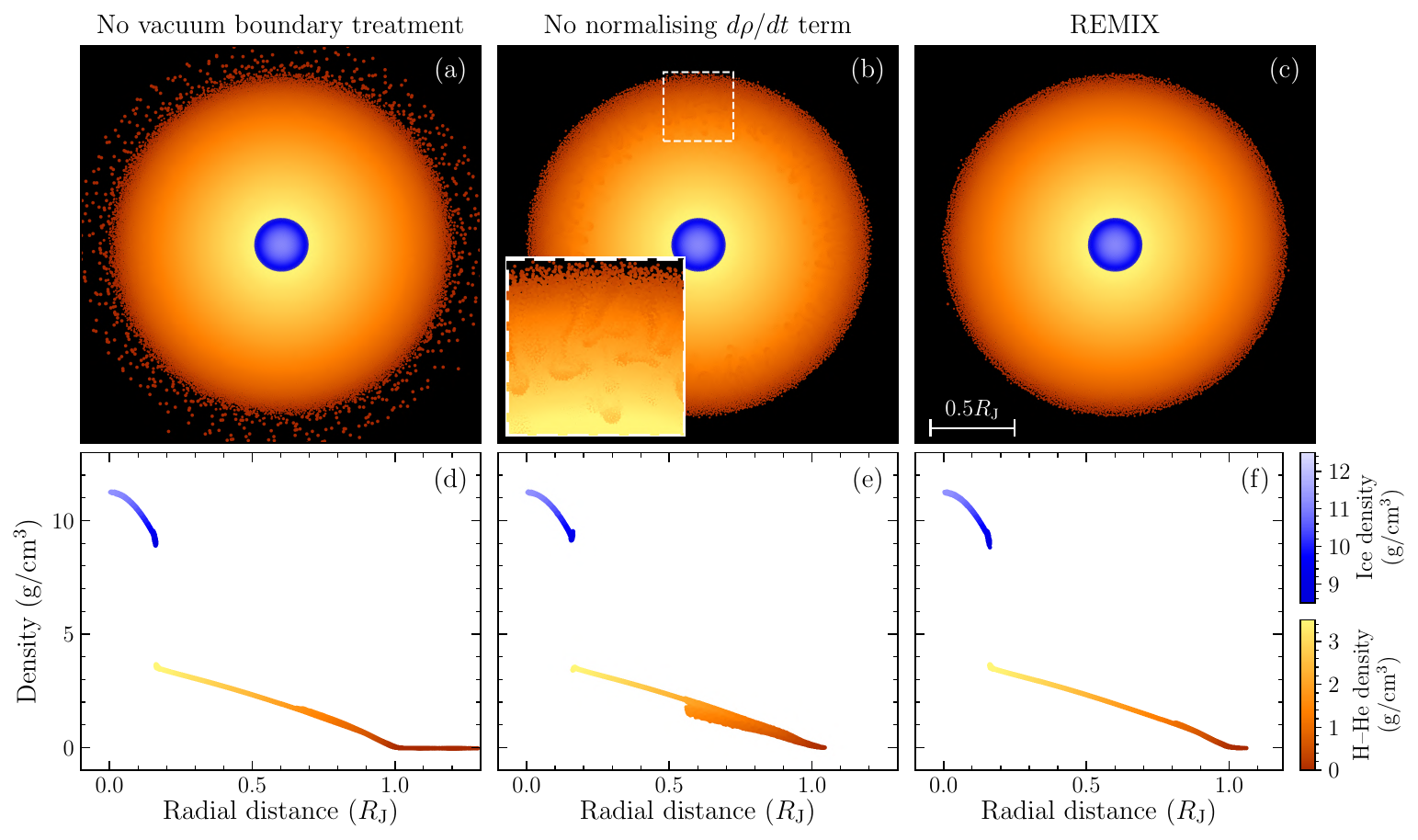}
}\hfill
	\vspace{-2em}
	\caption{Effect of the vacuum boundary treatment and the normalising term in simulations of a Jupiter-like planet in hydrostatic equilibrium. We plot snapshots (a--c) and radial density profiles (d--f) at time $t = 20,\!000\,$s. Columns show simulations with: the REMIX scheme but without the inclusion of the vacuum boundary treatment (a, d); the REMIX scheme but without the inclusion of the normalising term in the density evolution (b, e); and the full REMIX scheme with no modification (c, f). Particles are coloured by material and density. The inset in (b) shows a magnified view of instabilities forming near the vacuum boundary when the normalising term is not included. The colour scale of the inset has been slightly tweaked to increase the contrast around the instabilities. The evolution of these instabilities, without tweaked colours, is shown in Fig.~\ref{fig:jupiter_nonormterm}.}
 \label{fig:jupiter_methods}
\end{figure}

\begin{figure}[t]
	\centering
{\includegraphics[width=\textwidth, trim={2.5mm 0mm 2.5mm 0mm}, clip]{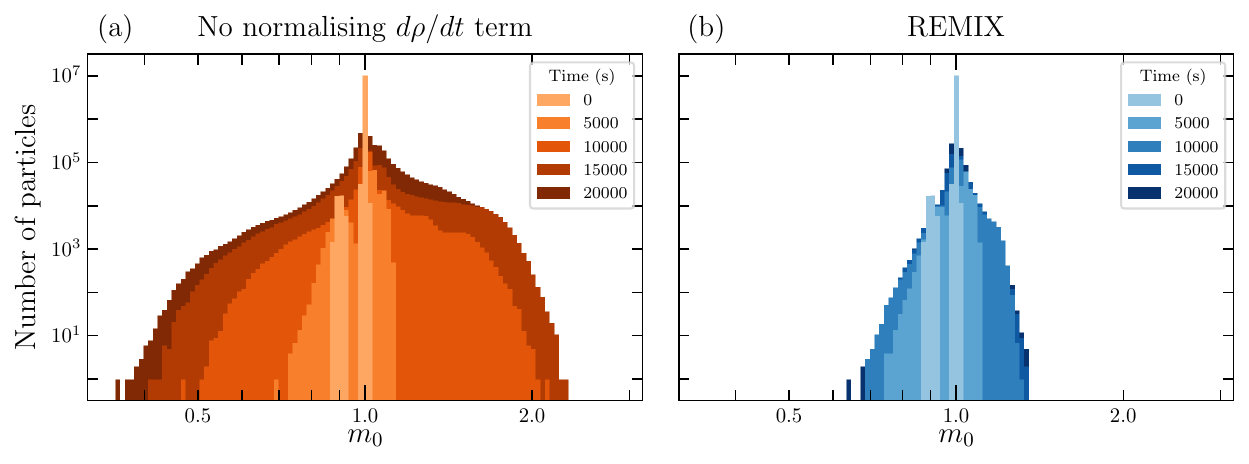}
}\hfill
	\vspace{-2.5em}
	\caption{The evolution of the distribution of the $m_0$ kernel geometric moment in simulations of a Jupiter-like planet. Plots show results from simulations with a resolution of $10^7$ particles using the REMIX scheme both (a) without and (b) with the normalising term in the density evolution. Only particles with a vacuum boundary switch $s_i > 0.9$ are plotted, to isolate particles that should have $m_0 \approx 1$ and filter out spikes near $m_0 = 0.5$.}
 \label{fig:jupiter_m0}
\end{figure}

Planets in hydrostatic equilibrium offer a test scenario to probe the interaction of our hydrodynamic methods with gravity in the context of a layered, multi-EoS structure with a free surface. This also acts as a useful validation test for potential science applications; REMIX has since been applied in high-resolution simulations of planetary giant impacts, demonstrating significant improvements compared with tSPH simulations \citep{sandnes2024no}. We also use this test to illustrate the importance of the inclusion of both the vacuum boundary treatment and the density evolution kernel normalising term in the REMIX scheme.

We apply our methods to an Earth-like and a Jupiter-like planet. The Earth-like planet represents a case in which materials have small variations of density within layers. The Jupiter-like case represents a scenario with relatively steep gradients of densities within material layers. This acts to assess stability against error-driven instabilities that can form due to these density gradients. These are only ``Earth-like'' and ``Jupiter-like'' because they are based on initial conditions for planetary giant impact simulations, which resemble the present day planets after the impact \citep{kegerreis2022immediate}. However, unlike in typical pre-impact ``settling'' simulations, where particle entropy can be fixed to prevent viscous heating \citep{kegerreis2020atmospheric}, here we use the full REMIX scheme with no modifications.

Initial hydrostatic equilibrium profiles and SPH particle placements are calculated using the publicly available code \woma \citep{kegerreis2019planetary,ruiz2021effect}. The Earth-like planet is constructed to satisfy the following conditions: two adiabatic layers consisting of a core of mass $0.27 \, M_\oplus$, where $M_\oplus = 5.97\times10^{24}$~kg, represented by particles with the ANEOS Fe\textsubscript{85}Si\textsubscript{15} (iron) EoS and a mantle of mass $0.62 \, M_\oplus$ with ANEOS forsterite (rock) \citep{stewart2020shock}; a surface pressure and temperature of $P_{\rm s} = 1 \times 10^5\,$Pa and $T_{\rm s} = 2000\,$K. The Jupiter-like planet is constructed to satisfy the following conditions: two adiabatic layers consisting of a core of mass $10 \, M_\oplus$ with the AQUA (ice) EoS \citep{haldemann2020aqua}, and a hydrogen--helium \citep{chabrier2021new} envelope of mass $298 \, M_\oplus$; a surface pressure and temperature of $P_{\rm s} = 1 \times 10^5\,$Pa and $T_{\rm s} = 165\,$K. In all our simulations, planets each consist of ${\sim} 10^7$ equal mass particles.

Fig.~\ref{fig:planet_profiles} shows radial density and pressure profiles of the two planets at a time $t = 10,\!000\,$s. We show profiles from simulations using both tSPH (a--d) and REMIX (e--h). The smoothing of density discontinuities in tSPH, and the corresponding pressure discontinuities, are clearly visible. We note that at this time particles have evolved to take up more relaxed positions, so density smoothing at the material interface, and in particularly at the vacuum boundary, are less extreme than in the initial condition configuration (\ref{app:planets}). However, these relaxed configurations typically yield large gaps between the different-density layers, so are a result of the surface tension rather than an indication that surface tension is reduced as the system relaxes. With REMIX, density discontinuities remain sharp, and pressures at the material boundaries are close to continuous.

We use this test to motivate the inclusion of the vacuum boundary treatment, detailed in \S\ref{subsec:remix_vacuum}. The vacuum boundary switch is able to accurately identify free surfaces based on $h_i\,|\mathbf{B_i}|$ (Eqn.~\ref{eq:vacswitch}), as shown in Fig.~\ref{fig:planet_vacterm} for the two example planets. In the Earth-like planet the outermost particles remain in an undisturbed shell, which all get identified as the vacuum boundary and no interior particles are flagged. In the Jupiter-like planet, however, the envelope density drops far lower before the outer edge, leading to steep local changes in density near the vacuum boundary. This leads to error-driven particle motion that disturbs the initial particle shells, demonstrated in the identification of the vacuum boundary by the switch function, which in this case extends smoothly to particles near the surface that are no longer neatly ordered in shells.

In Fig.~\ref{fig:jupiter_methods}(a) and (d), we show a cross-section and density profile from a REMIX simulation of the Jupiter-like planet at $t = 20,\!000\,$s without the vacuum boundary treatment in the kernel construction, although still included in the normalising term. In this case, linear-order reproducing kernels are used without modification for all particles across the simulation, including  particles near the free surface. Particles become unstable at the vacuum boundary, despite being set up to satisfy hydrostatic equilibrium, because bad estimates of pressure gradients lead particles to stream out from the surface. Similar behaviour is observed in equivalent simulations of the Earth-like planet.

In Fig.~\ref{fig:jupiter_methods}(b) and (e), we show similar results from a simulation with REMIX, but without the normalising term in the density evolution equation (\S\ref{subsec:remix_norm}). Here we see error-driven instabilities forming near the vacuum boundary. In \ref{app:planets}, we show the continued evolution of these instabilities to demonstrate how they continue to disturb the profile of the planet. These low density plumes that fall towards the planet's centre have a high local number density of particles, so should have higher densities. This disconnect between the particle density and the local distribution of mass in the simulation volume leads to a positive feedback effect, in which the falling plumes continue to accumulate particles without the density of particles evolving to reflect this, further driving the discontinuity downwards. In the full REMIX scheme, for which similar plots are shown in Fig.~\ref{fig:jupiter_methods}(c) and (f), we re-associate the evolved density to the mass distribution in simulation volume by the inclusion of the kernel normalising term, which prevents the formation of these instabilities. We show the direct effect of the normalising term in these simulations in Fig.~\ref{fig:jupiter_m0}. As in the KHI examples in Fig.~\ref{fig:kh_m0}, here we see how the normalising term acts to tie particle volume elements to the local distribution of particle masses. The signal velocity of the normalising term means that correction occurs over the timescale of particle motion. Therefore, particles are able to readjust to react to changes in density due to this term, while it still acts as an effective correction to accumulation of error, accumulated over timescales set by the velocity divergence estimate in the equations of motion. In regions where densities represent the local distribution of particle masses well, the normalising term has little effect on the hydrodynamics.

\section{Conclusions}\label{sec:conclusions}

We have presented a new formulation of smoothed particle hydrodynamics (SPH), REMIX (`Reduced Error MIXing'), that combines several novel and recently developed methods to address the well-known shortcomings of the traditional SPH formalism at density discontinuities.  By directly targeting sources of kernel smoothing error and discretisation error, this scheme dramatically reduces numerical effects that can otherwise lead to spurious surface tension-like effects and inhibit mixing. We demonstrate its effectiveness using 3D hydrodynamic tests in a broad range of scenarios and regimes. In addition to standard tests, REMIX can handle boundaries between dissimilar, stiff materials, and the particularly challenging case of density discontinuities in simulations with equal mass particles --- where both smoothing and discretisation errors are considerable.

The REMIX SPH scheme is based on thermodynamically consistent, conservative equations of motion, with free functions chosen to limit zeroth-order error. We use an evolved density estimate to avoid the kernel smoothing error in the standard SPH integral density estimate. To avoid potential accumulation of error in the evolved density estimate, such that densities would be no longer representative of the distribution of particle masses in the simulation volume, we introduce a new ``kernel normalising term''. Additionally, artificial diffusion, which is weak outside shocks, helps to smooth out accumulated noise in both particle densities and internal energies. To reduce discretisation error, we use linear-order reproducing kernels in the equations of motion. Since kernel densities are evolved in time, particle volume elements are not \textit{instantaneously} tied to simulation volume, despite the normalising term in the density evolution. Therefore, normalising the kernel to particle volume elements is an important step in calculating appropriate gradient estimates in the equations of motion. We introduce grad-$h$ terms to the kernels, adding completeness to the construction. Additionally, we present a method that identifies free surfaces and reverts our kernels to standard spherically symmetric functions, normalised to the continuum, to appropriately capture vacuum boundaries. We also use advanced artificial viscosity and diffusion schemes with linear reconstruction of quantities to particle midpoints, and a set of novel improvements to effectively switch between treatments for shock-capturing under compression and noise-smoothing in shearing regions.

REMIX shows a range of improvements compared with traditional SPH formulations, as we examined here with an extensive set of test cases. Our generalised error-reduction approach greatly improves the treatment of both static density discontinuities, as seen in the 3D square test with equal mass particles, and mixing and instability growth at evolving interfaces within a single ideal gas as well as between multiple materials, as demonstrated in fluid instability tests. This is achieved  without a need for a material-dependent approach in volume elements or density estimates, and without applying targeted corrections at material boundaries or needing particle mass ratios to be matched to density ratios across discontinuities. REMIX is able to capture shocks with reduced particle noise, as seen in the Sod shock tube, and can effectively simulate a system with gravity and emerging shocks where large amounts of energy are exchanged between different forms, as demonstrated in the Evrard collapse. Many aspects of REMIX combine to allow us to improve simulations of planetary bodies, including the evolved density estimate that corrects smoothing of density discontinuities; the vacuum boundary treatment that extends our methods to be able to deal with free surfaces; and the density evolution normalising term that ensures that particle densities are tied to the local distribution of masses.

REMIX is publicly available as a component of the open-source \swift code, at \url{www.swiftsim.com}.

\section*{Acknowledgements}
\noindent
We thank Joey Braspenning for providing the blob test initial conditions and useful suggestions. 
T.D.S. acknowledges support from Science and Technology Facilities Council (STFC)
grants ST/T506047/1 and ST/V506643/1.
J.A.K. acknowledges support from a NASA Postdoctoral Program Fellowship
administered by Oak Ridge Associated Universities.
The research in this paper made use of the \swift open-source simulation code
\citep{schaller2024swift}.
This work was supported by STFC % Science and Technology Facilities Council (STFC)
grants ST/P000541/1, ST/T000244/1 and ST/X001075/1,
and used the DiRAC@Durham facility managed by the
Institute for Computational Cosmology
on behalf of the STFC DiRAC HPC Facility (www.dirac.ac.uk).
This equipment was funded by BEIS via STFC capital grants
ST/K00042X/1, ST/P002293/1, ST/R002371/1, and ST/S002502/1,
Durham University and STFC operations grant ST/R000832/1.
DiRAC is part of the National e-Infrastructure.

\section*{Data availability}
\noindent
 The sizes of the simulation output files make it unfeasible for them to be made passively available, but they can be obtained from the corresponding author on reasonable request. Simulation initial conditions are available as part of the open-source \swift package.

\newpage
\appendix

\section{Notation}\label{app:notation}

\vspace{5pt}
\begin{tabularx}{\textwidth}{ l X }
    $A, \; \mathbf{A}$ & Scalar, vector \\[10pt]
    $A^{\alpha,\, ... \,, \, \omega}$ & Elements of a vector, matrix, or higher-order tensor. Greek letter superscripts correspond to spatial dimensions (e.g. the x, y, and z components of a 3D vector) and like indices are summed over\\[20pt]
    $A_i, \; A_j$ & Quantity associated with, or sampled at the position of: a particle $i$; a neighbour $j$ of particle $i$\\[10pt]
    $A_{ij}$ & Pairwise interaction associated with neighbour $j$ acting on particle $i$\\[10pt]
    $\mathbf{r}$ & Position vector at which we probe the continuum fluid\\[10pt]
    $\mathbf{r'}$ & Position vector integrated over  in the continuum limit, for convolutions with a kernel function\\[10pt]
    $\mathbf{r}_i$ & Position vector of particle $i$. When discretising the fluid, the notation changes $\mathbf{r} \Rightarrow \mathbf{r}_i$, as we probe the fluid at particle positions.\\[20pt]
    $\mathbf{r}_j$ & Position vector of particle $j$. When discretising the fluid, the notation changes $\mathbf{r'} \Rightarrow \mathbf{r}_j$ as, instead of integrating over the continuum, we sum over particle neighbours.\\[20pt]
    $\mathbf{r}_{ij}$ & $\mathbf{r}_i - \mathbf{r}_j$\\[10pt]
    $m_i$ & Mass of particle $i$\\[10pt]
    $V_i$ & Volume element of particle $i$\\[10pt]
    $H$ & Kernel compact support. The radial extent from the kernel centre above which a kernel function is zero\\[20pt]
    $h$ & Kernel smoothing length. We define the smoothing length as twice the standard deviation of the kernel\\[20pt]
    $\eta_{\rm kernel}$ & Parameter that scales the radial extent of the kernel\\[10pt]
    $W_{ij}$ & Kernel function centred at $\mathbf{r}_i$, sampled at $\mathbf{r}_j$. $W_{ij} \equiv W(\mathbf{r}_{ij},\, h_i) \equiv W(\mathbf{r}_{i} - \mathbf{r}_{j},\, h(\mathbf{r}_{i}))$\\[10pt]
    $\overline{W}_{ij}$ & Symmetrised kernel: $\overline{W}_{ij} \equiv \overline{W}(\mathbf{r}_i - \mathbf{r}_j,\, h(\mathbf{r}_i),\, h(\mathbf{r}_j)) \equiv \frac{1}{2}\left(W(\mathbf{r}_{ij},\, h_i) \,+\, W(\mathbf{r}_{ji},\, h_j)\right)$\\[10pt]
    $\hat{W}_{ij}$ & Kernel $W_{ij}$ normalised to volume elements $V_j$\\[10pt]
    $\mathcal{W}_{ij}$ & Linear-order reproducing kernel, constructed using $\overline{W}_{ij}$\\[10pt]
    $\widetilde{\mathcal{W}}_{ij}$ & Linear-order reproducing kernel with vacuum boundary treatment, constructed using $\overline{W}_{ij}$\\[10pt]
    $\langle A_i \rangle$, $\overline{A}_i$, $\hat{A}_i$ & Quantity calculated by kernel interpolation using $W_{ij}$, $\overline{W}_{ij}$, $\hat{W}_{ij}$, respectively\\[10pt]
    $\dfrac{dA}{d\mathbf{r}} \,, \;\dfrac{dA}{dr^\gamma}$ & Spatial derivative of $A$ (vector and its elements). In cases where relevant, includes grad-$h$ terms. Can be combined with the above notation that indicates the kernel used\\[20pt]
    $\nabla A, \; \partial^\gamma A$ & Gradient of $A$ (vector and its elements). In cases where relevant, does not include grad-$h$ terms\\[10pt]
    $\nabla_{\kappa} A, \; \partial_{\kappa}^\gamma A$ & Gradient of $A$ (vector and its elements), interpolated only based on particle neighbours of the same material. In cases where relevant, does not include grad-$h$ terms\\[10pt]
\end{tabularx}

\newpage
\section{REMIX SPH equations}\label{app:remix}

\begin{align}\label{eq:summary_final_drhodt}
\frac{d \rho_i}{d t} &=  \sum\limits_{j} m_j  \, \frac{\rho_i}{\rho_j} \, v_{ij}^{\alpha} \, \frac{1}{2}\left(\frac{d \widetilde{\mathcal{W}}}{dr^{\alpha}}\bigg|_{ij} - \frac{d \widetilde{\mathcal{W}}}{dr^{\alpha}}\bigg|_{ji}\right) \: + \: \left(\frac{d \rho_i}{d t}\right)_{\text{difn}}  \: + \: \left(\frac{d \rho_i}{d t}\right)_{\text{norm}} \;,\\
\label{eq:summary_final_a}
\frac{d v_i^{\alpha}}{d t} &=  -\sum\limits_{j} m_j \, \frac{P_i + Q_{ij} + P_j + Q_{ji}}{\rho_i \rho_j} \, \frac{1}{2}\left(\frac{d \widetilde{\mathcal{W}}}{dr^{\alpha}}\bigg|_{ij} - \frac{d \widetilde{\mathcal{W}}}{dr^{\alpha}}\bigg|_{ji}\right) \;,\\
\label{eq:summary_final_dudt}
\frac{d u_i}{d t} &=  \sum\limits_{j} m_j \, \frac{P_i + Q_{ij}}{\rho_i \rho_j} \, v_{ij}^{\alpha} \, \frac{1}{2}\left(\frac{d \widetilde{\mathcal{W}}}{dr^{\alpha}}\bigg|_{ij} - \frac{d \widetilde{\mathcal{W}}}{dr^{\alpha}}\bigg|_{ji}\right) \: + \: \left(\frac{d u_i}{d t}\right)_{\text{difn}} \;.
\end{align}

\vspace{5pt}
\noindent
\textit{B.1. Kernel gradients}

\vspace{-8pt}
\noindent\textcolor{black}{\rule{0.9\textwidth}{0.1mm}}
\vspace{-8pt}
\begin{equation}\label{eq:summary_vactreatment}
\frac{d \widetilde{\mathcal{W}}}{dr^{\gamma}}\bigg|_{ij} = s_i\frac{d \mathcal{W}}{dr^{\gamma}}\bigg|_{ij} + \left(1 - s_i\right)\frac{d W}{d r^{\gamma}}\bigg|_{ij} \;.
\end{equation}

\begin{align}\label{eq:summary_total_derivative_pair}
 \frac{d \mathcal{W}}{dr^{\gamma}}\bigg|_{ij}  &=  A_i B_i^{\alpha} \overline{W}_{ij} +
                                                    A_i \left(1 + B_i^{\alpha} r_{ij}^{\alpha}\right) \frac{d \overline{W}}{d r^{\gamma}}\bigg|_{ij} 
                                                    + \left(1 + B_i^{\alpha} r_{ij}^{\alpha}\right) \overline{W}_{ij} \frac{d A}{d r^{\gamma}}\bigg|_{i} + 
                                                    A_i r_{ij}^{\alpha} \overline{W}_{ij} \frac{d B^{\alpha}}{d r^{\gamma}}\bigg|_{i} \;. 
\end{align}

\vspace{5pt}
\noindent
\textit{B.1.1. Symmetrised kernels and their gradients}

\vspace{-8pt}
\noindent\textcolor{lightgray}{\rule{0.7\textwidth}{0.1mm}}
\vspace{-8pt}
\[
\begin{minipage}{\linewidth}
  \footnotesize
\begin{align}
%\hline\nonumber\\[-5pt]
&\overline{W} \equiv \frac{W_{ij} + W_{ji}}{2} \;, \\[5pt]
&\frac{d \overline{W}}{d r^{\gamma}}\bigg|_{ij} = \frac{1}{2}\left(\frac{\partial W}{\partial r^{\gamma}}\bigg|_{ij}  + \frac{\partial W}{\partial h}\bigg|_{ij} \partial_i^{\gamma} \hat{h} - \frac{\partial W}{\partial r^{\gamma}}\bigg|_{ji} \right) \;, 
&&\partial_i^{\gamma} \hat{h}  =  \sum_j (h_j - h_i) \, \partial_i^{\gamma} \hat{W}_{ij}  \frac{m_j}{\rho_j} \;,  \hfill
&&&\partial_i^{\gamma} \hat{W}_{ij} \equiv \frac{\partial_i^{\gamma} W_{ij}}{m_{0,\,i}} - \frac{W_{ij}}{m_{0,\,i}^2} \partial_i^{\gamma} m_{0} \;.
\end{align}\nonumber
\end{minipage}%
\]

\vspace{5pt}
\noindent
\textit{B.1.2. Linear-order reproducing kernel construction}

\vspace{-8pt}
\noindent\textcolor{lightgray}{\rule{0.7\textwidth}{0.1mm}}
\vspace{-8pt}
\[
\begin{minipage}{\linewidth}
  \footnotesize
\begin{align}
&A_i = \left( \overline{m}_{0,\,i} - \left(\overline{m}_{2,\,i}^{\, -1}\right)^{\alpha \beta} \overline{m}_{1,\,i}^{\, \alpha} \, \overline{m}_{1,\,i}^{\, \beta} \right)^{-1} \;, \hfill 
&&\frac{d A}{d r^{\gamma}}\bigg|_{i} = -A_i^2 \left(\frac{d \overline{m}_0}{d r^{\gamma}}\bigg|_{i} -  
2 \left(\overline{m}_{2,\,i}^{\,-1}\right)^{\alpha \beta}  \overline{m}_{1,\,i}^{\,\beta} \frac{d \overline{m}_1^{\alpha}}{d r^{\gamma}}\bigg|_{i} +
 \left(\overline{m}_{2,\,i}^{\,-1}\right)^{\alpha \phi} \frac{d \overline{m}_2^{\phi \psi}}{d r^{\gamma}}\bigg|_{i} \left(\overline{m}_{2,\,i}^{\,-1}\right)^{\psi \beta} \overline{m}_{1,\,i}^{\,\alpha} \overline{m}_{1,\,i}^{\,\beta}\right) \;,\\[5pt]
&B_i^{\alpha} = - \left(\overline{m}_{2,\,i}^{\, -1}\right)^{\alpha \beta} \overline{m}_{1,\,i}^{\, \beta} \;,\hfill  
&&\frac{d B^{\alpha}}{d r^{\gamma}}\bigg|_{i} = -\left(\overline{m}_{2,\,i}^{\,-1}\right)^{\alpha \beta}  \frac{d \overline{m}_1^{\beta}}{d r^{\gamma}}\bigg|_{i} + \left(\overline{m}_{2,\,i}^{\,-1}\right)^{\alpha \phi} \frac{d \overline{m}_2^{\phi \psi}}{d r^{\gamma}}\bigg|_{i} \left(\overline{m}_{2,\,i}^{\,-1}\right)^{\psi \beta} \overline{m}_{1,\,i}^{\,\beta} \;.\\[10pt]
%\hline \nonumber\\[-5pt]
&\overline{m}_{0,\,i} = \sum_j  \overline{W}_{ij} V_j \;,\hfill
&&\frac{d \overline{m}_0}{d r^{\gamma}}\bigg|_{i} = \sum_j  \frac{d \overline{W}}{d r^{\gamma}}\bigg|_{ij} V_j \;,\\[5pt] 
&\overline{m}_{1,\,i}^{\,\alpha} = \sum_j r_{ij}^{\alpha}  \overline{W}_{ij} V_j \;,\hfill 
&&\frac{d \overline{m}_1^{\alpha}}{d r^{\gamma}}\bigg|_{i} = \sum_j \left(r_{ij}^{\alpha}  \frac{d \overline{W}}{d r^{\gamma}}\bigg|_{ij} + \delta^{\alpha \gamma}  \overline{W}_{ij}\right) V_j \;,\\[5pt]
&\overline{m}_{2,\,i}^{\,\alpha \beta} = \sum_j r_{ij}^{\alpha} r_{ij}^{\beta}  \overline{W}_{ij} V_j \;, \hfill
&&\frac{d \overline{m}_2^{\alpha \beta}}{d r^{\gamma}}\bigg|_{i} = \sum_j \left(r_{ij}^{\alpha} r_{ij}^{\beta}  \frac{d \overline{W}}{d r^{\gamma}}\bigg|_{ij} + \left(r_{ij}^{\alpha} \delta^{\beta \gamma} + \delta^{\alpha \gamma} r_{ij}^{\beta} \right) \overline{W}_{ij} \right) V_j \;.
\end{align}\nonumber
\end{minipage}%
\]

\vspace{5pt}
\noindent
\textit{B.1.3 Vacuum boundary switch}

\vspace{-8pt}
\noindent\textcolor{lightgray}{\rule{0.7\textwidth}{0.1mm}}
\vspace{-8pt}
\begin{equation}\label{eq:summary_vacswitch}
s(h_i |\mathbf{B}_i|) = 
   \begin{cases}
      \exp{\left[ -\,\dfrac{\left(0.8 - h_i |\mathbf{B}_i|\right)^2}{0.08} \right] } & \text{for $h_i |\mathbf{B}_i| \ge 0.8$} \;,\\
      1 & \text{otherwise}\;.\\
    \end{cases} 
\end{equation}

\newpage

\noindent
\textit{B.2 Artificial viscosity and artificial diffusion}

\vspace{-8pt}
\noindent\textcolor{black}{\rule{0.9\textwidth}{0.1mm}}
\vspace{-8pt}
\begin{equation}\label{eq:summary_calcQ}
   Q_{ij} = \frac{1}{2}\left(a_{\mathrm{visc}} + b_{\mathrm{visc}}\mathcal{B}_{i}^{\text{visc}}\right)\rho_i \left(-\alpha c_{i} \mu_{ij} + \beta \mu_{ij}^2\right) \;,
\end{equation}

\noindent
with $\alpha = 1.5,$ $\;\beta = 3$, $\;a_{\mathrm{visc}} = 2/3,\;$ and $\;b_{\mathrm{visc}} = 1/3$.

\begin{equation}\label{eq:summary_calcmu}
\mu_{ij} = 
   \begin{cases}
   \dfrac {\tilde{\mathbf{v}}_{ij} \cdot \boldsymbol\eta_{ij}}{\boldsymbol\eta_{ij} \cdot \boldsymbol\eta_{ij} + \epsilon^2}& \text{for $\tilde{\mathbf{v}}_{ij} \cdot \boldsymbol\eta_{ij} < 0$}\;,\\
   0 & \text{otherwise}\;,
   \end{cases}
\end{equation}

\noindent
with $\epsilon = 0.1$ and $\boldsymbol\eta_{ij} \equiv (\mathbf{r}_i - \mathbf{r}_j) / h_i$.

 \begin{align}\label{eq:summary_u_diff}
\left(\frac{du_i}{dt}\right)_{\text{difn}} &=  \sum\limits_{j} \kappa_{ij} \left(a_{u} + b_{u} \mathcal{B}_{ij}^{\text{difn}} \right) v_{\text{sig}, ij} \: (\tilde{u}_{j} - \tilde{u}_{i}) \frac{m_j}{\rho_{ij}} \frac{1}{2}\left|\frac{d \widetilde{\mathcal{W}}}{d\mathbf{r}}\bigg|_{ij} - \frac{d \widetilde{\mathcal{W}}}{d\mathbf{r}}\bigg|_{ji}\right| \;,\\
\label{eq:summary_rho_diff}
\left(\frac{d\rho_i}{dt}\right)_{\text{difn}} &=  \sum\limits_{j} \kappa_{ij} \left(a_{\rho} + b_{\rho} \mathcal{B}_{ij}^{\text{difn}} \right) v_{\text{sig}, ij} \: (\tilde{\rho}_{j} - \tilde{\rho}_{i})  \frac{\rho_i}{\rho_j} \frac{m_j}{\rho_{ij}} \frac{1}{2}\left|\frac{d \widetilde{\mathcal{W}}}{d\mathbf{r}}\bigg|_{ij} - \frac{d \widetilde{\mathcal{W}}}{d\mathbf{r}}\bigg|_{ji}\right| \;,
\end{align}

\noindent
with $a_{u} = a_{\rho} = 0.05, \; b_{u} = b_{\rho} = 0.95, \; v_{\text{sig},\; ij} = \left|\tilde{\mathbf{v}}_{i} - \tilde{\mathbf{v}}_{j}\right|, \; \rho_{ij} \equiv (\rho_i + \rho_j) / 2, \,$ and $\;\kappa_{ij} = 1$ for particles of the same material and $\kappa_{ij} = 0$ otherwise.

\vspace{10pt}
\noindent
\textit{B.2.1. Quantities reconstructed to particle midpoints}

\vspace{-8pt}
\noindent\textcolor{lightgray}{\rule{0.7\textwidth}{0.1mm}}
\vspace{-8pt}
\[
\begin{minipage}{\linewidth}
  \footnotesize
\begin{align}
&\tilde{v}_{ij}^\alpha = v_{i}^\alpha + \frac{1}{2}\left(1 - \mathcal{B}_{i}^{\text{SL}}\right) \Phi_{v,\;ij} \left(r_j^\gamma - r_i^\gamma \right)  \partial_i^{\gamma}  \hat{v}^\alpha \;,
&&\partial_i^{\gamma}  \hat{v}^\alpha = \sum_j (v_j^\alpha - v_i^\alpha)  \, \partial_i^{\gamma} \hat{W}_{ij} \frac{m_j}{\rho_j} \;, \\[5pt]
&\tilde{u}_{i} = u_{i} + \frac{1}{2}\Phi_{u,\;ij} \left(r_j^\gamma - r_i^\gamma \right)  \partial_{\kappa, \,i}^{\gamma} \, \hat{u} \;,
&&\partial_{\kappa, \,i}^{\gamma} \, \hat{u} = \sum_j \kappa_{ij} \, (u_j - u_i)  \, \partial_i^{\gamma} \hat{W}_{ij} \frac{m_j}{\rho_j} \;,  \\[5pt]
&\tilde{\rho}_{i} = \rho_{i} + \frac{1}{2}\Phi_{\rho,\;ij} \left(r_j^\gamma - r_i^\gamma \right) \partial_{\kappa, \,i}^{\gamma} \, \hat{\rho} \;,
&&\partial_{\kappa, \,i}^{\gamma} \, \hat{\rho} = \sum_j \kappa_{ij}  \, (\rho_j - \rho_i)  \, \partial_i^{\gamma} \hat{W}_{ij}  \frac{m_j}{\rho_j} \;.
\end{align}\nonumber
\end{minipage}%
\]

\vspace{5pt}
\noindent
\textit{B.2.2. Slope limiter}

\vspace{-8pt}
\noindent\textcolor{lightgray}{\rule{0.7\textwidth}{0.1mm}}
\vspace{-8pt}
\[
\begin{minipage}{\linewidth}
  \footnotesize
  \begin{align}
  &\Phi_{ij} = 
   \begin{cases}
      0 & \text{for $A_{ij} < 0$}\;,\\
      \dfrac {4 A_{ij}}{(1 + A_{ij})^2} \exp{\left[-\left( \dfrac {\eta_{ij}^{\rm min} - \eta_{\text{crit}}}{0.2} \right)^2\right]} & \text{for $\eta_{ij}^{\rm min} < \eta_{\text{crit}}$}\;,\\
      \dfrac {4 A_{ij}}{(1 + A_{ij})^2} & \text{otherwise}\;, 
    \end{cases}       
   &&\eta_{\rm crit} = \frac{1}{h_i} \left(\frac{1}{\sum_j W_{ij}}\right)^{1/d} \equiv \frac{1}{\eta_{\rm kernel}} \;, \\[-5pt]\nonumber
\end{align}\nonumber
\end{minipage}%
\]
\vspace{-10pt}
\[
\begin{minipage}{\linewidth}
  \footnotesize
\begin{align}
& A_{v, \;ij} = \frac{\partial_i^\beta \hat{v}^\alpha (\mathbf{r}_j - \mathbf{r}_i)^\beta (\mathbf{r}_j - \mathbf{r}_i)^\alpha }{\partial_j^\gamma \hat{v}^\phi (\mathbf{r}_j - \mathbf{r}_i)^\gamma (\mathbf{r}_j - \mathbf{r}_i)^\phi} \;,
&&A_{u, \;ij} = \frac{\partial_{\kappa, \,i}^\alpha \, \hat{u} (\mathbf{r}_j - \mathbf{r}_i)^\alpha }{\partial_{\kappa, \,j}^\beta \, \hat{u} (\mathbf{r}_j - \mathbf{r}_i)^\beta} \;,
&&&A_{\rho, \;ij} = \frac{\partial_{\kappa, \,i}^\alpha \, \hat{\rho} (\mathbf{r}_j - \mathbf{r}_i)^\alpha }{\partial_{\kappa, \,j}^\beta \, \hat{\rho} (\mathbf{r}_j - \mathbf{r}_i)^\beta} \;.
\end{align}\nonumber
\end{minipage}%
\]

\vspace{5pt}
\noindent
\textit{B.2.3. Balsara switch}

\vspace{-8pt}
\noindent\textcolor{lightgray}{\rule{0.7\textwidth}{0.1mm}}
\vspace{-8pt}
\[
\begin{minipage}{\linewidth}
  \footnotesize
\begin{align}
&\mathcal{B}_{i}^{\text{visc}} \equiv \mathcal{B}_{i}^{\text{SL}} \equiv \mathcal{B}_i = \frac{\left| \nabla \cdot \mathbf{v}_i \right|}{\left| \nabla \cdot \mathbf{v}_i \right| + \left| \nabla \times \mathbf{v}_i \right| + 0.0001 c_{i} / h_i} \; ,
&&\mathcal{B}_{ij}^{\text{difn}} = \frac{\mathcal{B}_{i} + \mathcal{B}_{j}}{2}\;. 
\end{align}\nonumber
\end{minipage}%
\]

\vspace{5pt}
\noindent
\textit{B.3. Normalising term}

\vspace{-8pt}
\noindent\textcolor{black}{\rule{0.9\textwidth}{0.1mm}}
\vspace{-8pt}
\begin{equation}\label{eq:summary_drho_dt_normalisation_term}
    \left(\frac{d\rho_i}{dt}\right)_{\text{norm}} = \alpha_{\text{norm}} \, s_i \, (m_{0,\,i} - 1) \, \rho_{i} \sum\limits_{j}  v_{\text{norm}, \; ij} \:   \frac{m_j}{\rho_{ij}} \frac{1}{2}\left|\frac{d \widetilde{\mathcal{W}}}{d\mathbf{r}}\bigg|_{ij} - \frac{d \widetilde{\mathcal{W}}}{d\mathbf{r}}\bigg|_{ji}\right| \;,
\end{equation}

\noindent
with $\alpha_{\text{norm}} = 1, \; v_{\text{norm},\; ij} = \left|\mathbf{v}_{i} - \mathbf{v}_{j}\right|, \; \rho_{ij} \equiv (\rho_i + \rho_j) / 2, \,$ and $\;m_{0,\,i} = \sum_j  W_{ij} V_j\;$.

\newpage

\section{Traditional SPH formulations used for comparison simulations}\label{app:trad}

\begin{equation}\label{eq:summary_tsph_rho}
\langle\rho_i\rangle = \sum_j m_j W_{ij} \;,
\end{equation}

\begin{align}\label{eq:summary_tsph_a}
\frac{d v_i^{\alpha}}{d t} &=  -\sum\limits_{j} m_j \, \left( \frac{f_{ij} \, P_i}{\langle\rho_i\rangle^2} \, \frac{\partial W}{\partial r^{\alpha}}\bigg|_{ij} - \frac{f_{ji} \, P_j}{\langle\rho_j\rangle^2} \, \frac{\partial W}{\partial r^{\alpha}}\bigg|_{ji} \right)  \: + \: \left(\frac{d v_i^{\alpha}}{d t}\right)_{\text{visc}} \;,\\
\label{eq:summary_tsph_dudt}
\frac{d u_i}{d t} &=  \sum\limits_{j} m_j \, \frac{f_{ij} \, P_i}{\langle\rho_i\rangle^2} \,  v_{ij}^{\alpha} \, \frac{\partial W}{\partial r^{\alpha}}\bigg|_{ij}   \: + \: \left(\frac{d u_i}{d t}\right)_{\text{visc}}  \;.
\end{align}

\vspace{5pt}
\noindent
\textit{C.1 Gradient of smoothing length factor}

\vspace{-8pt}
\noindent\textcolor{black}{\rule{0.9\textwidth}{0.1mm}}
\vspace{-8pt}
\begin{equation}\label{eq:summary_tsph_n}
\langle n_i\rangle = \sum_j W_{ij} \;,
\end{equation}

\begin{equation}\label{eq:summary_tsph_omega}
f_{ij} = 1  - \frac{1}{m_j} \, \frac{h_i}{d \, \langle\rho_i\rangle} \frac{\partial \langle\rho\rangle}{\partial h}\bigg|_{i} \, \left(  1  + \frac{h_i}{d \, \langle n_i\rangle} \frac{\partial \langle n\rangle}{\partial h}\bigg|_{i} \right)^{-1} \;,
\end{equation}

\vspace{5pt}
\noindent
where $d$ is the number of spatial dimensions. 

\vspace{10pt}
\noindent
\textit{C.2 Artificial viscosity}

\vspace{-8pt}
\noindent\textcolor{black}{\rule{0.9\textwidth}{0.1mm}}
\vspace{-8pt}
\begin{align}\label{eq:summary_tsph_a_visc}
 \left(\frac{d v_i^{\alpha}}{d t}\right)_{\text{visc}} &= -\sum\limits_{j} \frac{\mathcal{B}_i + \mathcal{B}_j}{2} \, \left(-\alpha c_{ij} \, \mu_{ij} + \beta \mu_{ij}^2\right) \frac{m_j}{\langle\rho_{ij}\rangle}    \, \frac{1}{2} \,\left( f_{ij} \, \frac{\partial W}{\partial r^{\alpha}}\bigg|_{ij} - 
 f_{ji}\, \frac{\partial W}{\partial r^{\alpha}}\bigg|_{ji} \right)\;,\\
\label{eq:summary_tsph_dudt_visc}
\left(\frac{d u_i}{d t}\right)_{\text{visc}} &=  \frac{1}{2} \, \sum\limits_{j} \frac{\mathcal{B}_i + \mathcal{B}_j}{2} \, \left(-\alpha c_{ij} \, \mu_{ij} + \beta \mu_{ij}^2\right) \frac{m_j}{\langle\rho_{ij}\rangle}  \,  f_{ij} \, v_{ij}^{\alpha} \, \frac{\partial W}{\partial r^{\alpha}}\bigg|_{ij} \;,
\end{align}

\begin{equation}\label{eq:summary_tsph_calcmu}
\mu_{ij} = 
   \begin{cases}
   \dfrac {\mathbf{v}_{ij} \cdot \mathbf{r}_{ij}}{|\mathbf{r}_{ij}|}& \text{for $\mathbf{v}_{ij} \cdot \mathbf{r}_{ij} < 0$}\;,\\
   0 & \text{otherwise}\;,
   \end{cases}
\end{equation}

with $\alpha = 1.5, \; \beta = 3, \; \langle\rho_{ij}\rangle \equiv \left(\langle\rho_i\rangle + \langle\rho_j\rangle\right) / 2, \; c_{ij} \equiv (c_i + c_j) / 2, \; \mathbf{v}_{ij} \equiv \mathbf{v}_{i} - \mathbf{v}_{j}, \,$ and $\; \mathbf{r}_{ij} \equiv \mathbf{r}_{i} - \mathbf{r}_{j}.$

\begin{equation}\label{eq:summary_tsph_balsara}
\mathcal{B}_i = \frac{\left| \nabla \cdot \mathbf{v}_i \right|}{\left| \nabla \cdot \mathbf{v}_i \right| + \left| \nabla \times \mathbf{v}_i \right| + 0.0001 c_{i} / h_i} \;.
\end{equation}

\vspace{5pt}
\noindent
\textit{C.3 Artificial conduction}

\vspace{-8pt}
\noindent\textcolor{black}{\rule{0.9\textwidth}{0.1mm}}
%\vspace{-8pt}

\noindent
Only used where indicated. Based on the SPHENIX implementation \citep{borrow2022sphenix}. 

\vspace{5pt}

\begin{equation}\label{eq:summary_tsph_dudt_with_cond}
\frac{d u_i}{d t} =  \frac{P_i}{ \langle\rho_i\rangle^2} \, \sum\limits_{j} m_j \,  v_{ij}^{\alpha} \, \frac{\partial W}{\partial r^{\alpha}}\bigg|_{ij}   \: + \: \left(\frac{d u_i}{d t}\right)_{\text{visc}}  \: + \: \left(\frac{d u_i}{d t}\right)_{\text{cond}}  \;,
\end{equation}

\begin{equation}\label{eq:summary_tsph_dudt_cond}
\left(\frac{d u_i}{d t}\right)_{\text{cond}} = \sum\limits_{j} \, \alpha_{c, \, ij} \, v_{c, \, ij} \, \left(u_i - u_j\right) \frac{m_j}{\langle\rho_{ij}\rangle}    \, \frac{1}{2} \,\left( f_{ij} \, \frac{\partial W}{\partial r^{\alpha}}\bigg|_{ij} - 
 f_{ji}\, \frac{\partial W}{\partial r^{\alpha}}\bigg|_{ji} \right)\;,
\end{equation}

\begin{align}\label{eq:summary_tsph_alpha_cij}
 &\alpha_{c, \, ij} = \frac{P_i \, \alpha_{c, \, i} + P_j \, \alpha_{c, \, j}}{P_i + P_j} \;,
 && v_{c, \, ij} = \frac{1}{2} \, \left( \frac{|\mathbf{v}_{ij} \cdot \mathbf{r}_{ij}|}{|\mathbf{r}_{ij}|} + \sqrt{\frac{|P_i - P_j|}{\langle\rho_{ij}\rangle}}\right) \; ,
\end{align}

\noindent
where $\alpha_{c, \, i}$ is evolved in time as described by \citet{borrow2022sphenix}.

\newpage

\section{Kernel choice}\label{app:kernel}

\begin{figure}[t]
	\centering
{\includegraphics[width=\textwidth, trim={2.5mm 0mm 2.5mm 0mm}, clip]{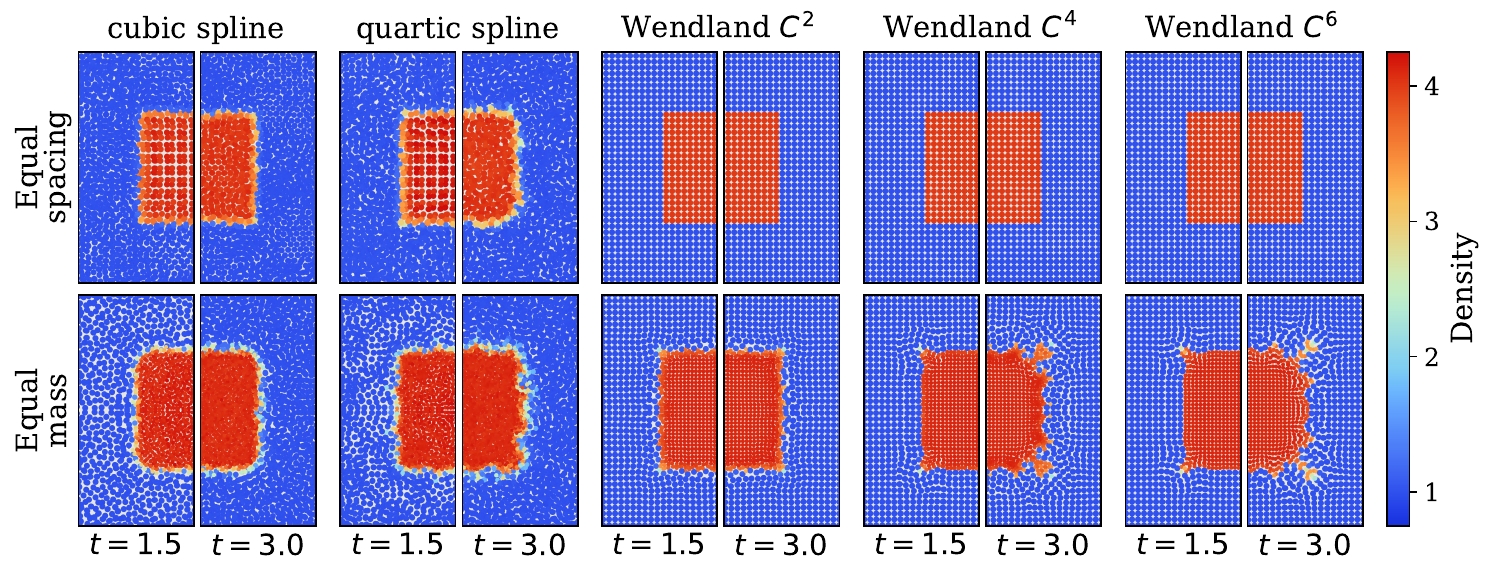}
}\hfill
	\vspace{-2em}
	\caption{The effect of kernel choice in square test simulations. We plot snapshots from REMIX square test simulations with equal initial particle spacing and equal particle mass, for 5 different kernel functions at two times. Particles are coloured by their density.}
	\label{fig:kernel_square}
\end{figure}

\begin{figure}[!t]
	\centering
{\includegraphics[width=\textwidth, trim={2.5mm 0mm 2.5mm 0mm}, clip]{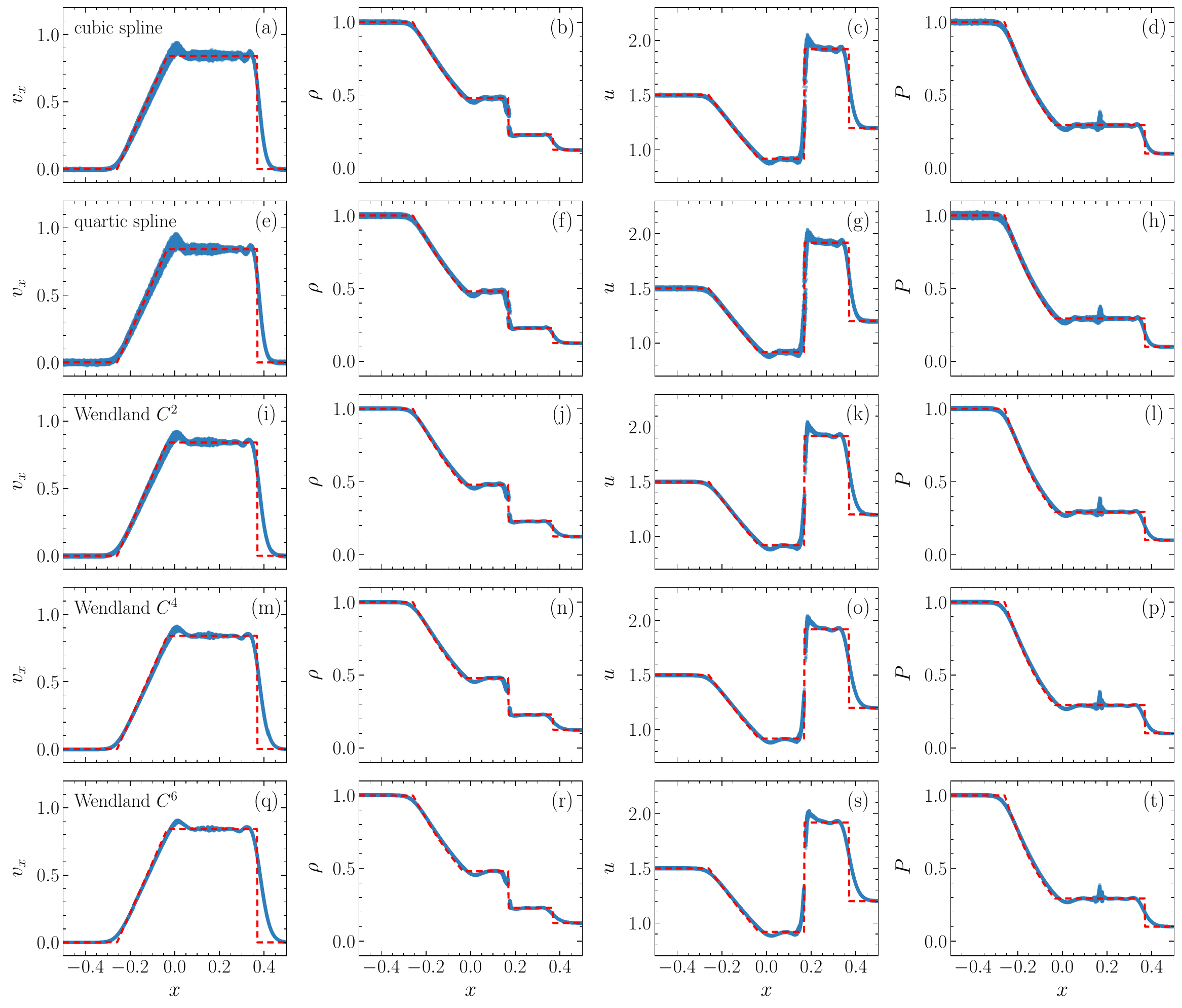}
}\hfill
	\vspace{-2em}
	\caption{The effect of kernel choice in REMIX Sod shock tube simulations. Velocity in the $x$-direction, $v_x$, density, $\rho$, specific internal energy, $u$, and pressure, $P$ plotted against $x$-position at time $t = 0.2$. Rows correspond to the 5 different kernel functions. The reference solution is plotted in red.}
	\label{fig:kernel_sod}
\end{figure}

\begin{figure}[t!]
	\centering
{\includegraphics[width=\textwidth, trim={2.5mm 0mm 2.5mm 0mm}, clip]{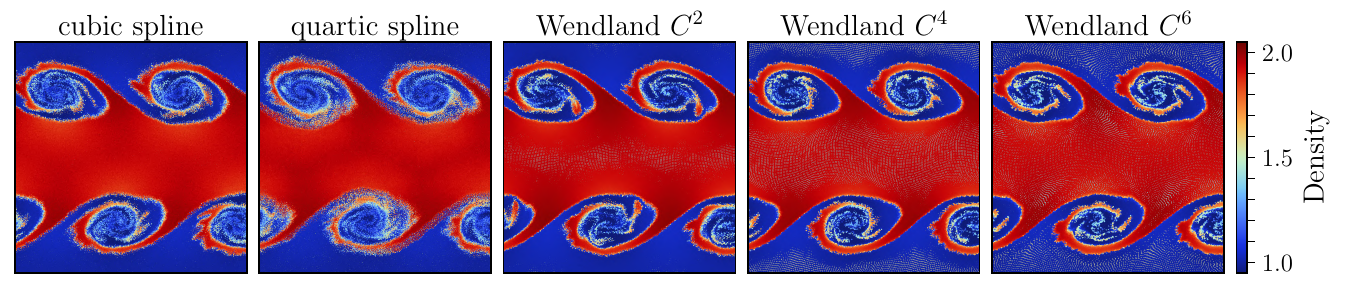}
}\hfill
	\vspace{-2em}
	\caption{The effect of kernel choice in REMIX ideal gas KHI simulation. Plotted at time $t = 2~\tau_{\rm KH}$, for 5 different kernel functions. Particles are coloured by their density. These simulations have a resolution of $N_1 = 128$, as described in \S\ref{subsubsec:kh_idg_discontinuous}.}
	\label{fig:kernel_kh_idg}
\end{figure}

\begin{figure}[t]
	\centering
{\includegraphics[width=\textwidth, trim={2.5mm 0mm 2.5mm 0mm}, clip]{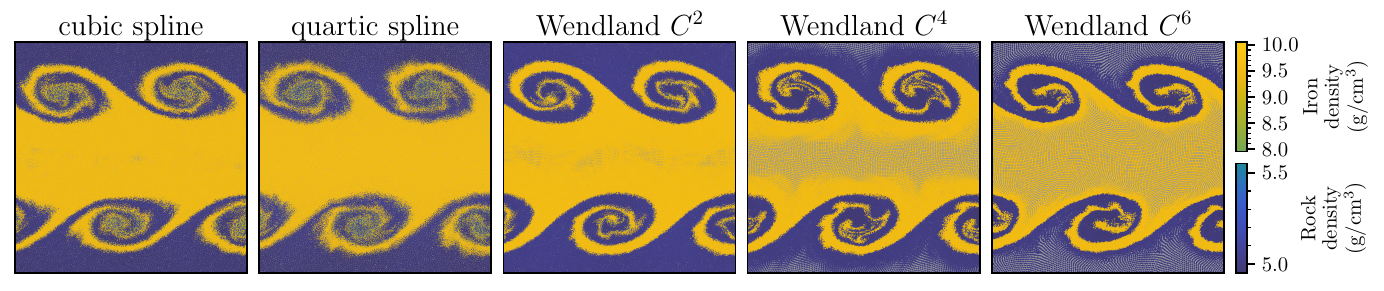}
}\hfill
	\vspace{-2em}
	\caption{The effect of kernel choice in REMIX KHI simulations between Earth-like iron \& rock. Plotted at time $t = 2~\tau_{\rm KH}$, for 5 different kernel functions. Particles are coloured by their material type and density. These simulations have a resolution of $N_1 = 128$, as described in \S\ref{subsubsec:kh_idg_discontinuous}.}
	\label{fig:kernel_kh_earth}
\end{figure}

Here we present results from square test (\S \ref{subsec:square}), Sod shock tube (\S \ref{subsec:sod}), ideal gas KHI (\S \ref{subsubsec:kh_idg_discontinuous}), and KHI with Earth-like iron \& rock  (\S \ref{subsec:kh_earth}) simulations to show the effect of smoothing kernel choice in REMIX simulations. We use five different kernel functions with corresponding $\eta_{\rm kernel}$: cubic spline with $\eta_{\rm kernel} = 1.292$ (${\sim}55$ neighbours); quartic spline with $\eta_{\rm kernel} = 1.203$ (${\sim}60$ neighbours); Wendland $C^2$ with $\eta_{\rm kernel} = 1.487$ (${\sim}100$ neighbours); Wendland $C^4$ with $\eta_{\rm kernel} = 1.643$ (${\sim}200$ neighbours); Wendland $C^6$ with $\eta_{\rm kernel} = 1.866$ (${\sim}400$ neighbours) \citep{dehnen2012improving}.

Square tests, presented in Fig.~\ref{fig:kernel_square}, show similar behaviour in the equal spacing scenario for all of these kernels, although with more noise in the lower-order kernels with fewer neighbours. For the higher-order Wendland kernels, results are very similar over these timescales, with very little particle motion. In the equal mass scenario, however, the use of either the Wendland $C^4$ or Wendland $C^6$ kernels leads to spurious behaviour at the corners. The particle noise in simulations with lower-order kernels is in fact helpful in disturbing the growth of these slowly evolving, error-driven features. We found that using grad-$h$ terms calculated directly from Eqn.~\ref{eq:smoothinglength} combined with a higher artificial viscosity helps the treatment of corners in the square tests with higher-order kernels. However, zeroth-order error in grad-$h$ terms calculated in that way leads to problematic behaviour in regions away from density discontinuities, and we choose to take a conservative artificial viscosity approach, keeping it low away from shocks. Therefore, we choose to calculate grad-$h$ terms as we describe in \S \ref{subsec:remix_linearkernels}. Again, lower-order kernels show more particle noise, however, the cross-section of the cube does not lose its square shape.

In the Sod shock (Fig.~\ref{fig:kernel_sod}) and both KHI tests (Figs.~\ref{fig:kernel_kh_idg} and \ref{fig:kernel_kh_earth}) We see a general trend of higher-order kernels reducing particle noise. However, these effects are minor compared with the primary improvements in all these simulations compared with traditional SPH equivalents.

Based on these simulations, we conclude that the Wendland $C^2$ kernel is a good compromise between accuracy and computational speed, which is why we use it for all simulations other than those presented in this section. Using a higher-order kernel only leads to small improvements in noise reduction in these tests and, in the case of the square test, gives worse results. We stress, however, that the behaviour at the corners of a 3D cube is not necessarily important for many science applications, if other benefits are offered in more typical configurations. The lower-order kernels lead to more particle noise, however, these still show significant improvements compared with traditional SPH simulations. This suggests that, for example a cubic spline kernel could be used with REMIX for applications where simulation run-speed is an important consideration.

\section{Choices made in linear-order reproducing kernel construction}\label{app:kernelconstruction}

\begin{figure}[t]
	\centering
{\includegraphics[width=\textwidth, trim={2.5mm 0mm 2.5mm 0mm}, clip]{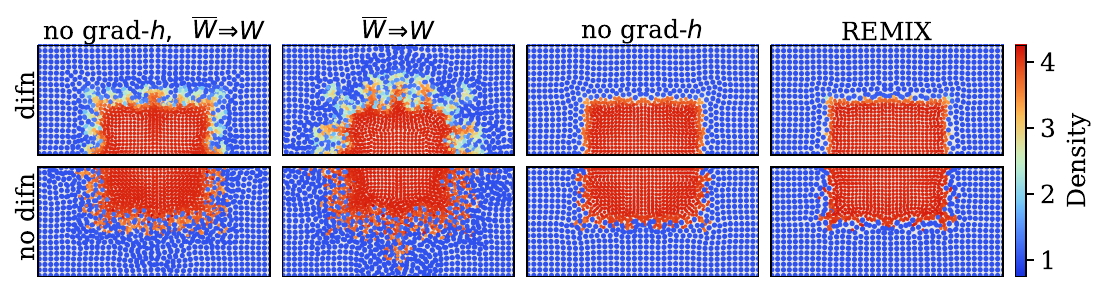}
}\hfill
	\vspace{-2em}
	\caption{The effect of choices made in the construction of linear-order reproducing kernels in square test simulations. Plots show a central cross-section from square test simulations with equal mass particles, using subtly different variations of REMIX. Snapshots are at time $t = 3.0$. We show the results with (top) and without (bottom) diffusion of density and internal energy, for the full REMIX scheme and with combinations of either no grad-$h$ terms and without using an averaged kernel in the construction of our linear-order reproducing kernels.}
	\label{fig:square_methods}
\end{figure}

In Fig.~\ref{fig:square_methods}, sensitivities to subtle choices in the kernel construction (\S\ref{subsec:remix_linearkernels}) are presented in the context of the square test with equal mass particles (\S \ref{subsec:square}). We include panels to show equivalent cases without artificial diffusion of density or internal energy in the model. This isolates the effect of kernel choice from stabilising effects caused by stronger diffusion for particles with higher relative speeds. 

We consider using $W_{ij}$ rather than the averaged kernel $\overline{W}_{ij} \equiv [W(\mathbf{r}_{ij}, h_i) + W(\mathbf{r}_{ji}, h_j)] / 2$. This permeates through the full reproducing kernel construction, including geometric moments and their gradients. Switching to $W_{ij}$ also increases grad-$h$ terms by a factor of 2, as seen in Eqn.~\ref{eq:dW_dr}. Using $\overline{W}_{ij}$ in the construction of the reproducing kernels is shown to be important in achieving good behaviour in the square test, with all $W_{ij}$ cases showing a disruption of the cube by particle motions at the interface. This demonstrates how additional error introduced in kernel antisymmetrisation (for conservation) in the equations of motion is sensitive to the base-kernel used in the construction of the reproducing kernels. Regardless of the kernel used in the construction, the cube would remain undisturbed if the antisymmetrisation step was not carried out.

The inclusion of grad-$h$ terms only leads to a small effect. We note again that we include these terms primarily for completeness of the methods, and since they have no negative impact on run speed or any other considerations.

\section{Choices made in artificial viscosity and diffusion construction}\label{app:viscdiff}

\begin{figure}[t]
	\centering
{\includegraphics[width=\textwidth, trim={2.5mm 0mm 2.5mm 0mm}, clip]{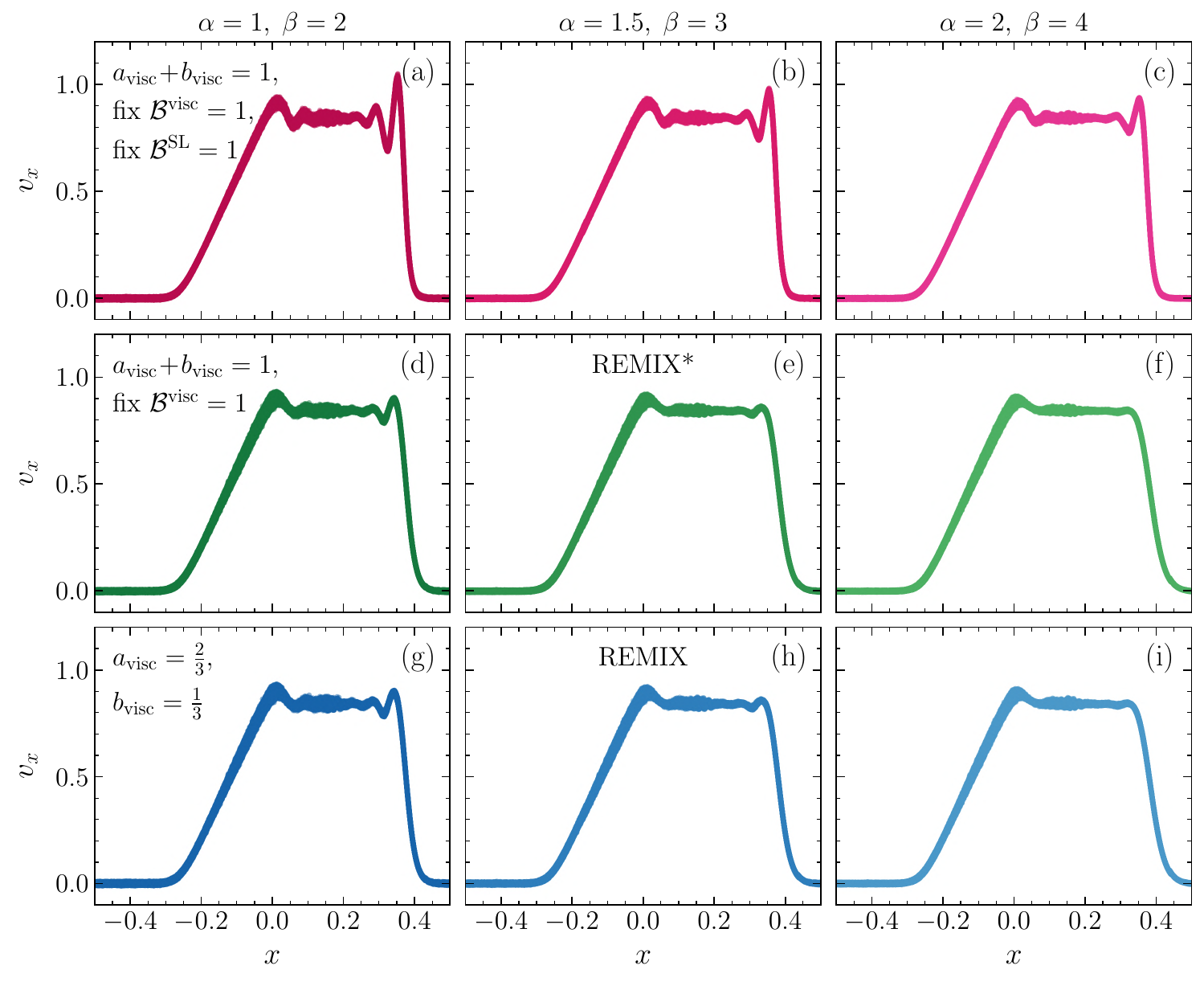}
}\hfill
	\vspace{-2.5em}
	\caption{The effect of artificial viscosity parameters in Sod shock tube simulations. Velocity in the $x$-direction, $v_x$, plotted against $x$-position from Sod shock tube simulations at time $t = 0.2$, using REMIX with variations in artificial viscosity formulation. The annotations ``REMIX'' and ``REMIX*'' correspond to panels showing the final REMIX scheme and the equivalent simplified shock case respectively.}
	\label{fig:sodshock_noB_visc}
\end{figure}

Here we show how Sod shock (\S\ref{subsec:sod}) and ideal gas Kelvin--Helmholtz instability (\S\ref{subsec:kh_idg}) simulations motivate the parameters used in the REMIX artificial viscosity and artificial diffusion schemes. We (1) demonstrate the effectiveness of a Balsara switch in the viscosity slope limiter in reducing oscillations in shocks; (2) motivate our choice of the standard viscosity constants, $\alpha$ and $\beta$; (3) motivate our choice of the parameters that distinguish between the treatment of artificial viscosity in shocks and shearing regions, $a_{\rm visc}$ and $b_{\rm visc}$; (4) motivate our choice of the parameters that distinguish between the treatment of artificial diffusion in shocks and shearing region,  $a_{\rm difn} \equiv a_{u} = a_{\rho}$ and $b_{\rm difn} \equiv b_{u} = b_{\rho}$. For simplicity, we use the same parameters for the diffusion of internal energy and density. Our approach to these artificial terms is a conservative one, in which we deliberately keep them weak, while still strong enough to give noticeable improvements.

The Sod shock and KHI examples were chosen since the artificial viscosity and diffusion play different roles in these two scenarios: we require strong viscosity and diffusion to accurately capture shocks, whereas in shearing regions they are only required to smooth out accumulated noise in particle velocities, densities, and internal energies. To isolate these elements of the construction we use simplified versions of the REMIX construction, in which we overwrite the Balsara switches that appear in Eqns.~\ref{eq:calcQ}--\ref{eq:rho_diff}, to 1 or 0 to isolate the treatment in shocks or shearing regions respectively: In the simplified shock case, we set $\mathcal{B} = 1$, such that the strength of viscosity or diffusion is parameterised by the respective $a + b$ in Eqns.~\ref{eq:calcQ}--\ref{eq:rho_diff}. In the simplified shear case, we set $\mathcal{B} = 0$, such that the strengths are set by $a$ only. Therefore, we use these to select values for $a$ and $b$ parameters, based on which the full scheme can then switch smoothly between these two simplified versions.

We first focus on the artificial viscosity scheme. In Fig.~\ref{fig:sodshock_noB_visc} we investigate the effect of changes in the viscosity treatment on Sod shock tube simulations. Rows correspond to: the simplified shock case with  $a_{\rm visc} + b_{\rm visc} = 1$ and Balsara switches in the viscosity slope limiter set to $\mathcal{B}^{\text{SL}} = 1$; the simplified shock case again with $a_{\rm visc} + b_{\rm visc} = 1$, but without fixing $\mathcal{B}^{\text{SL}}$; and the final REMIX scheme. Columns correspond to different choices of $\alpha$ and $\beta$, with a consistent $\beta = 2 \alpha$ in all cases here. We note that fixing this ratio leads to a degeneracy for $\alpha$ and $\beta$ with $a_{\rm visc} + b_{\rm visc}$ in the simplified shock case, and therefore we choose to set $a_{\rm visc} + b_{\rm visc} = 1$. 

We first note that the Balsara switch $\mathcal{B}^{\text{SL}}$ is effective in dissipating oscillations in the shock. The oscillations in the case that makes use of $\mathcal{B}^{\text{SL}}$ but has the weakest viscosity (Fig.~\ref{fig:sodshock_noB_visc}(d)) are smaller than even those in \ref{fig:sodshock_noB_visc}(c), where the $\alpha$ and $\beta$ factors are twice as large, but $\mathcal{B}^{\text{SL}} = 1$ is fixed. Using $\mathcal{B}^{\text{SL}}$ therefore allows us to target viscosity to shocks by more effectively switching off the linear reconstruction of velocities to particle midpoints. This in turn allows us to reduce the viscosity parameters so that we can construct a less dissipative artificial viscosity scheme. We note that Fig.~\ref{fig:sodshock_noB_visc}(c) is equivalent to the viscosity construction of \citet{frontiere2017crksph} and \citet{rosswog2020lagrangian}, other than the latter's use of quadratic reconstruction, which only makes a minor difference.

Next we consider how the Sod shock is sensitive to the primary viscosity parameters $\alpha$ and $\beta$ for $a_{\rm visc} + b_{\rm visc} = 1$, as shown in Fig.~\ref{fig:sodshock_noB_visc}(d)--(f). Increasing $\alpha$ and $\beta$ in this way uniformly increases the strength of the viscosity. With $\alpha=2$ and $\beta=4$, oscillations are effectively removed.  With $\alpha=1$ and $\beta=2$, oscillations in the shock are still large compared with the particle scatter. $\alpha=1.5$ and $\beta=3$, which we eventually use in REMIX, still has some oscillations, but these are small. The bottom row of figures demonstrates that our simplification of this shock case with  $\mathcal{B}^{\text{visc}} = 1$ is appropriate as we see little difference when comparing with a full REMIX-like construction.

\begin{figure}[t]
	\centering
{\includegraphics[width=\textwidth, trim={2.5mm 0mm 2.5mm 0mm}, clip]{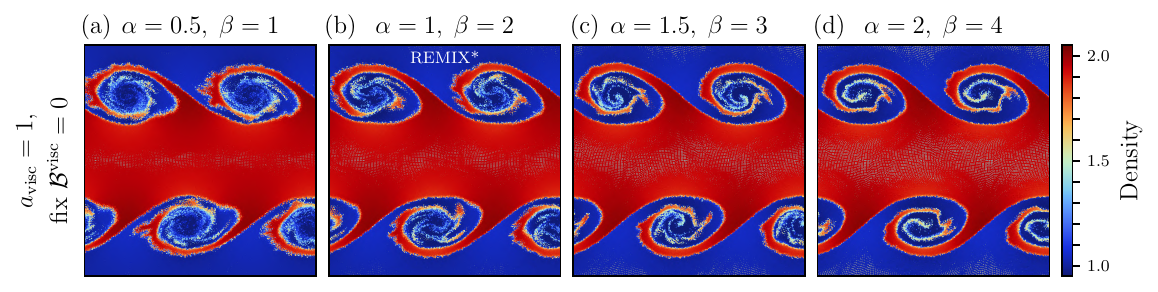}
}\hfill
	\vspace{-2em}
	\caption{The effect of artificial viscosity parameters in ideal gas KHI simulations with $\mathcal{B}^{\text{visc}}$ fixed to 0. Snapshots are from simulations with sharp initial discontinuities, at time $t = 2~\tau_{\rm KH}$, using REMIX with variations in artificial viscosity formulation. Particles are coloured by their density. The annotation ``REMIX*'' corresponds to the panel showing the simplified shear case that is equivalent to the final REMIX scheme. These simulations have a resolution of $N_1 = 128$, as described in \S\ref{subsubsec:kh_idg_discontinuous}.}
	\label{fig:kh_visc}
\end{figure}

\begin{figure}[t]
	\centering
{\includegraphics[width=\textwidth, trim={2.5mm 0mm 2.5mm 0mm}, clip]{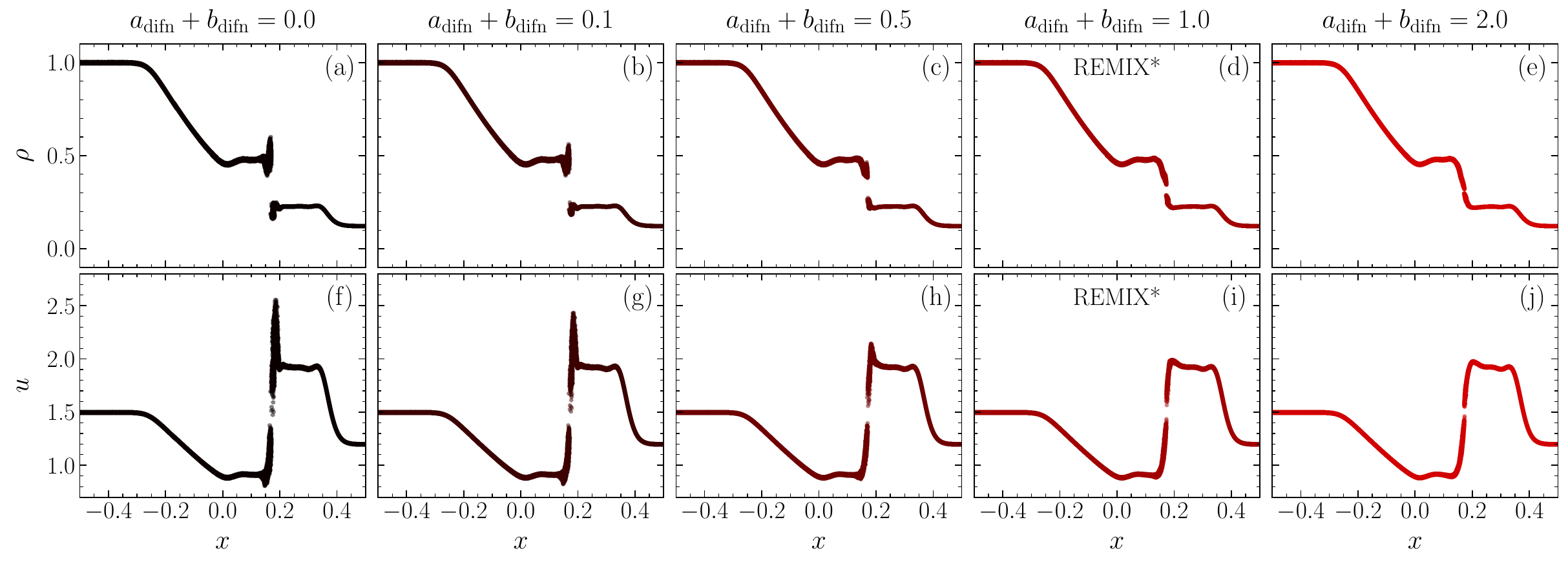}
}\hfill
	\vspace{-2em}
	\caption{The effect of artificial diffusion parameters in Sod shock tube simulations with $\mathcal{B}^{\text{difn}}$ fixed to 1. Density, $\rho$, and specific internal energy, $u$, plotted against $x$-position at time $t = 0.2$, using REMIX with variations in artificial diffusion formulation. The annotation ``REMIX*'' corresponds to the panels showing the simplified shock case that is equivalent to the final REMIX scheme.}
	\label{fig:sodshock_noB_diff}
\end{figure}

We use the ideal gas KHI in the simplified shear case, combined with the above Sod shock results, to make decisions for values of the REMIX viscosity parameters. In Fig.~\ref{fig:kh_visc} we see the effect of changing $\alpha$ and $\beta$  with $a_{\rm visc} = 1$ and $\mathcal{B}^{\text{visc}}$ fixed to 0. Again we note the degeneracy in increasing $\alpha$ and $\beta$, this time with increasing $a_{\rm visc}$. With increased artificial viscosity, the boundary of the spiralling KHI plume becomes more pronounced, with less mixing of particles across the interface. In the lowest viscosity case, the small-scale spirals are diffused and structure is not maintained.

For the REMIX scheme we take a conservative approach and choose a viscosity model that, with the assumptions of the simplified approaches considered here, switches between: \ref{fig:sodshock_noB_visc}(e) in shocks, in which oscillations have mostly, but not fully, been removed; to Fig.~\ref{fig:kh_visc}(b), in which small-scale KHI structure persists, but mixing on the particle scale is not strongly suppressed by artificial viscosity. We note that, in practice, the artificial viscosity will be slightly between the two cases we aim to switch between. This was kept in mind when making this choice. This corresponds to choices of $\alpha=1.5$, $\beta=3$ and $a_{\rm visc} = 2/3$, $b_{\rm visc} = 1/3$.

Next we consider the artificial diffusion model. We use the same parameter values in the artificial diffusion of density and internal energy, for simplicity. First we consider the Sod shock in the simplified shock case, however this time applied to the diffusion equations (Eqns.~\ref{eq:u_diff} and \ref{eq:rho_diff}). In Fig.~\ref{fig:sodshock_noB_diff}, we demonstrate our need for artificial diffusion of both density and internal energy: in the first column, we see the case of no diffusion leading to sizable spikes in these quantities. As the strength of artificial diffusion is increased, the spikes are smoothed. 

\begin{figure}[t]
	\centering
{\includegraphics[width=\textwidth, trim={2.5mm 0mm 2.5mm 0mm}, clip]{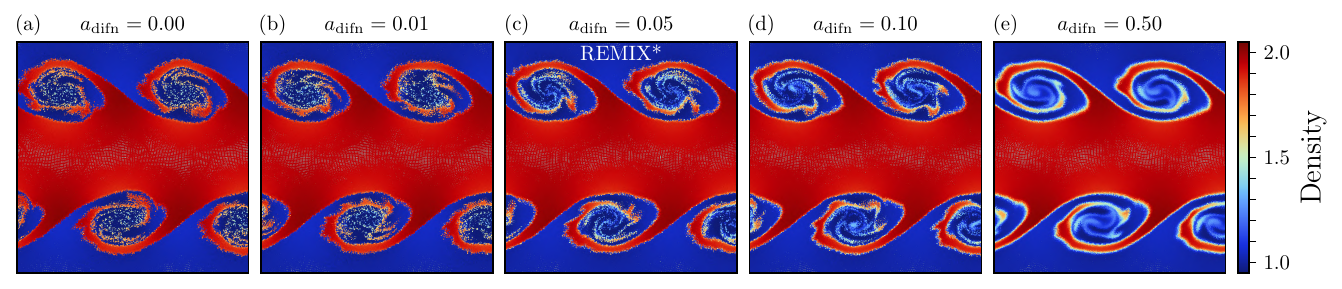}
}\hfill
	\vspace{-2em}
	\caption{The effect of artificial diffusion parameters in ideal gas KHI simulations with $\mathcal{B}^{\text{difn}}$ fixed to 0. Snapshots are from simulations with sharp initial discontinuities, at time $t = 2~\tau_{\rm KH}$, using REMIX with variations in artificial diffusion formulation. Particles are coloured by their densities. The annotation ``REMIX*'' corresponds to the panel showing the simplified shear case that is equivalent to the final REMIX scheme. These simulations have a resolution of $N_1 = 128$, as described in \S\ref{subsubsec:kh_idg_discontinuous}.}
	\label{fig:kh_cond}
\end{figure}

 \begin{figure}[b]
	\centering
{\includegraphics[width=\textwidth, trim={2.5mm 0mm 2.5mm 0mm}, clip]{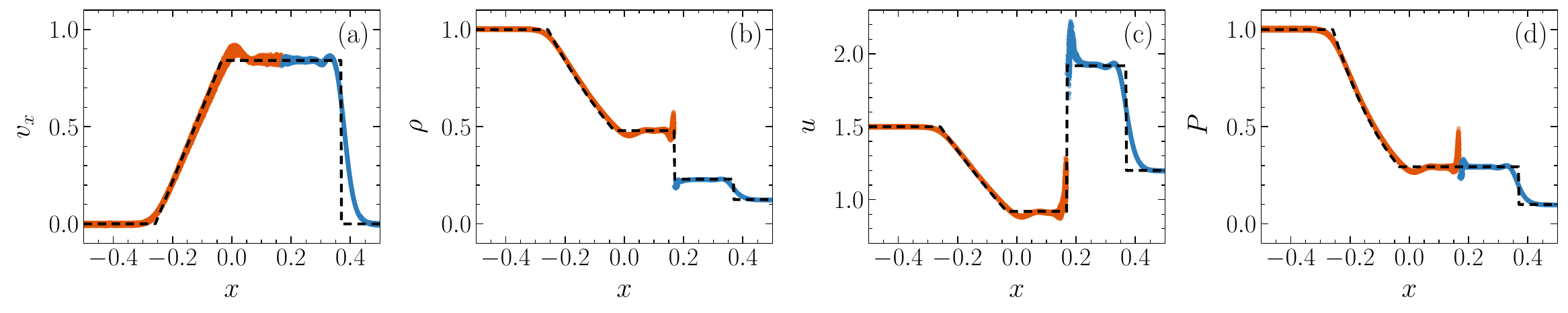}
}\hfill
	\vspace{-2em}
	\caption{REMIX Sod shock tube at time $t = 0.2$ in a simulation where artificial diffusion has been disabled between particles on opposite sides of $x=0$ in the initial conditions, to function as if representing distinct materials. Plots show velocity in the $x$-direction, $v_x$, density, $\rho$, specific internal energy, $u$, and pressure, $P$, of individual particles plotted against their $x$-position. Particles with initial positions $x<0$ are plotted in orange and those with $x>0$ in blue. The black, dashed line shows a reference solution, solved for directly by using a Riemann solver. All particles are plotted.}
	\label{fig:sodshock_mat_interface}
\end{figure}

In Fig.~\ref{fig:kh_cond}, we show snapshots from ideal gas KHI simulations in the simplified shear case. At high diffusion, the density discontinuity is smoothed and a sharp interface is not maintained. However, some diffusion is helpful in stabilising the evolution of the instability, as we see that with little or no diffusion the structure in the inner regions of the vortex is dominated by particle noise.

Again, for the REMIX scheme we take a conservative approach and choose a diffusion model that, with the assumptions of the simplified approaches considered here, switches between: \ref{fig:sodshock_noB_diff}(d) and (i) for shocks, where spikes in density and internal energy have mostly been removed; to Fig.~\ref{fig:kh_cond}(c), in which diffusion helps to stabilise the small-scale structure of the instability, but density discontinuities are allowed to persist. We note that, in practice, the artificial diffusion will be slightly between the two cases we aim to switch between. This was kept in mind when making this choice. This corresponds to a choice of  $a_{\rm difn} = 0.05$, $b_{\rm difn} = 0.95$. We also note that this can be seen as switching from a weak diffusion away from shocks with a factor of 0.05, similar to that of \citet{rosswog2020lagrangian}, to a stronger diffusion in shocks with a factor of 1, similar to \citet{price2018phantom}.

The Sod shock tube simulations we have considered above included particles of only a single ideal gas EoS. In REMIX, we deliberately do not allow artificial diffusion, of either density or internal energy, between particles of different EoS. Therefore, to test the effectiveness of the artificial diffusion scheme at material interfaces, we consider a Sod shock tube, still consisting of only a single ideal gas with $\gamma = 5/3$, but where particles on either side of $x=0$ in the initial conditions are treated as different materials by the artificial diffusion scheme. This choice of initial interface position leads to the largest differences with the standard ``single-material'' case. The results of this simulation at time $t = 0.2$ are shown in Fig.~\ref{fig:sodshock_mat_interface}, where artificial diffusion is disabled between particles of different colours. The density and internal energy profiles show close similarities with the simulations plotted in Fig.~\ref{fig:sodshock_noB_diff} for which artificial diffusion has been reduced globally. The $v_x$ plot shows similar results to the single-material case: without significant ringing and without additional artefacts at the material interface. The spike in internal energy, and the resulting effect it has in the pressure profile, constitute the most significant deviation from the reference solution, however, we note that the size of the internal energy spike is comparable with that in the tSPH scheme (Fig.~\ref{fig:sodshock}(c)).

\section{Further Kelvin--Helmholtz results and figures}\label{app:kh_secondary}

As discussed in \S\ref{subsubsec:kh_idg_discontinuous}, a discontinuous shearing interface will be unstable to the growth of perturbations of all wavenumbers \citep{robertson2010computational}. In simulations of Kelvin--Helmholtz instabilities, the wavenumbers of modes that are allowed to grow is limited by the numerical resolution; modes of wavelengths of the particle-separation scale and shorter will not be resolved. Here we demonstrate how error-seeded secondary modes will inevitably grow in KHI simulations with higher-resolution REMIX. We note that although this means that it is impossible to reach a converged solution in this scenario such that we cannot quantitatively judge the accuracy of these results, the presence of secondary modes is still a positive sign that spurious surface tension-like effects are not dominating behaviour at density discontinuities.

\begin{figure}[t]
	\centering
{\includegraphics[width=\textwidth, trim={2.5mm 0mm 2.5mm 0mm}, clip]{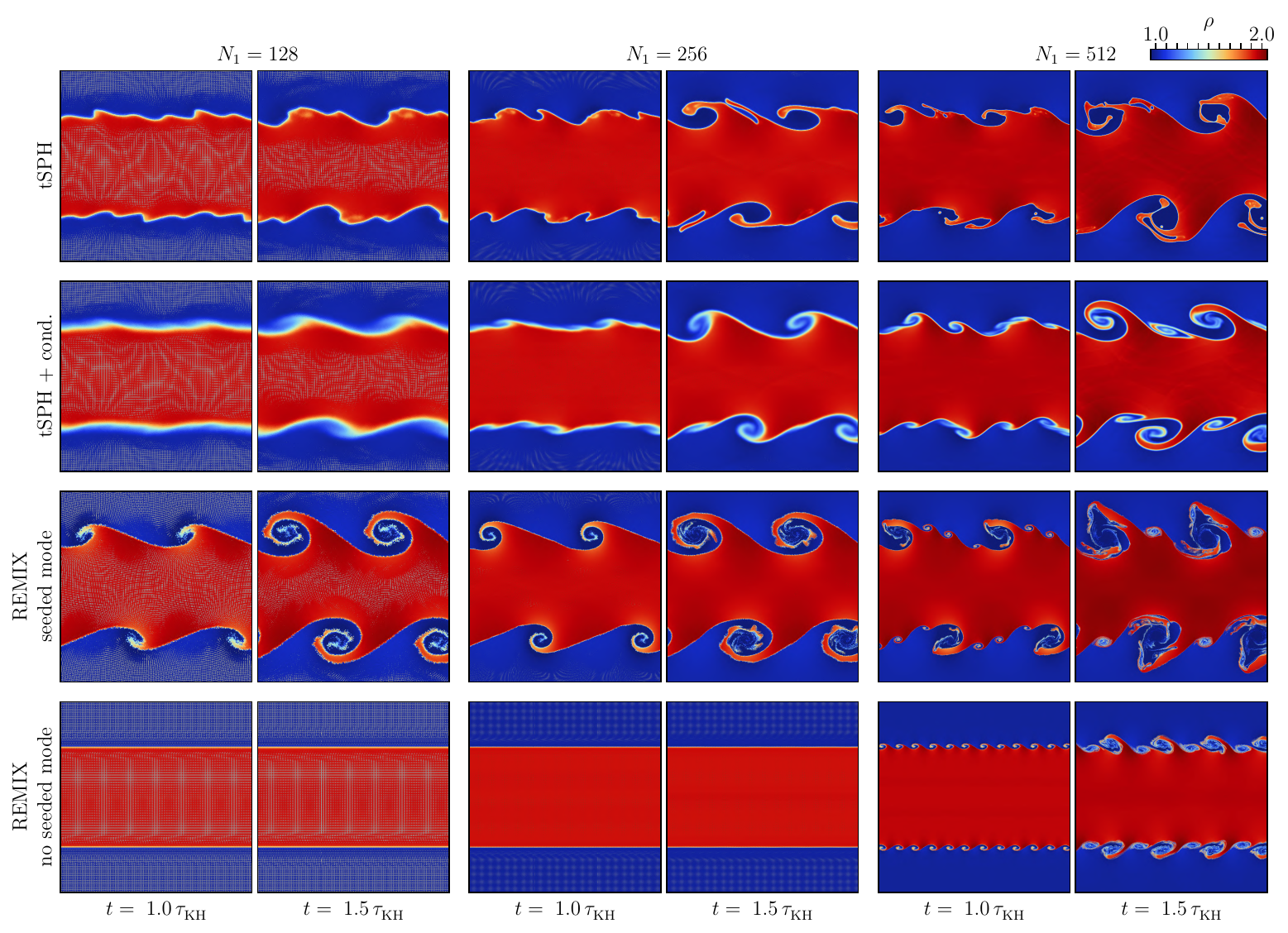}
}\hfill
	\vspace{-2em}
	\caption{The effect of resolution in ideal gas KHI simulations with sharp discontinuities that are unstable to perturbations of all wavelengths, carried out using tSPH, tSPH with conduction and REMIX. We also plot REMIX simulations without a deliberately seeded perturbation. The KHI is plotted at two times for simulations of three resolutions. Particles are coloured by their density.}
	\label{fig:kh_idg_resolutions}
\end{figure}

First, we consider ideal gas KHI simulations (\S\ref{subsubsec:kh_idg_discontinuous}). In Fig.~\ref{fig:kh_idg_resolutions}, we show snapshots from tSPH, tSPH with conduction, and REMIX simulations of KHIs at different resolutions. Additionally, in the bottom row of panels, we show how the shearing interface evolves with REMIX if no initial velocity perturbation is applied to the system. We see strong surface tension-like effects in the tSPH simulations for all resolutions. Artificial conduction is helpful as resolution is increased, however, the growth of the instability is still slow and the discontinuity becomes diffuse. Over these timescales, the $N_1=128$ and $N_1=256$ REMIX simulations are largely undisturbed by secondary modes, both in the cases with and without the seeded mode. In contrast, in the higher-resolution $N_1=512$ case, we see that secondary modes grow to greatly affect the evolution of the system, both with and without a deliberately seeded mode. The secondary modes grow over the same timescale in both these cases, demonstrating that these are purely seeded by error and noise in the numerical methods, rather than being associated with the growth of the primary mode. We note that over longer timescales than $\tau_{\rm KH}$, secondary modes will also grow in the lower-resolution simulations, although they do not greatly influence the early growth of the instability.

\begin{figure}[!b]
	\centering
{\includegraphics[width=\textwidth, trim={2.5mm 0mm 2.5mm 0mm}, clip]{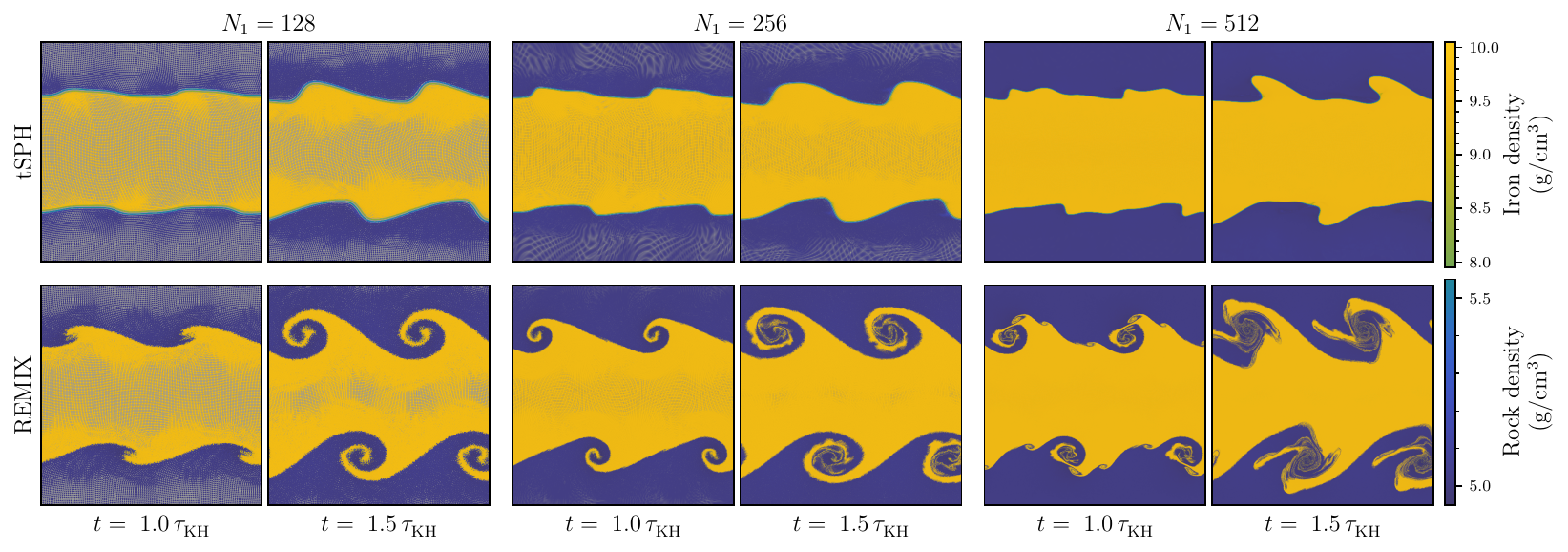}
}\hfill
	\vspace{-2em}
	\caption{The effect of resolution in Earth-like iron \& rock KHI simulations. Snapshots show two times for simulations of three resolutions, carried out using tSPH and REMIX. Particles are coloured by their material type and density.}
	\label{fig:kh_earth_resolutions}
\end{figure}

\begin{figure}[t]
	\centering
{
 \begin{center}
 \includegraphics[width=0.8\textwidth, trim={2.5mm 0mm 2.5mm 0mm}, clip]{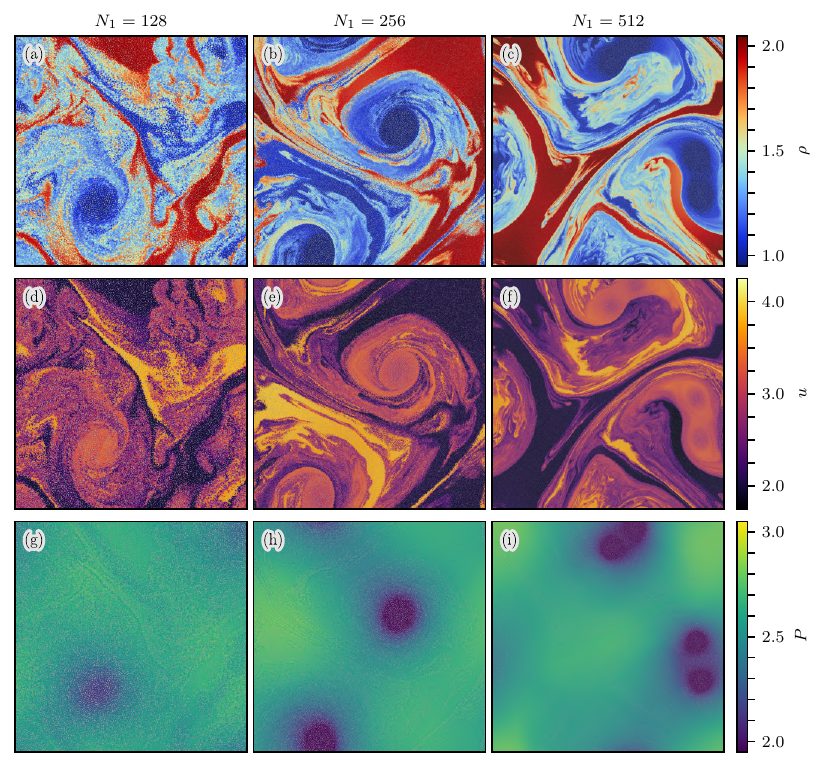}
  \end{center}
}\hfill
	\vspace{-2em}
	\caption{Ideal gas Kelvin--Helmholtz instabilities with sharp initial density and velocity profiles, and equal mass particles, at a later time of $t =8~\tau_{\rm KH}$. Columns correspond to simulations of different resolutions and rows show the density, $\rho$, specific internal energy, $u$, and pressure, $P$, of individual particles. The initial density ratio between the two regions is 1:2.}
	\label{fig:kh_idg_latetime}
\end{figure}

In Fig.~\ref{fig:kh_earth_resolutions}, we show similar snapshots from Earth-like KHI simulations at different resolutions with tSPH and REMIX, constructed equivalently to those presented in \S\ref{subsec:kh_earth}. Here surface tension-like effects are very strong in tSPH simulations at all resolutions. REMIX is able to deal with this challenging scenario, even at the lowest resolution simulated. Again, here we see error-seeded secondary modes in the $N_1 = 512$ REMIX simulation that grow to length scales where they play a significant role in the evolution of the system. In the lower-resolution simulations, these modes are not resolved and so the primary mode is undisturbed over these timescales.

Next, we consider the stability of REMIX KHI simulations over longer timescales. Snapshots at $\tau_{\rm KH} = 8$ from ideal gas KHI simulations with sharp interfaces at three different resolutions are shown in Fig.~\ref{fig:kh_idg_latetime}. The simulations shown are stable over these longer integration periods, with low pressure vortices being sustained, and the densities and internal energies of particles remaining well behaved and not freely evolving to unphysical values.

\section{Further planetary results and figures}\label{app:planets}

\begin{figure}[t]
	\centering
{\includegraphics[width=\textwidth, trim={2.5mm 2.5mm 2.5mm 2.5mm}, clip]{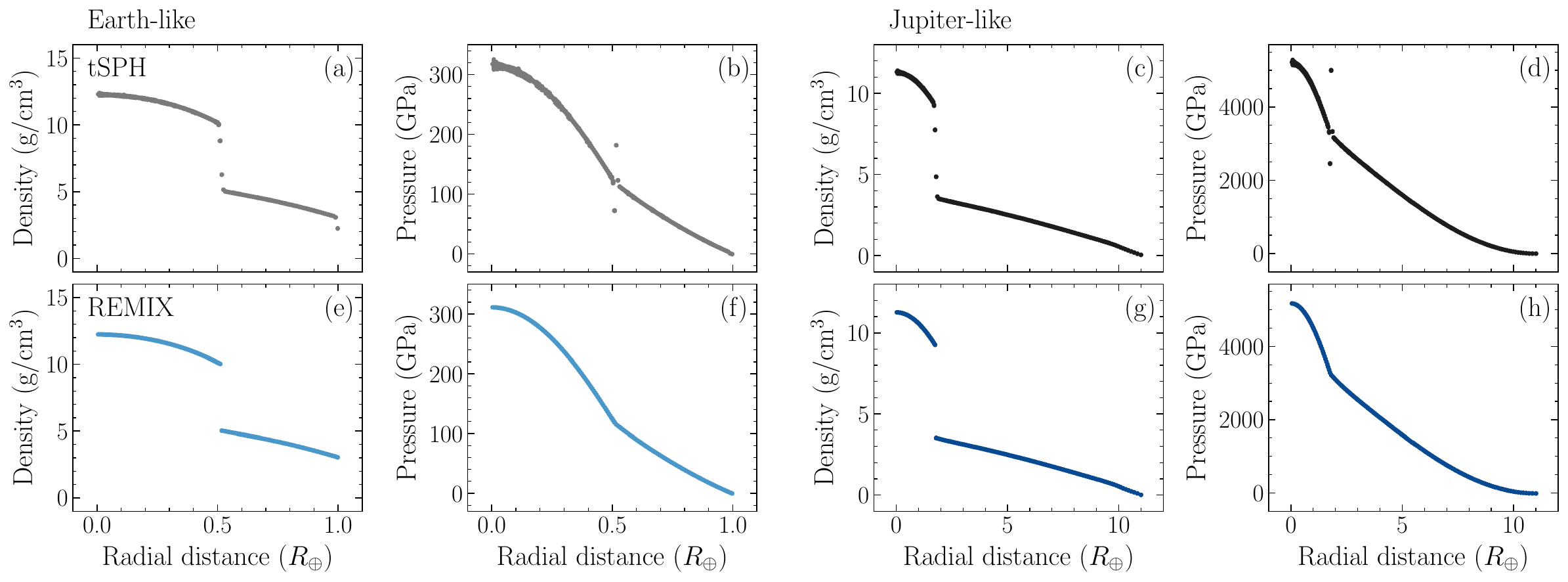}
}\hfill
	\vspace{-2em}
	\caption{Radial profiles of density and pressure for Earth-like (a, b, e, f) and Jupiter-like (c, d, g, h) planets at the initial time $t = 0$. Plots show profiles from simulations using tSPH (a--d) and REMIX (e--h).}
	\label{fig:planet_profiles_0}
\end{figure}

Here we present further results regarding the planetary settling examples (\S\ref{subsec:planet}) to expand on the issues faced in these simulations. First, in Fig.~\ref{fig:planet_profiles_0}, we show radial profiles, equivalent to those presented in Fig.~\ref{fig:planet_profiles}, but at the start of the simulation rather than at a time when particles have evolved from their initial configuration. Here we notice in the tSPH case that the blips in pressure are even larger, and that there is also noticeable smoothing of the density field at the vacuum boundary in the Earth-like planet. At the later times, plotted in Fig.~\ref{fig:planet_profiles}, particles have evolved to closer to equilibrium configurations where these issues do not appear to be extreme. However, we note that in a more kinematically interesting simulation with particles approaching the material interface, they will encounter the erroneous pressures that lead to surface-tension like effects. Therefore, it is not the case that as the planets relax, they reach a state where spurious surface tension-like forces disappear. We note that both material interfaces and the vacuum boundary are corrected, trivially at this initial time, by the evolved density estimate in REMIX.

\begin{figure}[t]
	\centering
{\includegraphics[width=\textwidth, trim={2.5mm 0mm 2.5mm 0mm}, clip]{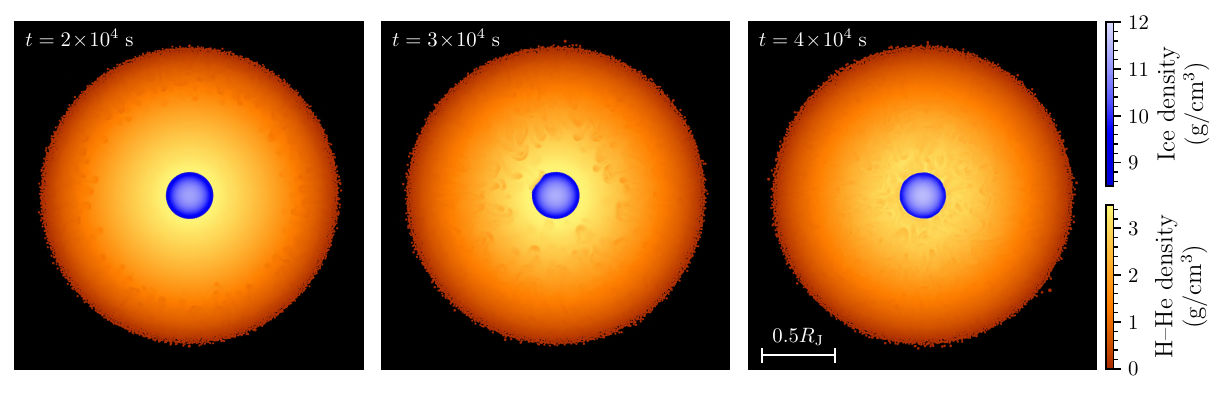}
}\hfill
	\vspace{-2em}
	\caption{Snapshots showing the evolution of spurious instabilities in a Jupiter-like planet from a simulation without the density evolution normalising term.}
	\label{fig:jupiter_nonormterm}
\end{figure}

In Fig.~\ref{fig:jupiter_nonormterm}, we show the evolution of a simulation of a Jupiter-like planet with REMIX, but without the kernel normalising term. These demonstrate how the instabilities seen in Fig.~\ref{fig:jupiter_methods} continue falling inwards at later times. As seen in Fig.~\ref{fig:jupiter_m0}, the densities of particles without the normalisation term are not tied to the distribution of mass in the simulation volume, leading to error driven instabilities. These grow to fully disturb the profile of the hydrogen--helium envelope, but are fully avoided in the full REMIX scheme.

\section{MFM and MFV comparisons}\label{app:mfm_mfv}

Here we present square test and KHI simulations, with initial conditions matching those presented in \S\ref{subsec:square} and \S\ref{subsubsec:kh_idg_discontinuous}, carried out using meshless finite-mass (MFM) and  meshless finite-volume (MFV) methods \citep{hopkins2015new}. We use the MFM and MFV implementations included in the \swift code. The main comparisons made in this study have been with traditional forms of SPH, where the primary aim has been to justify the choices made in the construction of the REMIX scheme and how each of these choices combine to improve on the more traditional approach. The aim of this section is to briefly demonstrate how these alternative methods deal with these particular 3D, equal-mass particle tests that focus on density discontinuities. 
%A more comprehensive comparison of REMIX with different, sophisticated hydrodynamics methods to applications like these would be desirable for future work.

Both MFM and MFV are Lagrangian hydrodynamics methods that make use of Riemann solvers and improved gradient estimates \citep{vila1999particle, gaburov2011astrophysical, hopkins2015new}. Where MFV allows the masses of particles to change based on mass fluxes, in MFM, the velocities of the effective surfaces between particles for flux calculations are set to move with (and thus cancel out) the flux of mass. Therefore, MFM is more easily adapted to simulations with multiple materials, since particles do not exchange mass and material, and so the material type of individual particles can remain fixed throughout the simulation. Although taking a different approach in their derivation from SPH schemes, the constituent equations of MFM and MFV \citep{hopkins2015new} remain similar to those of modern SPH schemes.

\begin{figure}[t]
	\centering
 {\includegraphics[width=\textwidth, trim={2.5mm 0mm 2.5mm 0mm}, clip]{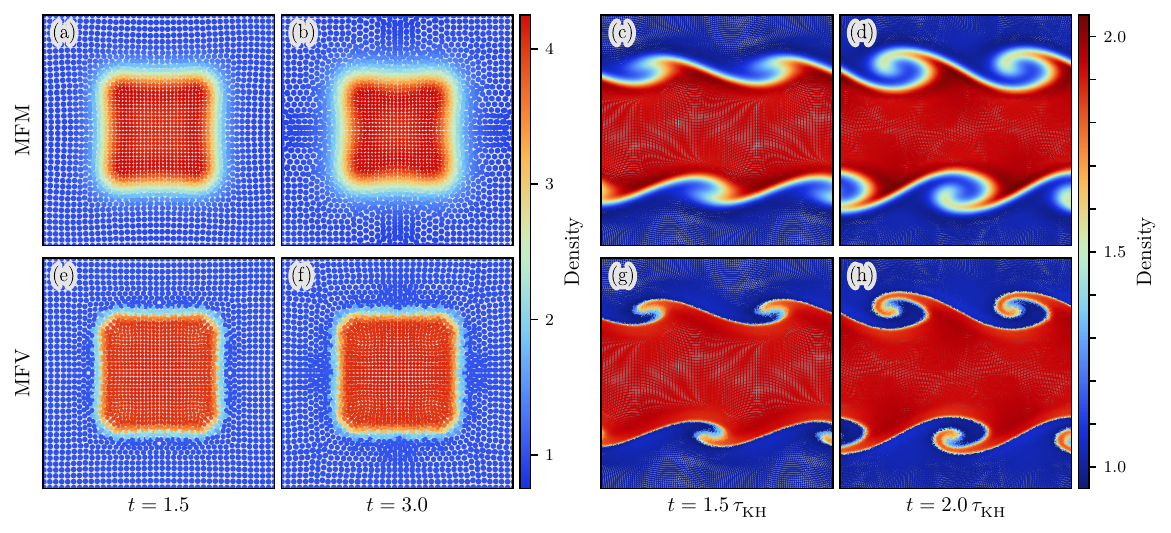}
 }\hfill
	\vspace{-2em}
	\caption{Meshless finite-mass (MFM; (a--d)) and finite-volume (MFV; (e--h)) simulations of a 3D square test (a, b, e, f) and a 3D Kelvin--Helmholtz instability with equal mass particles and initially sharp discontinuities (c, d, g, h). Snapshots at two times are plotted for each simulation. Individual particles are plotted and coloured by their densities. Kelvin--Helmholtz instability simulations have a resolution of $N_1 = 128$, as described in \S\ref{subsubsec:kh_idg_discontinuous}.}
	\label{fig:mfm_mfv}
\end{figure}

Fig.~\ref{fig:mfm_mfv}(a--d) show results from square test and KHI simulations carried out using MFM. In both simulations, discontinuities do not remain sharp and the evolution of both scenarios is more similar to the tSPH simulations with artificial conduction, shown in Figs.~\ref{fig:square} and \ref{fig:kh_idg}, than to the equivalent REMIX simulations. With MFV (Fig.~\ref{fig:mfm_mfv}(e--h)), sharper interfaces are maintained, however, there is still some smoothing of particle densities. In the REMIX simulations presented in \S\ref{subsec:square} and \S\ref{subsubsec:kh_idg_discontinuous} the discontinuities are sharper than with both these simulation methods, the cube more closely retains its original shape, and the KHI evolves over shorter timescales, indicating some suppression is still impactful for MFM and MFV.

\bibliographystyle{elsarticle-num-names} 
\bibliography{bibliography.bib}

\end{document}